\documentclass[mathpazo]{cicp}

\usepackage{geometry}
\usepackage{amsthm, amsmath, amssymb, amsfonts, amsbsy}
\usepackage{mathrsfs} 
\usepackage{stmaryrd} 
\usepackage{mathtools} 
\allowdisplaybreaks
\usepackage{indentfirst}
\usepackage{graphicx}
\usepackage{tikz-cd}
\usepackage{makeidx}
  \makeindex

\usepackage{pdfpages} 
\usepackage{longtable}
\usepackage{multirow}
\usepackage{xcolor}
  \definecolor{b-}{rgb}{0, 0, 0.5}
  \definecolor{r|b-}{rgb}{0.75, 0, 0.25}
  \definecolor{r+}{rgb}{1, 0.8, 0.8}
  \definecolor{g|b-}{rgb}{0, 0.75, 0.25}
\usepackage{ulem}
\newcommand{\mydel}[1]{}
\newcommand{\myadd}[1]{{#1}}
\usepackage[
  colorlinks,
  urlcolor = black,
  linkcolor = b-,
  anchorcolor = r|b-,
  citecolor = g|b-,
  hypertexnames = false 
]{hyperref}
\usepackage{extramarks}
\usepackage[fit]{truncate}
\usepackage{fancyhdr}
\usepackage{float}
\usepackage{silence}
\renewcommand{\(}{\left(}
\renewcommand{\)}{\right)}
\renewcommand{\[}{\left\lbrack}
\renewcommand{\]}{\right\rbrack}
\newcommand{\lsbsb}{\left\llbracket}
\newcommand{\rsbsb}{\right\rrbracket}
\newcommand{\lbk}{\left\lbrace}
\newcommand{\rbk}{\right\rbrace}

\newcommand{\lmdl}{\left\vert}
\newcommand{\rmdl}{\right\vert}

\newcommand{\tp}{^{T}} 

\newcommand{\bigO}[1]{\mathcal{O} \( {#1} \)} 

\newcommand{\mypower}[2]{%
  \ifthenelse{
    \( \equal{#2}{0} \)
  }{}{%
    \ifthenelse{
      \( \equal{#2}{1} \)
    }{#1}{{#1} ^{#2}}
  }
}
\newcommand{\mykgms}[3]{%
  \ifthenelse{
    \( \equal{#1}{0} \) \AND \( \equal{#2}{0} \) \AND \( \equal{#3}{0} \)
  }{%
    1}{}
  \ifthenelse{
    \( \equal{#1}{0} \) \AND \( \equal{#2}{0} \) \AND \( \NOT \equal{#3}{0} \)
  }{%
    \mypower{\textup{s}}{#3}}{}
  \ifthenelse{
    \( \equal{#1}{0} \) \AND \( \NOT \equal{#2}{0} \) \AND \( \equal{#3}{0} \)
  }{%
    \mypower{\textup{m}}{#2}}{}
  \ifthenelse{
    \( \equal{#1}{0} \) \AND \( \NOT \equal{#2}{0} \) \AND \( \NOT \equal{#3}{0} \)
  }{%
    \mypower{\textup{m}}{#2} \cdot \mypower{\textup{s}}{#3}}{}
  \ifthenelse{
    \( \NOT \equal{#1}{0} \) \AND \( \equal{#2}{0} \) \AND \( \equal{#3}{0} \)
  }{%
    \mypower{\textup{kg}}{#1}}{}
  \ifthenelse{
    \( \NOT \equal{#1}{0} \) \AND \( \equal{#2}{0} \) \AND \( \NOT \equal{#3}{0} \)
  }{%
    \mypower{\textup{kg}}{#1} \cdot \mypower{\textup{s}}{#3}}{}
  \ifthenelse{
    \( \NOT \equal{#1}{0} \) \AND \( \NOT \equal{#2}{0} \) \AND \( \equal{#3}{0} \)
  }{%
    \mypower{\textup{kg}}{#1} \cdot \mypower{\textup{m}}{#2}}{}
  \ifthenelse{
    \( \NOT \equal{#1}{0} \) \AND \( \NOT \equal{#2}{0} \) \AND \( \NOT \equal{#3}{0} \)
  }{%
    \mypower{\textup{kg}}{#1} \cdot \mypower{\textup{m}}{#2} \cdot \mypower{\textup{s}}{#3}}{}
}
\newcommand{\e}[1]{\times 10 ^{#1}}

\counterwithin{equation}{section}
\counterwithin{figure}{section}
\counterwithin{table}{section}

\usepackage{lineno}

\begin{document}
\title{Proper Orthogonal Decomposition-based Model-Order Reduction for Smoothed Particle Hydrodynamics Simulation --- Mass-Spring-Damper System}


\author[Fang et.~al.]{
  Lidong Fang\affil{1}, Zilong Song\affil{2}, Kirk Fraser\affil{3}, and Huaxiong Huang\affil{4}\affil{5}\comma\corrauth}
\address{
  \affilnum{1}\ School of Mathematics, Shanghai University of Finance and Economics, Shanghai, 200433, China. \\
  \affilnum{2}\ Department of Mathematics and Statistics, Utah State University, Logan, 84322, USA. \\
  \affilnum{3}\ National Research Council Canada, Saguenay, G7H 8C3, Canada. \\
  \affilnum{4}\ Zu Chongzhi Center, Duke Kunshan University, 8 Duke Ave., Suzhou, 215316, China. \\
  \affilnum{5}\ Department of Mathematics and Statistics, York University, Toronto, M3J 1P3, Canada. \\
  }
\emails{{\tt fanglidong@sufe.edu.cn} (L.~Fang), {\tt zilong.song@usu.edu} (Z.~Song), {\tt kirk.fraser@cnrc-nrc.gc.ca} (K.~Fraser), {\tt hhuang@yorku.ca} (H.~Huang)}

\begin{abstract}
  Model Order Reduction (MOR) based on Proper Orthogonal Decomposition (POD) and Smooth Particle Hydrodynamics (SPH) has proven effective in various applications. Most MOR methods utilizing POD are implemented within a pure Eulerian framework, while significantly less attention has been given to POD in a Lagrangian context.
  In this paper, we present the POD-MOR of SPH simulations applied to a mass-spring-damper system with two primary objectives:
  1. To evaluate the performance of the data-driven POD-MOR approach.
  2. To investigate potential methods for accelerating POD-MOR computations.
  Although the mass-spring-damper system is linear, its SPH implementations are nonlinear, and POD-MOR does not automatically lead to faster computations. Our findings indicate that (1) the POD-MOR effectively reduces the degrees of freedom in the SPH simulations by capturing the essential modes, and (2) in various cases, the acceleration of POD-MOR can be achieved without compromising accuracy. We hope that our results will motivate further investigations into the design of POD-MOR algorithms for nonlinear Lagrangian systems.
\end{abstract}

\pac{02.70.-c, 46.15.-x, 47.11.-j, 47.85.Dh}

\maketitle

\section{Introduction}


Model order reduction (MOR) is important in engineering and science because it allows for efficient simulations by significantly lowering the degrees of freedom (DoFs) while keeping the reduction error small \cite{lassila2014model, lucia2004reduced, peherstorfer2015dynamic}. Physical simulations can be expensive due to their complexity, especially in cases like parameter studies or design optimization that require multiple simulations, where MOR is particularly helpful \cite{copeland2022reduced}.

The most common MOR method uses projection techniques \cite{benner2015survey}, which approximate state variables in a lower-dimensional space \cite{mojgani2017lagrangian}. A typical projection-based MOR has two parts: an offline phase that reduces dimensions by identifying the key modes in the dynamics, and an online phase that calculates approximate solutions \cite{magargal2022lagrangian}. One leading method for offline dimensional reduction is Proper Orthogonal Decomposition (POD), see \cite{taira2017modal} for other methods. In the online phase, either Galerkin projection or Petrov-Galerkin projection can be used \cite{rowley2004model, rapun2010reduced, carlberg2017galerkin, choi2019space, carlberg2011efficient}.

The POD method is based on singular value decomposition that can automatically find the most important spatial modes in a dataset, which form the reduced basis for the exact solution. By capturing the key features of the system, POD allows for the representation of complex phenomena with far fewer DoFs. This not only reduces computational costs but also improves understanding of the underlying physics \cite{chatterjee2000introduction, berkooz1993proper, kerschen2005method}. The effectiveness of POD is especially clear when applied to the heat equation on a fixed grid. This method works well in this context due to the fast decay of high-frequency modes, so that the number of DoFs (i.e., the number of important low-frequency modes) can be effectively reduced. Another example is the complex Friction Stir Welding (FSW) problem solved using the finite difference method in an Eulerian framework, where POD can be applied to achieve MOR and improve computational efficiency \cite{cao2022machine}.


Recently, the mesh-free Lagrangian framework has been used to solve a variety of problems, particularly those with complex boundary conditions and large deformations. This makes mesh-free methods within the Lagrangian framework, such as Smoothed Particle Hydrodynamics (SPH) \cite{monaghan1992smoothed, monaghan2005smoothed, liu2010smoothed}, well-suited for FSW simulations. However, it is still not clear how POD-based MOR (POD-MOR) performs in the SPH simulations under the Lagrangian framework.

In \cite{mojgani2017lagrangian}, the first application of POD-MOR for mesh-based Lagrangian simulations is presented, showing better performance than the Eulerian framework in some one-dimensional convection-dominated solutions, as it is difficult to efficiently represent these solutions using global Eulerian spatial and temporal basis functions. In one spatial dimension, the algorithm requires interpolation back to the Eulerian framework when grid entanglement occurs due to the mesh-based approach. Another recent study that applies POD to a finite element method in a mesh-based Lagrangian framework is \cite{copeland2022reduced}, where techniques like time-windowing and hyper-reduction are used to speed up the computation of reduced models in two and three dimensions.

Mesh-free methods like SPH are quite different from above mesh-based methods and pose extra computational challenges because of the mesh distortion and remeshing. Few studies have focused on the POD-MOR of SPH simulations. A recent study \cite{magargal2022lagrangian} runs SPH simulations for a fluid dynamic system and constructs POD modes by developing a post-processing Lagrangian-to-Eulerian mapping. After this mapping, the interpolated Eulerian modes can capture coherent model structures effectively. The authors note that the irregular SPH solutions make it hard for traditional decomposition methods to find coherent and low-dimensional structures in the dynamic system.

A recent work \cite{fang0000proper} has studied a 2D Friction Stir Spot Welding problem by SPH simulation, and then the POD-MOR is applied to reduce the DoF.  With the same DoFs, POD-MOR shows significantly lower errors, compared with uniform reduction of particle numbers in traditional coarsening methods. Some POD modes are illustrated in the Lagrangian framework, showing that POD can achieve efficient and accurate representations of physical phenomena. However, that work mainly focuses on reducing DoFs in the offline phase, rather than improving computational time in the online phase. The possibility of speeding up POD-MOR of SPH simulation has not yet been explored due to the complexity of the system. This motivates us to a systematic study of the effectiveness and speed-up based POD-MOR for SPH simulations, over a simpler setting. This will provide insights into the challenges and techniques for the speed-up using POD-MOR for other SPH simulations.

In this paper, we examine both aspects of POD-MOR: whether reducing DoFs maintains accuracy and whether computational costs can be lowered using POD for a Lagrangian SPH formulation of a 1D mass-spring-damper system. Our findings suggest that both can be achieved under certain conditions.

The manuscript is arranged as follows. In \autoref{sec:model}, the formulation of the 1D mass-spring-damper system, its SPH formulation as well as the POD-MOR and acceleration will be given.
\autoref{sec:numerical} shows numerical results with various initial conditions and parameter settings, explaining the effective of POD-MOR and its acceleration. Finally, the conclusions are provided in \autoref{sec:conclusion}.

\section{Model}
\label{sec:model}

\subsection{Mass-spring-damper Model}
\label{sec:model__MSD}

\begin{figure}[!htp]
  \centering
  \includegraphics[width = 0.8\textwidth]{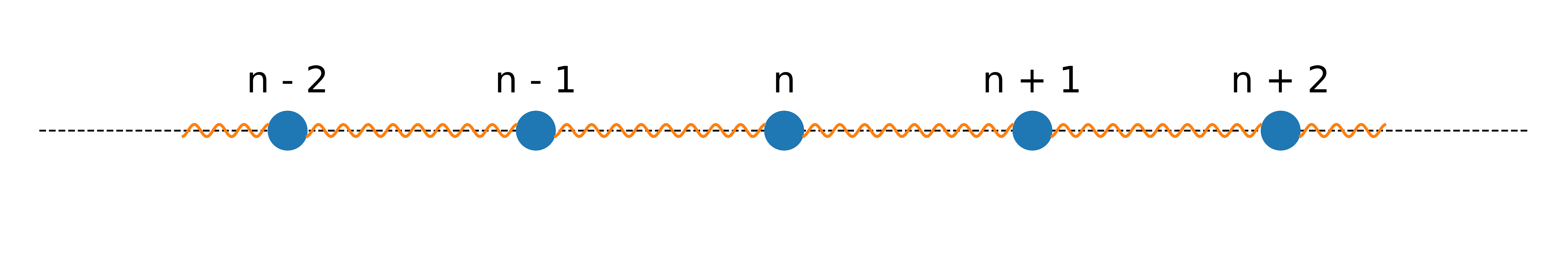}
  \caption{Illustration of the mass points (in blue) and springs (in orange). The mass points are labeled by numbers.}
  \label{fig:model__spring__illustration}
\end{figure}

We consider a one-dimensional mass-spring-damper system in the Lagrangian framework, \myadd{which is more straightforward than that in the Eulerian framework}.
The system consists of a one-dimensional chain where alternating springs and mass points are connected. It assumes motion occurs in one spatial direction.
Each mass point experiences forces from the two adjacent springs --- one on the left and one on the right --- creating a coupled structure, see \autoref{fig:model__spring__illustration} for illustration. 
The damping force acts on each mass point and is proportional to its velocity, but in the opposite direction.

Now, we model this system mathematically. For a mass point $n$, with mass $m _{n} \mydel{\in \mathbb{R} _{+} \[ \mykgms{1}{0}{0} \]}$, we define its position at time $t$ as $y _{n} \mydel{\in \mathbb{R} \[ \mykgms{0}{1}{0} \]}$.
The governing equation of motion is given by
\begin{align}
  m _{n} y _{n} '' \( t \)
  & =
  -
  d _{n} \( t \) y _{n} ' \( t \)
  +
  k _{n - \frac{1}{2}} y _{n - 1} \( t \)
  -
  \[ k _{n - \frac{1}{2}} + k _{n + \frac{1}{2}} \] y _{n} \( t \)
  +
  k _{n + \frac{1}{2}} y _{n + 1} \( t \)
  +
  m _{n} f _{n} \( t \),
  \label{eq:model__spring__Y_discrete}
\end{align}
where $d _{n} \mydel{\in \mathbb{R} _{+} \[ \mykgms{1}{0}{-1} \]}$ is the damping coefficient,
$k _{n + \frac{1}{2}} \mydel{\in \mathbb{R} _{+} \[ \mykgms{1}{0}{-2} \]}$ is the spring constant between particles $n$ and $n + 1$,
and $f _{n} \mydel{\in \mathbb{R} \[ \mykgms{0}{1}{-2} \]}$ is the external force (per mass).
This system can also be regarded as a finite difference discretization (in space) of an elastic solid.

\myadd{Although it is possible to model an elastic solid in the Eulerian framework, in the current work, we aim to investigate the POD-MOR for a Lagrangian formulation so that the Lagrangian framework will be used in the following modeling.} Motivated by \autoref{eq:model__spring__Y_discrete}, we will study its continuum limit (see \autoref{sec:appendix_equivalence} for more details) \myadd{in terms of position and velocity, which is then} written as a system in terms of density and velocity --- more suitable for SPH setup --- in the following. 
The spatial domain for $x$ is between $0 \mydel{\[ \mykgms{0}{1}{0} \]}$ and $L \mydel{\in \mathbb{R} _{+} \[ \mykgms{0}{1}{0} \]}$, and the temporal domain for $t$ is between $0 \mydel{\[ \mykgms{0}{0}{1} \]}$ and $T \mydel{\in \mathbb{R} _{+} \[ \mykgms{0}{0}{1} \]}$.
The mass density field $\rho \( x, t \) \mydel{: \Omega \times \Theta \to \mathbb{R} _{+} \[ \mykgms{1}{-3}{0} \]}$ and the velocity field $v \( x, t \) \mydel{: \Omega \times \Theta \to \mathbb{R} \[ \mykgms{0}{1}{0} \]}$ satisfy
\begin{align}
  \partial _{t} \rho \( x, t \)
  & =
  - \[ \rho \( x, t \) \] ^{2} \[ \rho _{0} \( x \) \] ^{-1} \partial _{x} v \( x, t \)
  ,
  \label{eq:model__spring__VD__D}
  \\
  \rho _{0} \( x \)
  \partial _{t} v \( x, t \)
  & =
  \partial _{x} \[ c \( x, t \) \rho _{0} \( x \) \[ \rho \( x, t \) \] ^{-1} - c \( x, t \) \]
  +
  \rho _{0} \( x \) b \( x, t \)
  ,
  \label{eq:model__spring__VD__V}
\end{align}
where $c \mydel{: \Omega \times \Theta \to \mathbb{R} _{+} \[ \mykgms{1}{-1}{-2} \]}$ is the Young's modulus, $b := -dv + f$ is the body force per mass, $d \mydel{: \Omega \times \Theta \to \mathbb{R} _{+} \[ \mykgms{0}{0}{-1} \]}$ is the damping coefficient, and $f\mydel{: \Omega \times \Theta \to \mathbb{R} \[ \mykgms{0}{1}{-2} \]}$ is the external force per mass.

For the above PDE system, we associate the following initial conditions and the periodic boundary conditions for $\rho$ and $v$.
\begin{align}
  \rho \( x, 0 \) = \rho _{0} \( x \),
  \quad
  v \( x, 0 \) = v _{0} \( x \),
  \quad
  \rho \( 0, t \) = \rho \( L, t \),
  \quad
  v \( 0, t \) = v \( L, t\).
  \label{eq:model__spring__VD__BC}
\end{align}

To solve this model numerically, standard methods such as finite difference or finite elements can be used. 
In this paper, we use them as a model problem to explore SPH formulation and POD-based model reduction techniques.
Any numerical ODE schemes can be used for temporal integration, and the time step size $\Delta t$ is chosen based on the Courant–Friedrichs–Lewy (CFL) condition and the speed of sound:
\begin{align}
  v _{\textup{sound}} 
  = 
  \sqrt{\frac{\partial p}{\partial \rho}}
  =
  \sqrt{\frac{c \rho _{0}}{\rho ^{2}}}
  .
\end{align}
Here $p := c - c \rho _{0} \rho ^{-1}$ can be regarded as a potential term defined from \autoref{eq:model__spring__VD__V}.

\subsection{SPH in Lagrangian Framework}

SPH is a mesh-free particle method based on the Lagrangian formulation and has been widely applied in various areas of engineering and science. In the 1D mass-spring-damper system within the Lagrangian framework, the velocity $v$ and density $\rho$ need to be solved for each particle through the governing equations in the SPH formulation.

We follow the standard SPH formulation, which consists of two main steps.
The first step, kernel approximation, involves representing a function and its derivatives in continuous form as integrals over $\Omega$.
The second step, particle approximation, rewrites these integrals over the computational domain as summations over discrete particles, indexed by $\Lambda := \lbk 1, 2, \cdots, N \rbk$.
This approach allows us to derive the particle evolution equations as a dynamical system for velocity $v _{i}$ and density $\rho _{i}$ for all particles $i \in \Lambda$ in the 1D mass-spring-damper system.

In the kernel approximation step, we define a 1D smoothing kernel function as a widely-used cubic spline function \cite{monaghan2005smoothed} $W _{h}\mydel{: \mathbb{R} \[ \mykgms{0}{1}{0} \] \to \mathbb{R} \[ \mykgms{0}{1}{0} ^{-3} \]}$, with the smoothing length $h \mydel{\in \mathbb{R} _{+} \[ \mykgms{0}{1}{0} \]}$.
For all $x \mydel{\in \mathbb{R} \[ \mykgms{0}{1}{0} \]}$, we have:
\begin{align}
  W _{h} \( x \)
  & :=
  \frac{2}{3h a ^{2}}
  \left\{
  \begin{aligned}
    & 
    1 - \frac{3}{2} \[ \frac{\lmdl x \rmdl}{h} \] ^{2} + \frac{3}{4} \[ \frac{\lmdl x \rmdl}{h} \] ^{3}, 
    && 
    0 \leq \frac{\lmdl x \rmdl}{h} \leq 1,
    \\
    & 
    \frac{1}{4} \[ 2 - \frac{\lmdl x \rmdl}{h} \] ^{3}, 
    && 
    1 \leq \frac{\lmdl x \rmdl}{h} \leq 2,
    \\
    & 
    0, 
    && 
    2 \leq \frac{\lmdl x \rmdl}{h}.
  \end{aligned}
  \right.
  \label{eq:model__SPH__cubicspline}
\end{align}
Here, $a = 1 \mydel{\[ \mykgms{0}{1}{0} \]}$ is the unit length in the spatial domain.
The kernel function $W _{h}$ has a finite support of interval of length $2h$ and approaches a Delta function as $h \to 0$.

The corresponding derivative is
\begin{align}
  \nabla W _{h} \( x \)
  & =
  \frac{2}{3h ^{2} a ^{2}}
  \left\{
  \begin{aligned}
    & 
    -3 \[ \frac{x}{h} \] + \frac{9}{4} \[ \frac{x}{h} \] ^{2}, 
    && 
    0 \leq \frac{x}{h} \leq 1,
    \\
    & 
    -\frac{3}{4} \[ 2 - \frac{x}{h} \] ^{2}, 
    && 
    1 \leq \frac{x}{h} \leq 2,
    \\
    & 
    -3 \[ \frac{x}{h} \] - \frac{9}{4} \[ \frac{x}{h} \] ^{2}, 
    && 
    -1 \leq \frac{x}{h} \leq 0,
    \\
    & 
    \frac{3}{4} \[ 2 + \frac{x}{h} \] ^{2}, 
    && 
    -2 \leq \frac{x}{h} \leq -1,
    \\
    & 
    0, 
    && 
    2 \leq \frac{\lmdl x \rmdl}{h}.
  \end{aligned}
  \right.
\end{align}

For an arbitrary Lagrangian particle function $f _{i} \( t \)$, $i \in \Lambda$, we define the SPH approximation as:
\begin{align}
  \lsbsb f \rsbsb _{i} \( t \)
  :=
  \sum _{j \in \Lambda}
  \frac{m _{j}}{\rho _{j} \( 0 \)} 
  f _{j} \( t \) 
  W _{h} \( x _{ij} \( t \) \).
\end{align}
The first Lagrangian spatial derivative is given by:
\begin{align}
  \lsbsb \nabla f \rsbsb _{i} \( t \)
  :=
  \sum _{j \in \Lambda}
  \frac{m _{j}}{\rho _{j} \( 0 \)}
  f _{j} \( t \)
  \nabla W _{h} \( x _{ij} \( t \) \),
\end{align}
Here, we denote $x _{ij} \( t \) := x _{i} \( t \) - x _{j} \( t \)$, for any particle function $g$. 

Now, the 1D mass-spring-damper governing equations \autoref{eq:model__spring__VD__D} and \autoref{eq:model__spring__VD__V} can be expressed in SPH formulation, where neighbor search is utilized to determine the neighbors within kernel support for each particle. We use the notation $\sum _{j \sim i}$ to indicate the summation over $j \in \Lambda$ such that $\lmdl x _{ij} \rmdl \le 2h$, ensuring that the kernel function is nonzero. We implement the SPH simulation in the Lagrangian framework, where the smoothed particles are fixed on their reference position $\dot{x} _{i} = 0$. That means the kernel function $W _{ij}$ and its derivative $\nabla W _{ij}$ are time-independent since particle positions of $\lbk x _{i} \rbk _{i \in \Lambda}$ are fixed. In the following, we denote $W _{ij} := W \( x _{ij} \)$, and $\nabla W _{ij} := \nabla W \( x _{ij} \)$.

\myadd{Here is the SPH formulation for the mass-spring-damper system.}
From \autoref{eq:model__spring__VD__D}, the update of particle density is:
\begin{align}
  \dot{\rho} _{i}
  \approx
  - \frac{\rho _{i} ^{2}}{\rho _{i} \( 0 \)} \sum _{j \sim i} \frac{m _{j}}{\rho _{j} \( 0 \)} v _{j} \nabla W _{ij}
  .
  \label{eq:model__SPH__D}
\end{align}
From \autoref{eq:model__spring__VD__V}, the update of particle velocity is:
\begin{align}
  \dot{v} _{i}
  & 
  \approx
  -
  \frac{1}{\rho _{i} \( 0 \)} \[ \sum _{j \sim i} \frac{m _{j}}{\rho _{j} \( 0 \)} c _{j} \[ 1 - \frac{\rho _{j} \( 0 \)}{\rho _{j}} \] \nabla W _{ij} \]
  -
  d _{i} v _{i}
  +
  f _{i}
  .
  \label{eq:model__SPH__V}
\end{align}
The periodic boundary condition is applied for SPH particles. When a particle exits one side of the region, i.e., $x _{i} < 0$ or $x _{i} > L$, it reappears on the opposite side, i.e., $x _{i} \leftarrow x _{i} \pm L$, carrying the same density and velocity. When determining the neighbors of a particle, we also consider the particles across the boundary as if the spatial domain $\[ 0, L \]$ is periodically extended from both sides.

\subsection{POD}


The projection-based MOR typically consists of two phases: the offline dimensional compression phase, which identifies key features in the dynamics from the data, and the online evolution phase, where state variables are approximated in a low-dimensional subspace, allowing for the calculation of approximate solutions \cite{magargal2022lagrangian}.

Assuming we are solving a discrete time-dependent problem, the numerical solution $\mathbf{w} ^{m}$ at time step $t ^{m}$ ($m = 1, 2, \cdots, M$) serves as an approximation of the true solution $w \( \mathbf{x}, t ^{m} \)$,
\begin{align}
  \mathbf{w} ^{m} = \( w _{1} ^{m}, \cdots, w _{N} ^{m} \) \tp \approx \( w \( x _{1}, t ^{m} \), \cdots, w \( x _{N}, t ^{m} \) \) \tp \in \mathbb{R} ^{N},
\end{align}
where $N$ is the DoF of the discretized problem and $\mathbf{x} = \( x _{1}, \cdots, x _{N} \)$ is the discretized spatial coordinate. To explain the projection-based MOR, the numerical scheme is assumed to be one-step for simplicity, 
\begin{align}
  \mathbf{R} \( \mathbf{w} ^{m}, \mathbf{w} ^{m - 1} \) = \mathbf{0}, 
  \quad 
  m = 1, \cdots, M,
\end{align}
but the generalization to multi-step schemes can be done similarly. 


During the offline stage, we generate a matrix $\mathbf{U} \in \mathbb{R} ^{N \times k}$, $k \ll N$, and its columns form a basis for the solution subspace. This allows us to approximate the numerical solution $\mathbf{w} ^{m}$ with the reduced-order solution $\tilde{\mathbf{w}} _{\mathbf{U}} ^{m}$ via
\begin{align}
  \mathbf{w} ^{m} 
  \approx 
  \tilde{\mathbf{w}} _{\mathbf{U}} ^{m} 
  = 
  \mathbf{U} \cdot \mathbf{U} \tp \cdot \mathbf{w} ^{m} 
  = 
  \mathbf{U} \cdot \mathbf{a} ^{m}
  , 
  \quad 
  m = 1, \cdots, M
  ,
\end{align}
where $\mathbf{a} ^{m} \in \mathbb{R} ^{k}$ is the reduced coordinates in the subspace basis. The POD is used to construct the subspace basis $\mathbf{U} \in \mathbb{R} ^{N \times k}$ from the data of snapshots \myadd{---} matrix $\mathbf{X} = \( \mathbf{w} ^{0}, \cdots, \mathbf{w} ^{M} \)$. This is done by  applying singular value decomposition (SVD) to $\mathbf{X}$,
\begin{align}
  \mathbf{X} = \hat{\mathbf{U}} \cdot \hat{\boldsymbol{\Sigma}} \cdot \hat{\mathbf{V}} \tp,
  \label{eq:model__POD__SVD}
\end{align}
and $\mathbf{U}$ is defined as the first $k$ columns of $\hat{\mathbf{U}}$, since the singular values in $\hat{\boldsymbol{\Sigma}}$ are ordered by significance. The above process is also equivalent to an optimization problem to find the best low-rank approximation for $\mathbf{X}$ \cite{chatterjee2000introduction, berkooz1993proper, kerschen2005method}.
The POD method automatically identifies the most dominant spatial modes so that it will provide deeper insights into the underlying physics.

In the online stage, since we have an over-determined system for solving the reduced coordinates, we multiply a  basis matrix $\boldsymbol{\Phi} \in \mathbb{R} ^{N \times k}$ to get
\begin{align}
  \boldsymbol{\Phi} \tp \cdot \mathbf{R} \( \tilde{\mathbf{w}} _{\mathbf{U}} ^{m}, \tilde{\mathbf{w}} _{\mathbf{U}} ^{m - 1} \) 
  =
  \boldsymbol{\Phi} \tp \cdot \mathbf{R} \( \mathbf{U} \cdot \mathbf{a} ^{m}, \mathbf{U} \cdot \mathbf{a} ^{m - 1} \)
  = 
  \mathbf{0}
  .
  \label{eq:model__POD__general}
\end{align}
Here, we use the Galerkin projection, i.e., $\boldsymbol{\Phi} = \mathbf{U}$. 

For a linear system, the above equation gives a low-dimensional system. But for general nonlinear systems, directly computing (\autoref{eq:model__POD__general}) requires the calculation of vectors in high-dimensional space, for example, the calculation of $\mathbf{U} \cdot \mathbf{a} ^{m - 1} \in \mathbb{R} ^{N}$.
This prevents the speed-up of calculation, even though the dimension has been reduced. For SPH simulations, the scheme operator $\mathbf{R}$ is more complicated as the use of kernel functions in mesh-free settings introduces extra nonlinearity. In this work, the original POD-MOR implementation is the \autoref{eq:model__POD__general} applying to the governing equations \autoref{eq:model__SPH__D} and \autoref{eq:model__SPH__V} for $\rho$ and $v$ respectively. 
\myadd{In the next section, we will propose an acceleration scheme for this POD-MOR.}

\subsection{POD Acceleration of SPH in Lagrangian Framework}
\label{sec:acceleration}

To accelerate POD-MOR, we assume that the particles remain close to their reference positions during the evolution. 
We first apply the approximation with density linearization
\begin{align}
\label{eq:density_linearization}
  1 - \frac{\rho _{j} \( 0 \)}{\rho _{j}}
  \approx
  \frac{\rho _{j}}{\rho _{j} \( 0 \)} - 1
  ,
\end{align}
which holds because $\rho _{j} \( t \) \approx \rho _{j} \( 0 \)$ for $t \in \Theta$.

Next, we stack all particle velocities $\lbk v _{i} \rbk _{i \in \Lambda}$ into a vector $\mathbf{v} \in \mathbb{R} ^{N} \mydel{\[ \mykgms{0}{1}{-1} \]}$, and denote the corresponding POD matrix and reduced coordinate as $\mathbf{U} ^{\mathbf{v}} \in \mathbb{R} ^{N \times k}$ and $\mathbf{a} ^{\mathbf{v}} \in \mathbb{R} ^{k} \mydel{\[ \mykgms{0}{1}{-1} \]}$. Similarly, we define $\boldsymbol{\rho}$, $\mathbf{U} ^{\boldsymbol{\rho}}$, and $\mathbf{a} ^{\boldsymbol{\rho}}$ for density. Thus, we have:
\begin{align}
  \mathbf{v} \approx \mathbf{U} ^{\mathbf{v}} \cdot \mathbf{a} ^{\mathbf{v}}
  ,
  \quad
  \boldsymbol{\rho} \approx \mathbf{U} ^{\boldsymbol{\rho}} \cdot \mathbf{a} ^{\boldsymbol{\rho}}
  .
\end{align}
From \autoref{eq:model__SPH__V} and \autoref{eq:model__SPH__D}, we derive the following evolutions:
\begin{align}
  \dot{\mathbf{a}} ^{\mathbf{v}}
  & =
  -
  \mathbf{U} ^{\mathbf{v}} {} \tp
  \cdot
  \mathbf{A}
  \cdot
  \mathbf{U} ^{\boldsymbol{\rho}} \cdot \mathbf{a} ^{\boldsymbol{\rho}}
  +
  \mathbf{U} ^{\mathbf{v}} {} \tp
  \cdot
  \mathbf{B}
  \cdot
  \mathbf{1}
  -
  \mathbf{U} ^{\mathbf{v}} {} \tp
  \cdot
  \mathbf{D}
  \cdot
  \mathbf{U} ^{\mathbf{v}}
  \cdot
  \mathbf{a} ^{\mathbf{v}}
  +
  \mathbf{U} ^{\mathbf{v}} {} \tp
  \cdot
  \mathbf{f}
  .
  \label{eq:model__acceleration__V}
  \\
  \dot{\mathbf{a}} ^{\boldsymbol{\rho}}
  & =
  - 
  \mathbf{U} ^{\boldsymbol{\rho}} {} \tp
  \cdot
  \mathbf{C}
  \cdot
  \mathbf{U} ^{\mathbf{v}} \cdot \mathbf{a} ^{\mathbf{v}}
  .
  \label{eq:model__acceleration__D}
\end{align}
Here for $i, j \in \Lambda$,
\begin{align}
  \mathbf{A} _{ij} & := \frac{m _{j} c _{j}}{\rho _{i} \( 0 \) \rho _{j} \( 0 \) ^{2}} \nabla W _{ij},
  \\
  \mathbf{B} _{ij} & := \frac{m _{j} c _{j}}{\rho _{i} \( 0 \) \rho _{j} \( 0 \)} \nabla W _{ij},
  \\
  \mathbf{C} _{ij} & := \frac{\rho _{i} ^{2}}{\rho _{i} \( 0 \)} \frac{m _{j}}{\rho _{j} \( 0 \)} \nabla W _{ij},
  \label{eq:model__acceleration__DSquare}
  \\
  \mathbf{D} _{ij} & := d _{i} \delta _{ij},
\end{align}
where $\delta$ is the Kronecker delta function,
$\mathbf{1} \in \mathbb{R} ^{N}$ is the vector filled with ones, and $\mathbf{f} \in \mathbb{R} ^{N} \mydel{\[ \mykgms{0}{1}{-2} \]}$ is a vector that contains all external force (per mass).

The evolution is already represented by reduced coordinates $\mathbf{a} ^{\mathbf{v}}$ and $\mathbf{a} ^{\boldsymbol{\rho}}$, allowing the temporal integration to be implemented using any numerical ODE scheme, for example, the explicit Euler method.
In practice, most calculations that require full coordinates can be precomputed at the beginning of the iteration, except for the term $\mathbf{U} ^{\mathbf{v}} {} \tp \cdot \mathbf{f}$ in \autoref{eq:model__acceleration__V} and $\rho _{i} ^{2}$ in \autoref{eq:model__acceleration__DSquare}.
For these two terms, the freezing coefficient method can be applied, meaning they can be updated every $n _{\textup{freeze}}$ time steps during the numerical scheme to enhance efficiency.

\section{Numerical Results}
\label{sec:numerical}

We set up the 1D mass-spring-damper system described in \autoref{sec:model__MSD}, with $L = 1 \[ \mykgms{0}{1}{0} \]$ and $T = 10 \[ \mykgms{0}{0}{1} \]$.
The SPH simulation, along with its POD-MOR solutions, is implemented.
In this section, we investigate POD-MOR error (POD error), the selection of POD data, and the POD acceleration via techniques in Section \autoref{sec:acceleration}.


\subsection{\texorpdfstring{\myadd{Verification of POD-MOR in a simple setting}}{A Mass-spring System without Damping}}
\label{sec:numerical__SPH_PBC_test28}

In the first example, we set the damping coefficient as $d \equiv 0 \[ \mykgms{0}{0}{-1} \]$, the Young's modulus as $c \equiv 0.1 \[ \mykgms{1}{-1}{-2} \]$, and the external force as $f \( t \) \equiv 0 \[ \mykgms{0}{1}{-2} \]$. We study a case with multi-mode initial velocity and uniform initial density
\begin{align}
  v _{0} \( x \) = 1 \e{-3} \sum _{l = 1} ^{3} l \sin \( 2 \pi l \frac{x}{L} \), 
  \quad 
  \rho _{0}(x) = 1
  ,
  \label{eq:numerical__Lagrangian_PBC_test28_V_initial}
\end{align}
with units $\[ \mykgms{0}{1}{-1} \]$ and $\[ \mykgms{1}{-3}{0} \]$. These standard units will be tacitly used in other quantities and in the following sections. 

\begin{figure}[!htp]
  \centering
  \includegraphics[width = 0.48\textwidth]{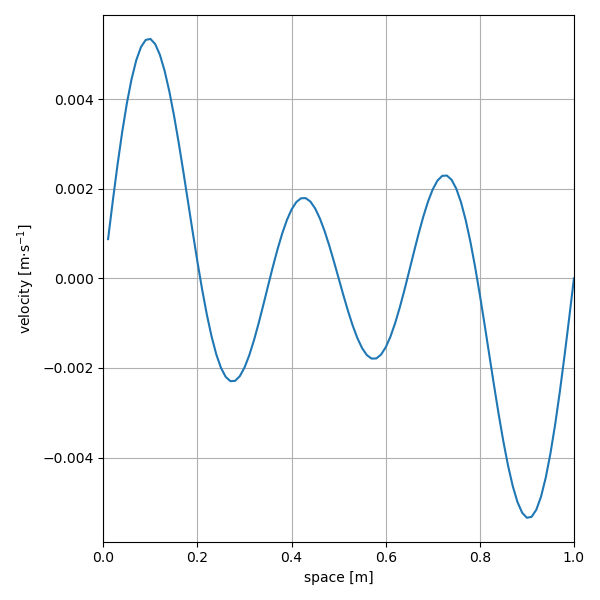}
  \includegraphics[width = 0.48\textwidth]{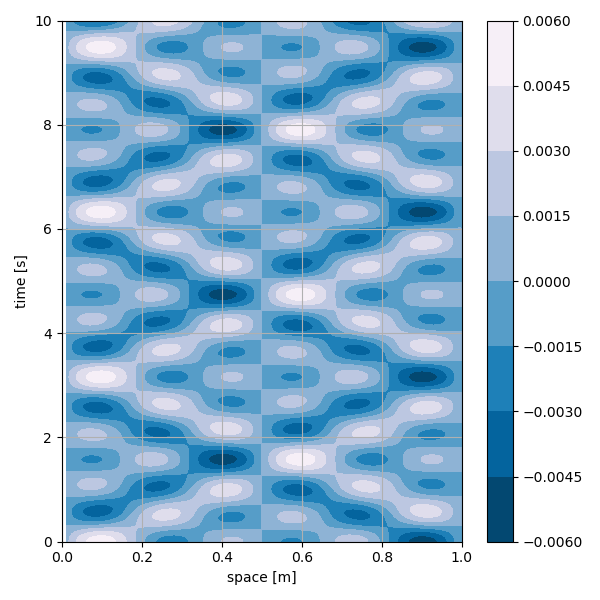}
  \caption{Initial condition (left figure) and profile of the reference/analytic velocity (right figure).}
  \label{fig:numerical__SPH_PBC_test28_POD_V_profile}
\end{figure}

The analytic solution is available, which is expressed as:
\begin{align}
  v _{\textup{analytic}} \( x, t \) 
  & = 
  1 \e{-3} \sum _{l = 1} ^{3} l \sin \( 2 \pi l \frac{x}{L} \) \cos \( 2 \pi l \frac{t}{L} \sqrt{\frac{c}{\rho _{0}}} \)
  ,
  \label{eq:numerical__SPH_PBC_test28_analytic_V}
  \\
  \rho _{\textup{analytic}} \( x, t \) 
  & =
  \frac{\rho _{0}}{1 + \partial _{x} \int _{0} ^{t} v _{\textup{analytic}} \( x, s \) \mathrm{d} s}
  \label{eq:numerical__SPH_PBC_analytic_D}
  .
\end{align}
We regard this analytic solution as the reference solution for this example, i.e., $v _{\textup{ref}} = v _{\textup{analytic}}$ and $\rho _{\textup{ref}} = \rho _{\textup{analytic}}$. \autoref{fig:numerical__SPH_PBC_test28_POD_V_profile} shows the initial velocity $v _{0} \( x \)$ and the profile of the analytic velocity $v _{\textup{analytic}} \( x, t \)$.

The SPH simulation evolves following equations \autoref{eq:model__SPH__V} and \autoref{eq:model__SPH__D}.
To compare the SPH solutions with the reference solution, we use the relative $L ^{2}$ norm. For example, the relative $L ^{2}$ error in velocity is defined as:
\begin{align}
  \frac{\int _{\Theta} \int _{\Omega} \[ v _{\textup{SPH}} \( x, t \) - v _{\textup{ref}} \( x, t \) \] ^{2} \mathrm{d} x \mathrm{d} t}{\int _{\Theta} \int _{\Omega} \[ v _{\textup{ref}} \( x, t \) \] ^{2} \mathrm{d} x \mathrm{d} t}
  .
\end{align}

We set the spatial step size as $\Delta x = 1 \e{-2}$ (i.e., corresponding to $N = 100$ smoothed particles) and set the SPH smoothing length as $h = 1.04 \Delta x$. Note that $h = 1.2 \Delta x$ is often used in the literature (such as \cite{fraser2017robust}), but in \autoref{sec:appendix_derivative} we showed that this will cause significant inconsistency in our current context, which is verified in numerical tests in \autoref{sec:numerical__SPH_PBC_test25}. The SPH performs better with $h = 1.04 \Delta x$ here. We use the explicit Euler scheme for time integration with a temporal step size of $\Delta t = 2 \e{-4}$. The relative $L ^{2}$ error in velocity is approximately $4.520 \e{-2}$, while the relative $L ^{2}$ error in density is around $3.651 \e{-4}$. 
The relative error of density is significantly smaller than that of velocity, since the variation of velocity is relatively large and the density remains around the initial density $\rho _{0} = 1$.



\subsubsection{\texorpdfstring{A Baseline \myadd{POD-MOR} Case}{A Baseline POD-MOR Case}}

We apply the POD-MOR to the SPH simulation. In total, $100$ snapshots are uniformly collected from $t \in \[ 0, 1.6 \]$ with spacing $0.016$ from the SPH solution. 
In this example, the solution is periodic with a temporal period $T' = L \sqrt{\rho _{0} / c} = \sqrt{10}$, and is symmetric with respect to $T' / 2$. Therefore, data collection is more efficient if we gather POD data for $t \lesssim T' / 2$.

\begin{figure}[!htp]
  \centering
  \includegraphics[width = 0.8\textwidth]{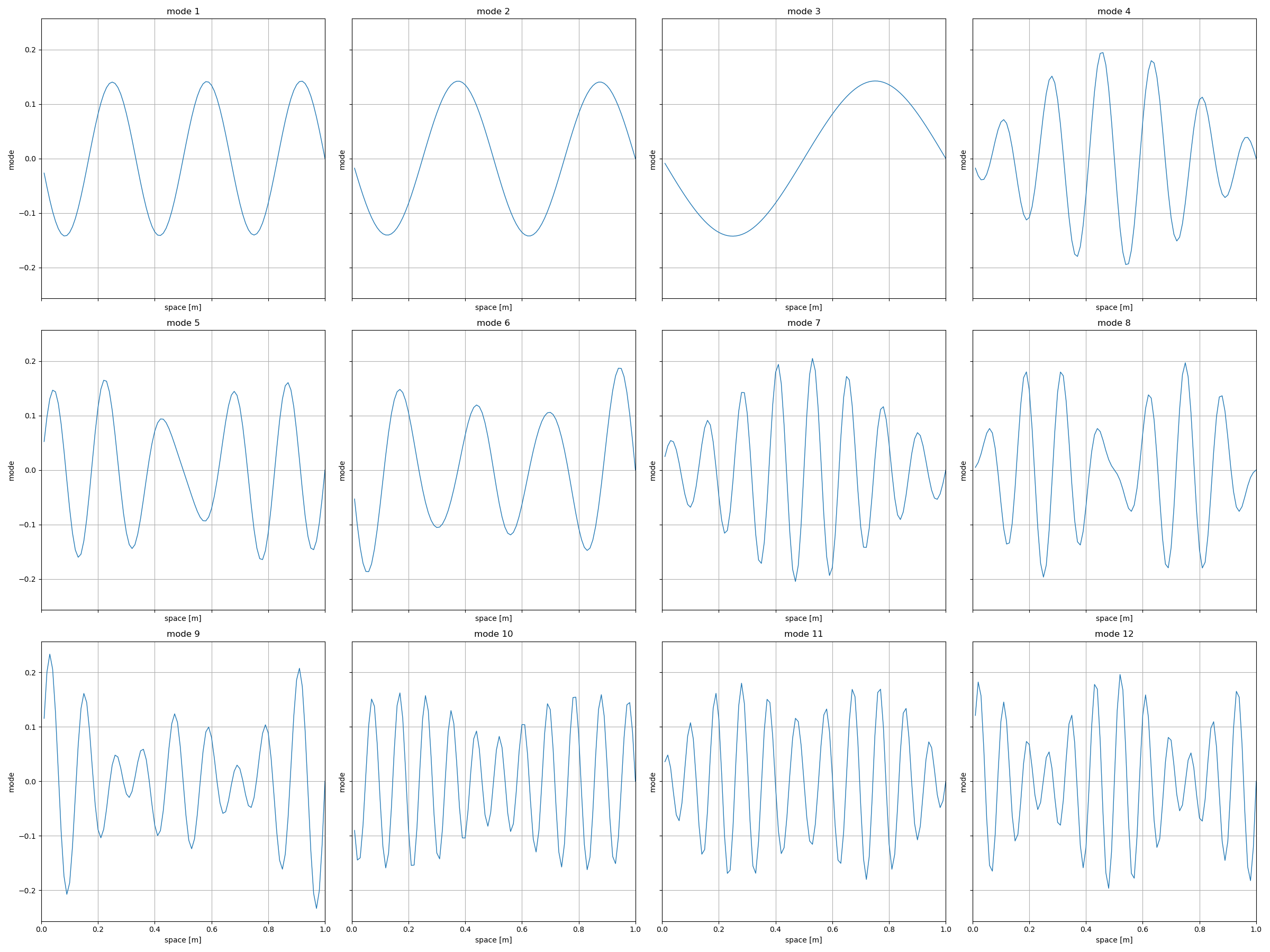}
  \includegraphics[width = 0.8\textwidth]{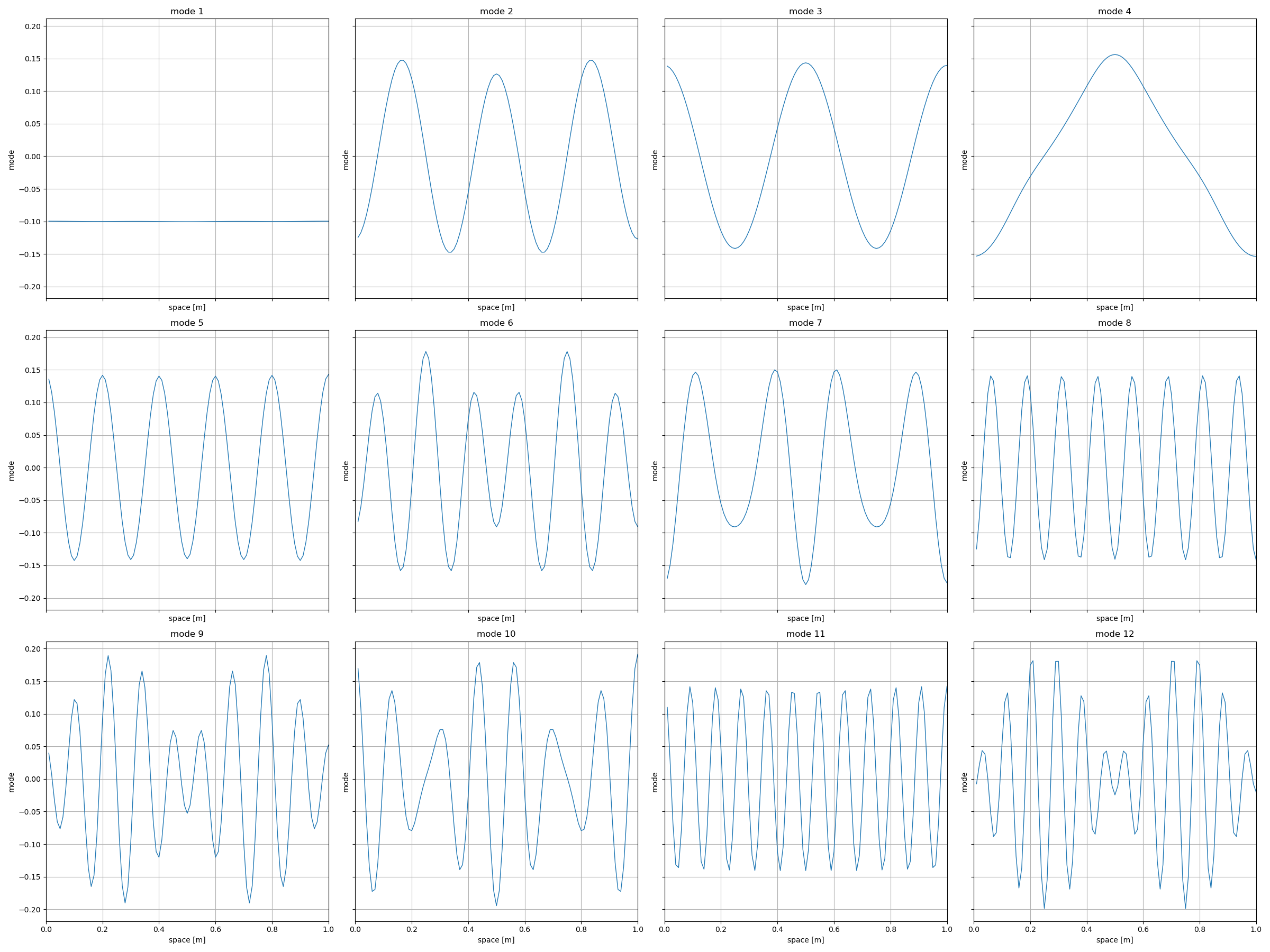}
  \caption{The dominant POD modes of velocity (upper 12 figures) and density (lower  12 figures).}
  \label{fig:numerical__SPH_PBC_test28_POD_nx100_nt5k_nn1d06_PodDataSph100First160m_modes}
\end{figure}

\begin{figure}[!htp]
  \centering
  \includegraphics[width = 0.48\textwidth]{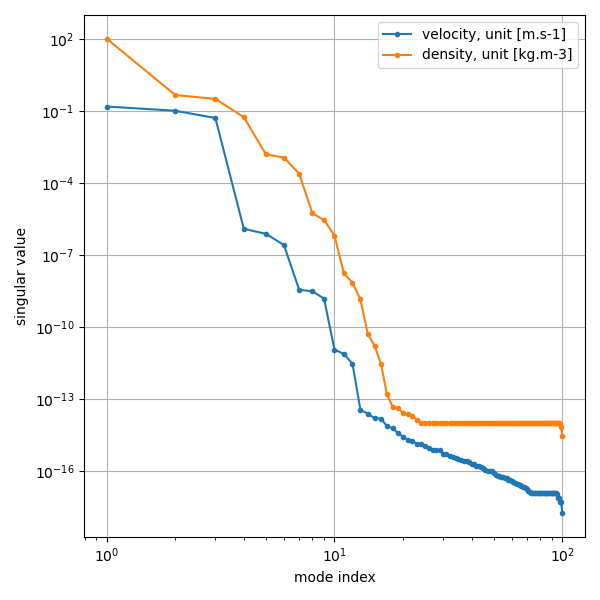}
  \includegraphics[width = 0.48\textwidth]{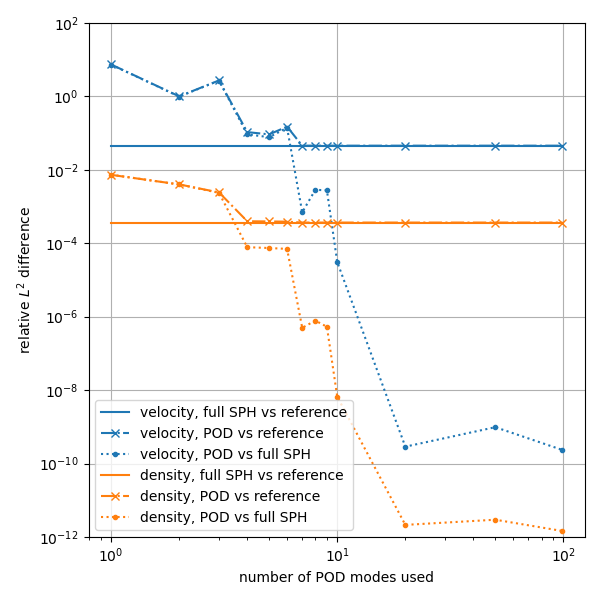}
  \caption{
  The singular value of the POD modes (left figure) and the relative $L ^{2}$ error (right figure) among the reference solution, the full SPH solution, and the POD-MOR of SPH simulation with different number of POD modes used, where
  $100$ snapshots are uniformly collected from $t \in \[ 0, 1.6 \]$ for the POD data.}
  \label{fig:numerical__SPH_PBC_test28_POD_nx100_nt5k_nn1d06_PodDataSph100First160m_error}
\end{figure}

We plot the dominant POD modes of velocity and density in \autoref{fig:numerical__SPH_PBC_test28_POD_nx100_nt5k_nn1d06_PodDataSph100First160m_modes}, ordered according to the magnitude of their singular values (the diagonal values of $\hat{\boldsymbol{\Sigma}}$ in \autoref{eq:model__POD__SVD}) shown in \autoref{fig:numerical__SPH_PBC_test28_POD_nx100_nt5k_nn1d06_PodDataSph100First160m_error}(left). Both the velocity and density modes exhibit rapid decay in their singular values, indicating that the first few modes are dominant.


The first three velocity modes have dominant singular values and look like pure harmonics, while most of the subsequent modes are more irregular. We observe that singular values of velocity modes decay in groups of three, which corresponds to the presence of three modes in the initial velocity (and the analytic solution \autoref{eq:numerical__SPH_PBC_test28_analytic_V}). For the density modes, the first mode is nearly uniform and has the largest singular value. According to the analytic solution of density \autoref{eq:numerical__SPH_PBC_analytic_D}, the density is almost constant and its variation depends on the spatial derivative of velocity. This results in a zeroth harmonic with a large singular value in density, while the remaining modes can be regarded as perturbations due to the velocity, also appear in groups of three.

\autoref{fig:numerical__SPH_PBC_test28_POD_nx100_nt5k_nn1d06_PodDataSph100First160m_error}(right) shows the relative $L ^{2}$ error of the POD-MOR solution based on \autoref{eq:model__POD__general}, in comparison with both the full SPH solution and reference solution. The difference between the POD-MOR solution and the full SPH solution represents the POD error. The errors for both velocity and density decrease as the number of POD modes increases, reflecting an expected trend. When more than $7$ POD modes are utilized, the POD error becomes smaller than the SPH numerical error, i.e., the error between the full SPH solution and the reference solution, indicating that the MOR is effective around 7 POD modes in this example. 

\subsubsection{\texorpdfstring{Selection of Snapshots --- Required number of Snapshots}{Selection of Snapshots --- Minimum Amount of Snapshots}}
\label{section3_1_2}

We investigate how POD-MOR works with different POD data. One key question is: how many POD snapshots are required for a successful POD-MOR? In general, it depends on the initial conditions and exact solutions. We take the previous example for illustration. 

From the singular values in \autoref{fig:numerical__SPH_PBC_test28_POD_nx100_nt5k_nn1d06_PodDataSph100First160m_error}, it is shown that the first three velocity modes and the first density mode are the principal modes. But, three additional density modes are necessary to capture the variation of density and achieve a reasonable POD error. This means the minimum amount of POD data consists of three velocity snapshots and four density snapshots. For instance, we select velocity and density snapshots at times $t = 0, 0.4, 0.8$, and an additional density snapshot at $t = 1.2$.

\begin{figure}[!htp]
  \centering
  \includegraphics[width = 0.6\textwidth]{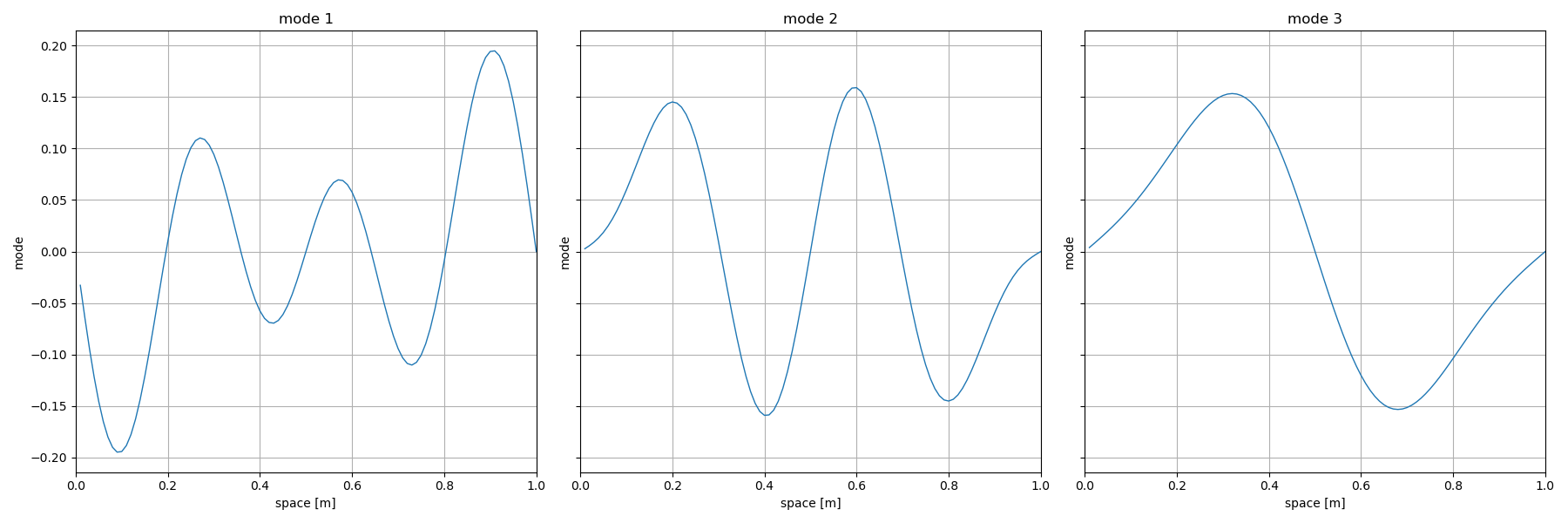}
  \includegraphics[width = 0.8\textwidth]{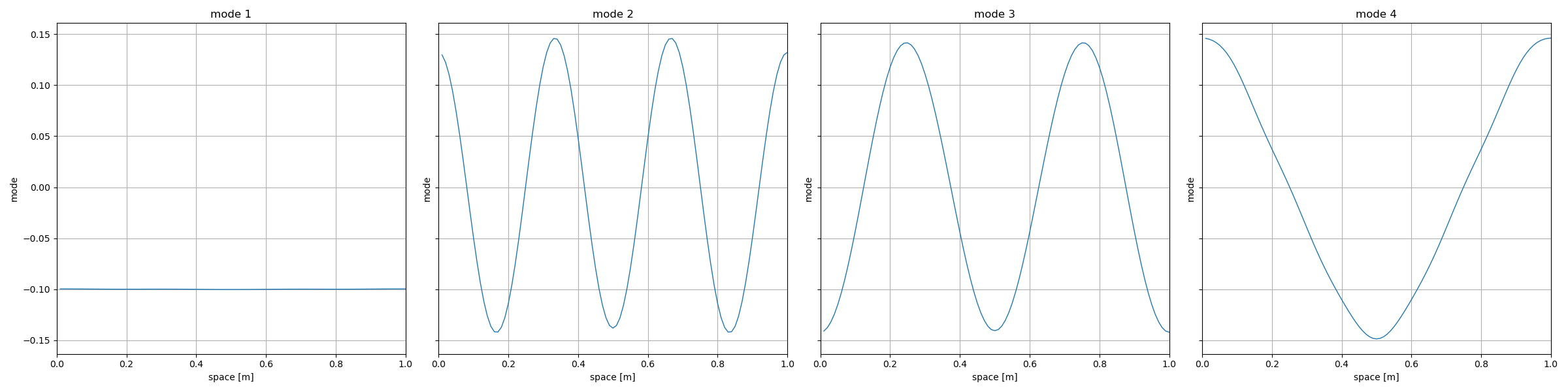}
  \caption{The POD modes of velocity (upper figure) and density (lower figure).}
  \label{fig:numerical__SPH_PBC_test28_POD_nx100_nt5k_nn1d06_PodDataSph3First120mSph4First160m_modes}
\end{figure}

\begin{figure}[!htp]
  \centering
  \includegraphics[width = 0.48\textwidth]{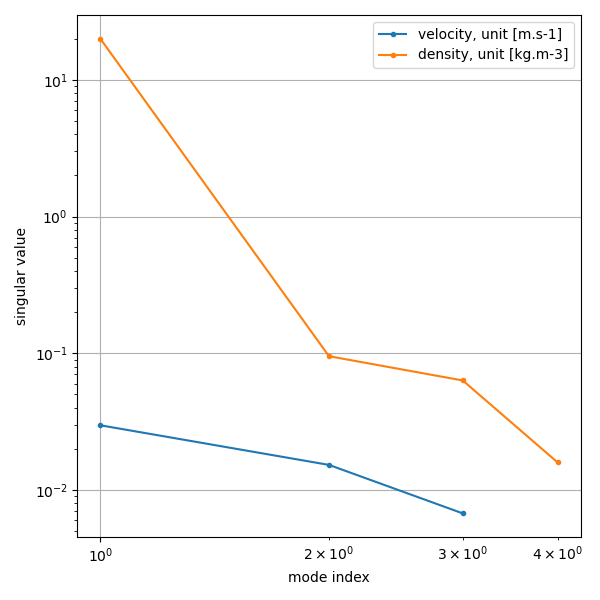}
  \includegraphics[width = 0.48\textwidth]{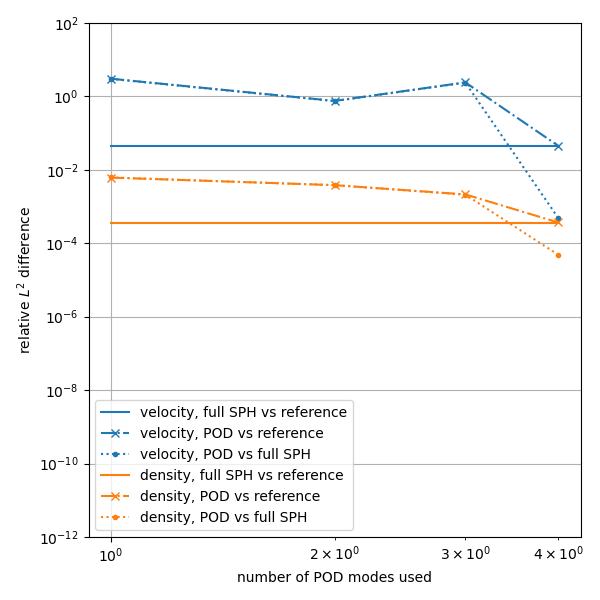}
  \caption{The singular value of the POD modes (left figure) and the relative $L ^{2}$ error (right figure) among the reference solution, the full SPH solution, and the POD-MOR of SPH simulation with different number of POD modes used, where $3$ velocity snapshots and $4$ velocity snapshots are collected for the POD data.}
  \label{fig:numerical__SPH_PBC_test28_POD_nx100_nt5k_nn1d06_PodDataSph3First120mSph4First160m_error}
\end{figure}

We present the POD modes of velocity and density in \autoref{fig:numerical__SPH_PBC_test28_POD_nx100_nt5k_nn1d06_PodDataSph3First120mSph4First160m_modes}, along with their singular values in decreasing order and the relative $L ^{2}$ error of the POD-MOR \autoref{eq:model__POD__general} in \autoref{fig:numerical__SPH_PBC_test28_POD_nx100_nt5k_nn1d06_PodDataSph3First120mSph4First160m_error}. It shows that with these snapshots and POD modes, the POD approximation already achieved high accuracy.  Compared to the total number of particles, $N = 100$, the number of modes and snapshots required can be significantly smaller, which could depend on the number of modes in the initial condition.


In \autoref{fig:numerical__SPH_PBC_test28_POD_nx100_nt5k_nn1d06_PodDataSph3First120mSph4First160m_modes}, we observe that the first three velocity modes are not closely aligned with the first three harmonics shown in the analytic solution, even though the POD-MOR works very well. In fact, if we exclude the error from the POD data and use data from the reference solution, the velocity data can be represented as (up to a constant) $A = S C$, where
\begin{align}
  S 
  & =
  \[ \sin \( 2 \pi \frac{\mathbf{x}}{L} \), \sin \( 4 \pi \frac{\mathbf{x}}{L} \), \sin \( 6 \pi \frac{\mathbf{x}}{L} \) \]
  \in \mathbb{R} ^{N \times 3}
  \\
  C 
  & =
  \[ \begin{array}{ccc} \cos \( 2 \pi \frac{t _{1}}{T'} \) & \cos \( 2 \pi \frac{t _{2}}{T'} \) & \cos \( 2 \pi \frac{t _{3}}{T'} \) \\ 2 \cos \( 4 \pi \frac{t _{1}}{T'} \) & 2 \cos \( 4 \pi \frac{t _{2}}{T'} \) & 2 \cos \( 4 \pi \frac{t _{3}}{T'} \) \\ 3 \cos \( 6 \pi \frac{t _{1}}{T'} \) & 3 \cos \( 6 \pi \frac{t _{2}}{T'} \) & 3 \cos \( 6 \pi \frac{t _{3}}{T'} \) \end{array} \]
  \in \mathbb{R} ^{3 \times 3}
  .
\end{align}
Here, $\mathbf{x}$ is the column vector containing the spatial grid points,
and $t _{1} = 0$, $t _{2} = 0.4$, $t _{3} = 0.8$.
The three columns of discrete sine vectors are orthogonal to each other, so we can easily show that the POD modes of velocity are $S U$, where 
\begin{align}
  U
  \approx
  \[ \begin{array}{ccc} -0.142 & 0.251 & 0.958 \\ -0.528 & 0.799 & -0.287 \\ -0.837 & -0.546 & 0.019 \end{array} \]
  \in \mathbb{R} ^{3 \times 3}
\end{align}
is computed from the decomposition of $C$.
In general, the matrix $U$ is not diagonal, thus, the velocity POD modes are a combination of harmonic modes in reference solution. However, 
using the mixture of harmonic modes as POD modes will not affect the POD-MOR when all the modes are included, since the column space remains unchanged.

In this example, the minimum number of snapshots are $3$ and $4$ for velocity and density. It is also worth noting that the velocity modes in \autoref{fig:numerical__SPH_PBC_test28_POD_nx100_nt5k_nn1d06_PodDataSph100First160m_modes} are much closer to harmonics.
This occurs because we selected more snapshots ($100$ snapshots) that nearly uniformly cover $T' / 2$, which is sufficient due to the symmetry in the temporal domain. In this case, the matrix $C \in \mathbb{R} ^{3 \times 100}$ has rows that are almost orthogonal to each other, resulting a nearly diagonal $U$.

\subsubsection{\texorpdfstring{Selection of Snapshots --- Small Temporal Region}{Selection of Snapshots --- Small Temporal Region}}
\label{section3_1_3}

In this section, we investigate the requirements for the distribution of POD snapshots from the SPH simulation. Specifically, we explore whether it is essential for the snapshots to be widely distributed or if they can be selected from a small temporal region.

\begin{figure}[!htp]
  \centering
  \includegraphics[width = 1.0\textwidth]{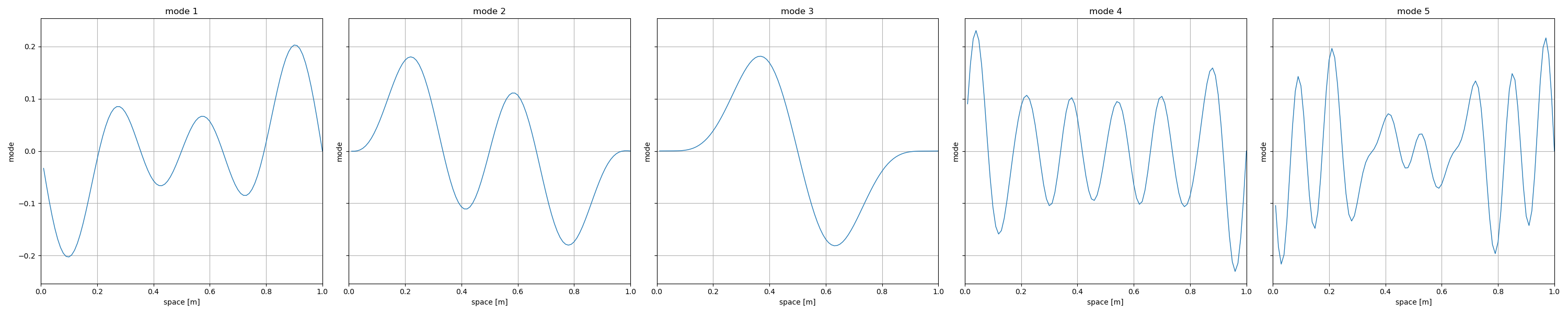}
  \includegraphics[width = 1.0\textwidth]{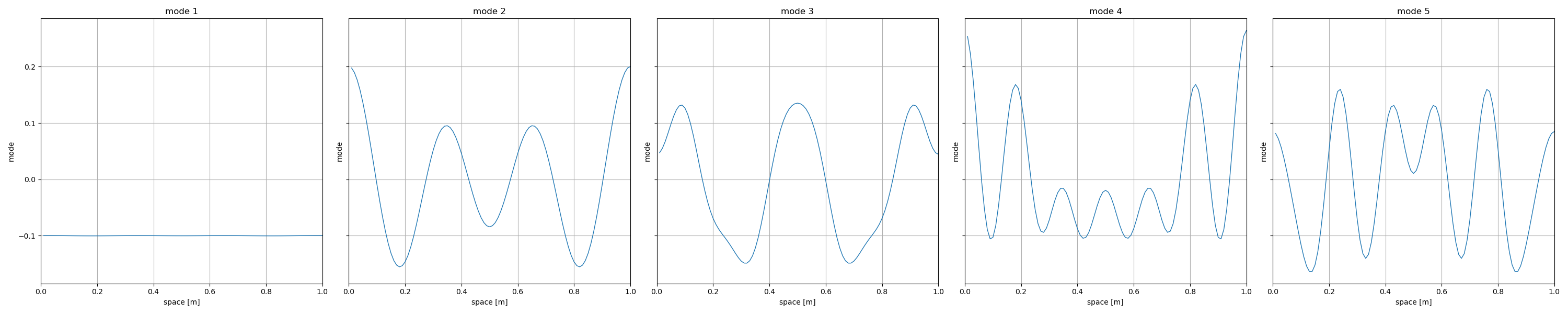}
  \caption{The dominant POD modes of velocity (upper figure) and density (lower figure).}
  \label{fig:numerical__SPH_PBC_test28_POD_nx100_nt5k_nn1d06_PodDataSph5First10m_modes}
\end{figure}

\begin{figure}[!htp]
  \centering
  \includegraphics[width = 0.48\textwidth]{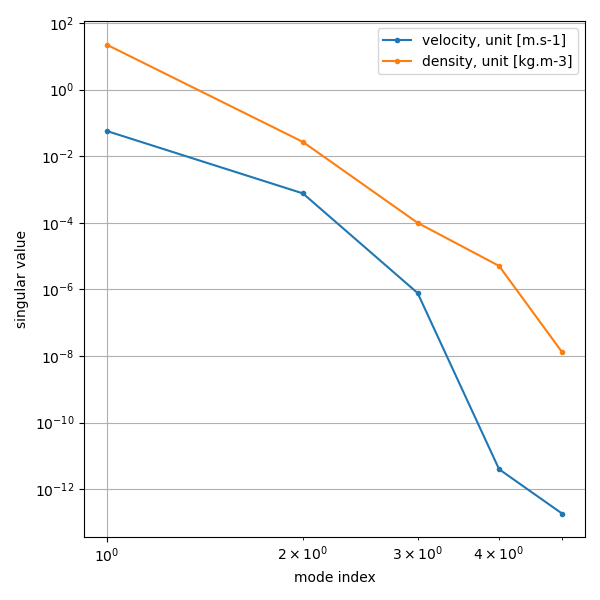}
  \includegraphics[width = 0.48\textwidth]{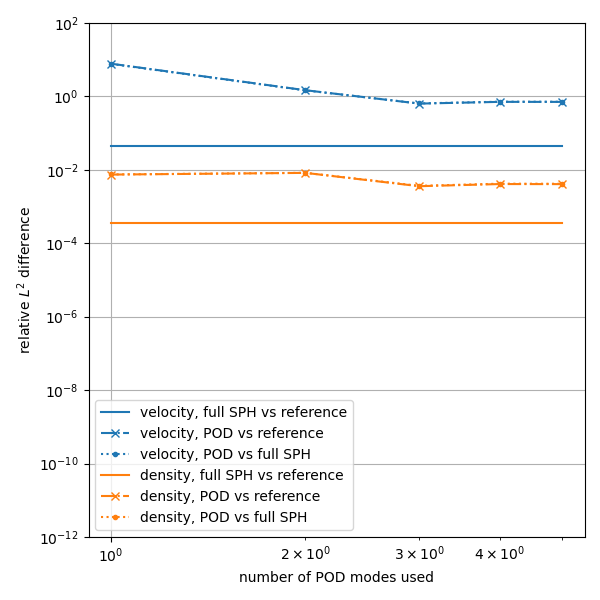}
  \caption{The singular value of the POD modes (left figure) and the relative $L ^{2}$ error (right figure) among the reference solution, the full SPH solution, and the POD-MOR of SPH simulation with different number of POD modes used, where
  $5$ snapshots are uniformly collected from $t \in \[ 0, 0.1 \]$ for the POD data.}
  \label{fig:numerical__SPH_PBC_test28_POD_nx100_nt5k_nn1d06_PodDataSph5First10m_error}
\end{figure}

\textbf{\myadd{(1) The Case with Fewer Snapshots.}} 
We select five velocity snapshots and five density snapshots uniformly from the SPH solution in $t \in \[ 0, 0.1 \]$, more precisely at $t = 0, 0.02, 0.04, 0.06, 0.08$. The POD modes of velocity and density are shown in \autoref{fig:numerical__SPH_PBC_test28_POD_nx100_nt5k_nn1d06_PodDataSph5First10m_modes}. Additionally, we present the corresponding singular values of the POD modes and the relative $L ^{2}$ error of the POD-MOR \autoref{eq:model__POD__general} in \autoref{fig:numerical__SPH_PBC_test28_POD_nx100_nt5k_nn1d06_PodDataSph5First10m_error}. It is important to note that the collected snapshots are taken from the beginning of the evolution, which does not cover a significant portion of the simulation (compared to $T' / 2$). 
We observe that the singular values decay much faster than in the previous case, as the POD primarily captures information near the initial state, resulting in a larger POD error.
In contrast, the previous case included data covering nearly half of the temporal period, highlighting the necessity for POD snapshots to span a broader range of the temporal domain if the number of snapshots is not large. 
Given that the SPH simulation is periodic in time, it is beneficial for the POD data to encompass that period.


\begin{figure}[!htp]
  \centering
  \includegraphics[width = 0.8\textwidth]{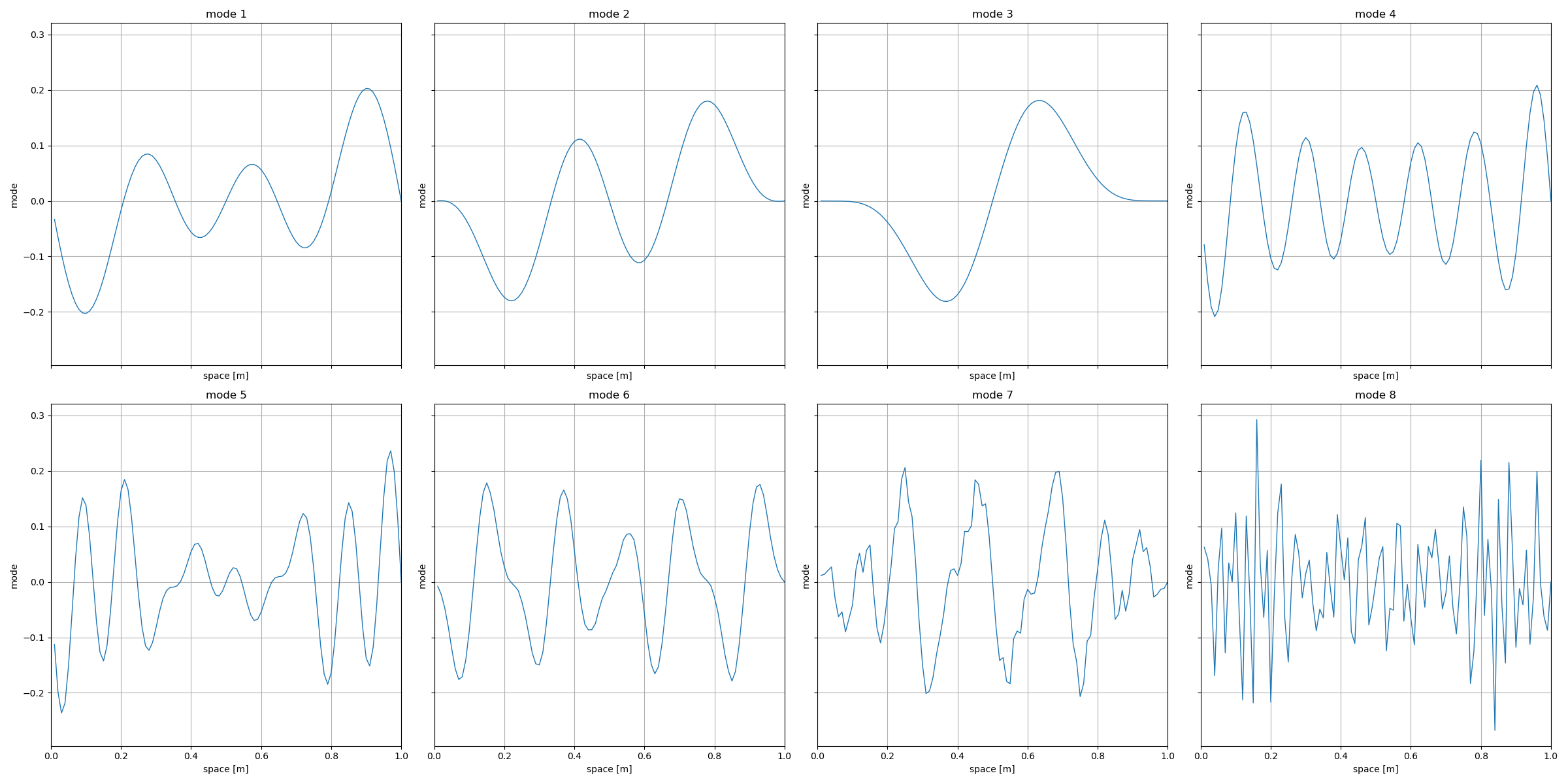}
  \includegraphics[width = 0.8\textwidth]{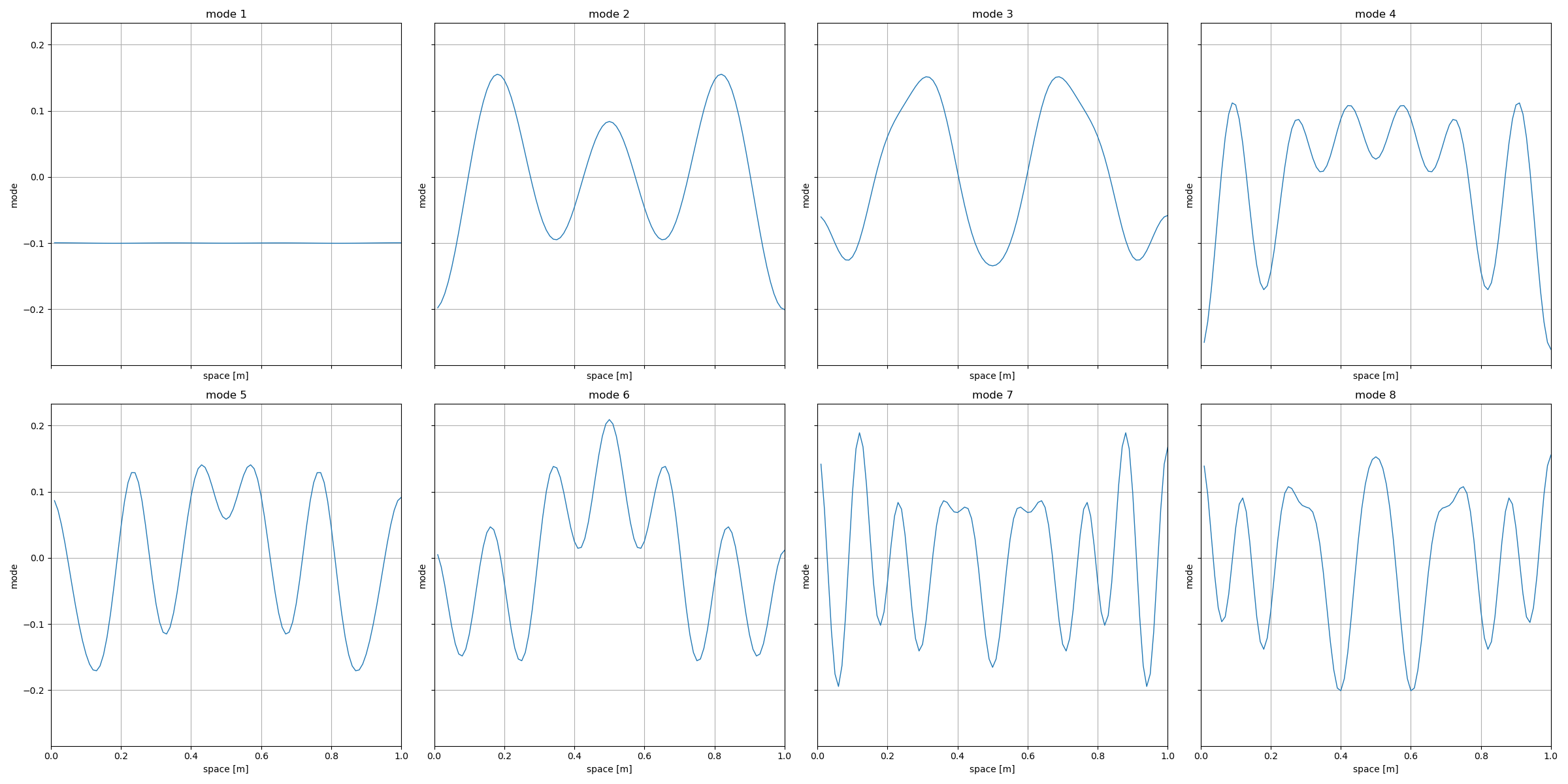} \caption{The dominant POD modes of velocity (upper figure) and density (lower figure).}
  \label{fig:numerical__SPH_PBC_test28_POD_nx100_nt5k_nn1d06_PodDataSph100First10m_modes}
\end{figure}

\begin{figure}[!htp]
  \centering
  \includegraphics[width = 0.48\textwidth]{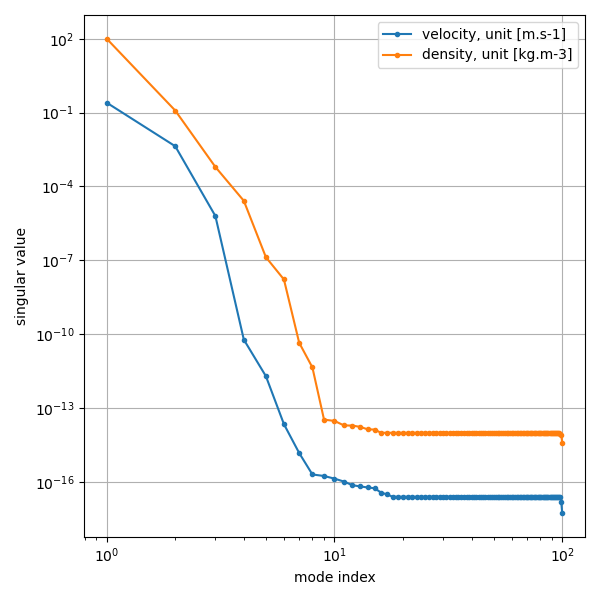}
  \includegraphics[width = 0.48\textwidth]{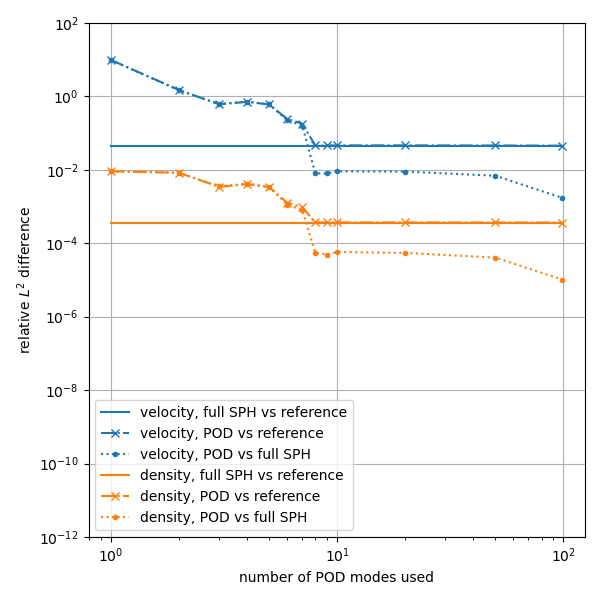}
  \caption{
  The singular value of the POD modes (left figure) and the relative $L ^{2}$ error (right figure) among the reference solution, the full SPH solution, and the POD-MOR of SPH simulation with different number of POD modes used, where
  $100$ snapshots are uniformly collected from $t \in \[ 0, 0.1 \]$ for the POD data.}
  \label{fig:numerical__SPH_PBC_test28_POD_nx100_nt5k_nn1d06_PodDataSph100First10m_error}
\end{figure}

\textbf{\myadd{(2) The Case with More Snapshots.}} 
We study the case when more snapshots are collected within a short time period and see if it can achieve successful POD-MOR.
In this case, $100$ snapshots are uniformly collected from $t \in \[ 0, 0.1 \]$ with spacing $t = 0.001$ of the SPH solution. The dominant POD modes of velocity and density are presented in \autoref{fig:numerical__SPH_PBC_test28_POD_nx100_nt5k_nn1d06_PodDataSph100First10m_modes}, along with the singular values of the POD modes and the relative $L ^{2}$ error of the original POD-MOR \autoref{eq:model__POD__general} in \autoref{fig:numerical__SPH_PBC_test28_POD_nx100_nt5k_nn1d06_PodDataSph100First10m_error}. The errors in \autoref{fig:numerical__SPH_PBC_test28_POD_nx100_nt5k_nn1d06_PodDataSph100First10m_error} show that the POD-MOR is effective when we use at least $8$ modes. 
Although we know there are only three essential velocity modes in the analytic solution, the POD-MOR cannot accurately identify them when using POD data from a short range, so the inclusion of more modes is necessary to reduce the POD error. 
This case demonstrates that it is still feasible to apply POD-MOR with limited data from a short temporal range, provided we gather more POD data and utilize more POD modes. 
This flexibility can be advantageous for large systems, where conducting SPH simulations over extended periods can be costly.


\begin{figure}[!htp]
  \centering
  \includegraphics[width = 0.8\textwidth]{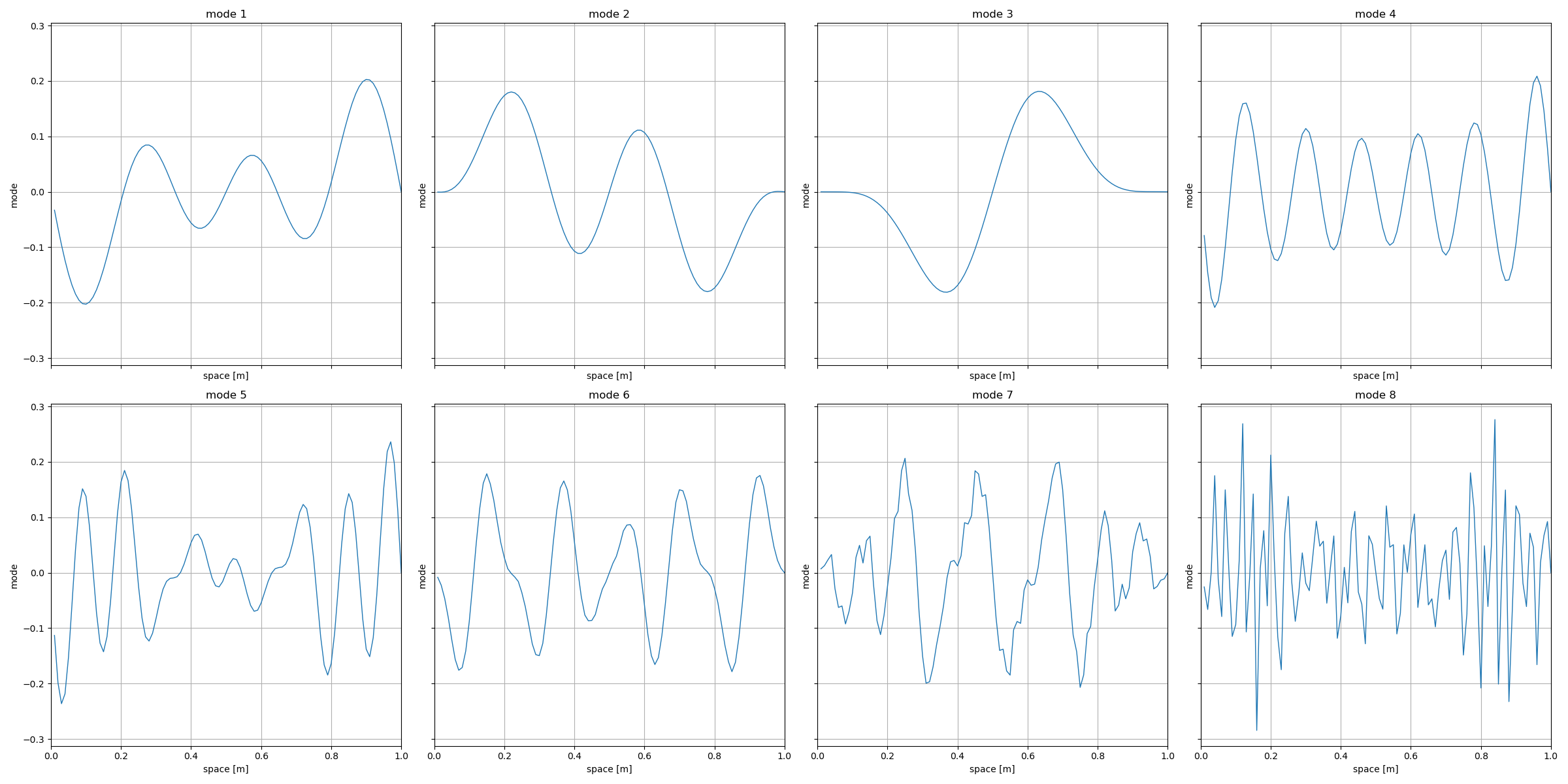}
  \includegraphics[width = 0.8\textwidth]{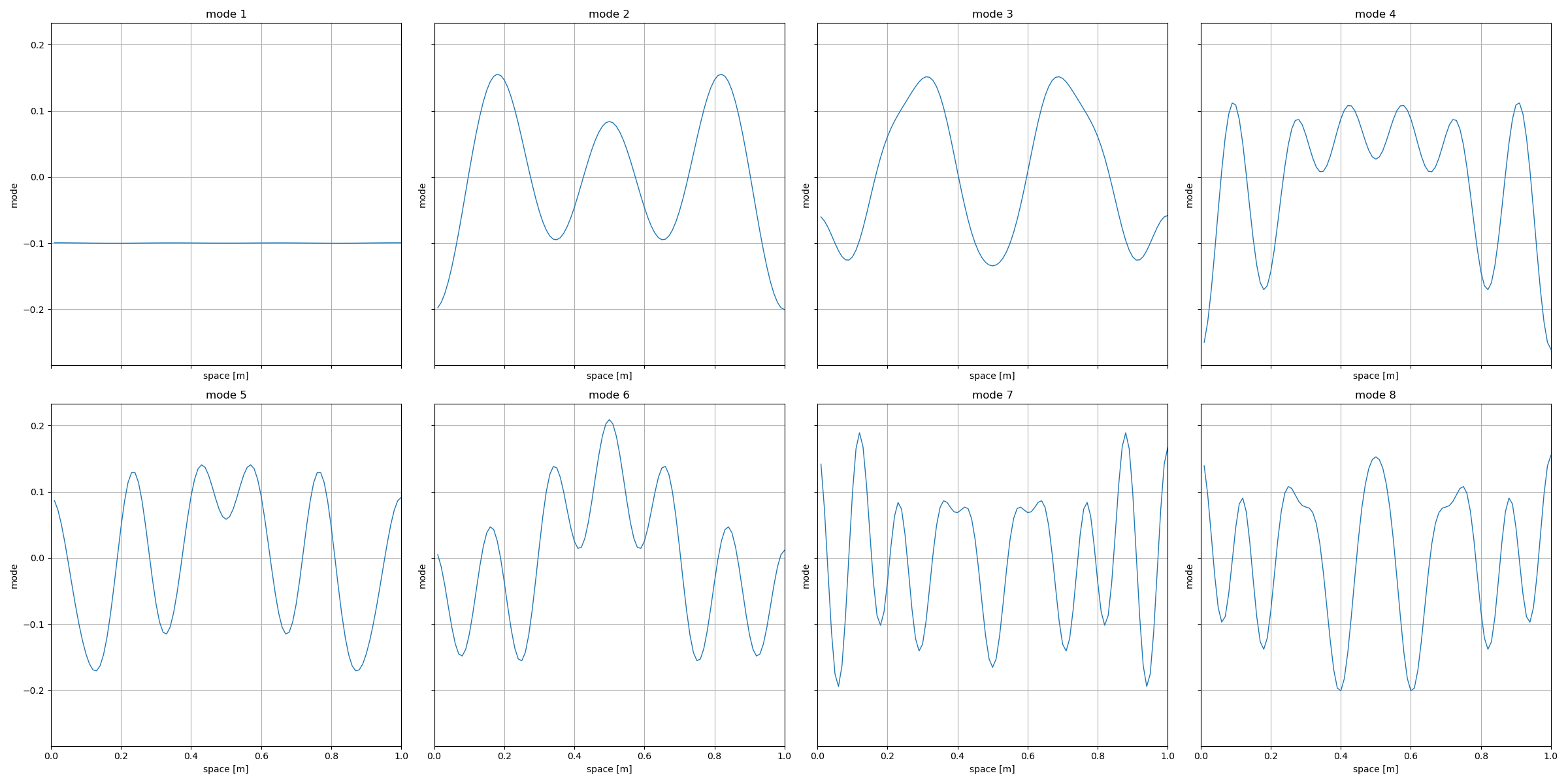}
  \caption{The dominant POD modes of velocity (upper figure) and density (lower figure).}
  \label{fig:numerical__SPH_PBC_test28_POD_nx100_nt5k_nn1d06_PodDataSph100First10mRand_modes}
\end{figure}

\begin{figure}[!htp]
  \centering
  \includegraphics[width = 0.48\textwidth]{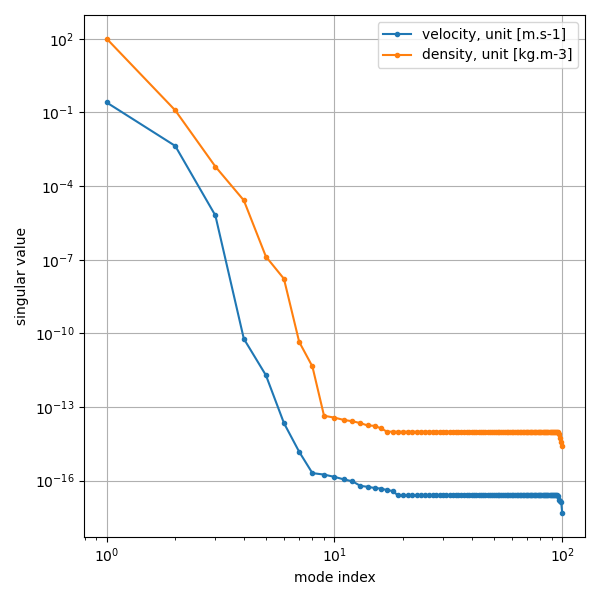}
  \includegraphics[width = 0.48\textwidth]{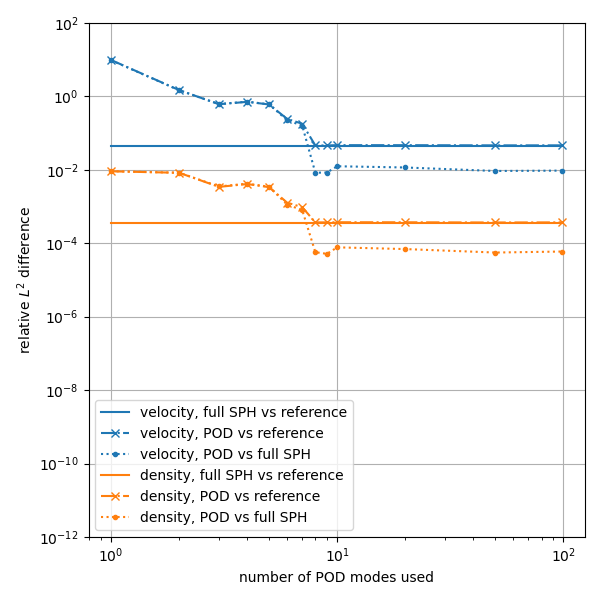}
  \caption{
  The singular value of the POD modes (left figure) and the relative $L ^{2}$ error (right figure) among the reference solution, the full SPH solution, and the POD-MOR of SPH simulation with different number of POD modes used, where
  $100$ snapshots are randomly collected from $t \in \[ 0, 0.1 \]$ for the POD data.}
  \label{fig:numerical__SPH_PBC_test28_POD_nx100_nt5k_nn1d06_PodDataSph100First10mRand_error}
\end{figure}

\textbf{\myadd{(3) The Case with Randomly-Spaced Snapshots.}}
We study the case \myadd{with randomly-spaced snapshots} and see if the non-uniform distribution of snapshots affects the effectiveness of POD-MOR. In this case, $100$ snapshots are randomly collected from $t \in [ 0, 0.1 ]$ of the SPH solution. We plot the dominant POD modes of velocity and density in \autoref{fig:numerical__SPH_PBC_test28_POD_nx100_nt5k_nn1d06_PodDataSph100First10mRand_modes}, and also display the singular values of the POD modes along with the relative $L ^{2}$ error of the POD-MOR \autoref{eq:model__POD__general} in \autoref{fig:numerical__SPH_PBC_test28_POD_nx100_nt5k_nn1d06_PodDataSph100First10mRand_error}. Compared with errors in \autoref{fig:numerical__SPH_PBC_test28_POD_nx100_nt5k_nn1d06_PodDataSph100First10m_error} from the uniform-spacing case, the random spacing case performs similarly.


\noindent \textbf{\myadd{Remark:}} \myadd{Although we have discussed various strategies for selecting POD data and POD modes here, some recent work \cite{duan2023feasibility} has introduced some measures for choosing the POD modes.  That work studies the number of modes required for POD-MOR to be effective, by considering singular values and the differences between training and testing data (extra data set). Because of its complexity and unclear applicability here, we still adopt the current strategy for POD data and then choosing POD modes based on the decay of singular values.}

\subsection{\texorpdfstring{\myadd{Results without Damping}}{More Initial Modes without Damping}}
\label{sec:numerical__SPH_PBC_test23}

In this section, we study a more complicated example where the initial velocity contains more modes. The initial velocity is
\begin{align}
  v _{0} \( x \) = 4 \e{-3} \frac{x}{L} \[ 1 - \frac{x}{L} \] \sum _{l = 1} ^{10} \sin \( 2 \pi l \frac{x}{L} + l ^{2} \).
  \label{eq:numerical__Lagrangian_PBC_test23_V_initial}
\end{align}
Other settings are the same as in the previous section, i.e., $\rho _{0} \( x \) = 1$, $d = 0$, $c = 0.1$, and $f \( t \) = 0$. Another example with Gaussian-type initial velocity is investigated in \autoref{sec:numerical__SPH_PBC_test29}.

\begin{figure}[!htp]
  \centering
  \includegraphics[width = 0.48\textwidth]{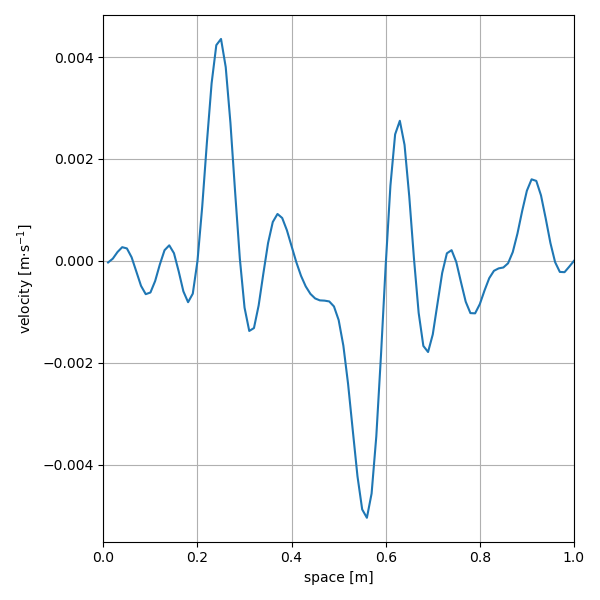}
  \includegraphics[width = 0.48\textwidth]{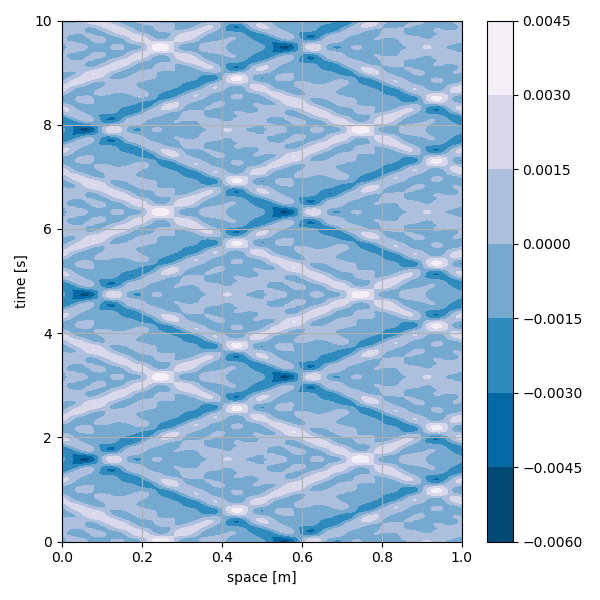}
  \caption{Initial condition (left figure) and profile of the analytic velocity (right figure).}
  \label{fig:numerical__SPH_PBC_test23_POD_V_profile}
\end{figure}

In general, the high spatial frequency requires a finer grid in space, which also need a finer grid in time due to the CFL condition. The reference solution of 
\autoref{eq:model__spring__VD__V} and \autoref{eq:model__spring__VD__D} is generated using a first-order finite difference method in space with $\Delta x = 1.25 \e{-3}$ and a second-order Runge-Kutta scheme in time with $\Delta t = 2.5 \e{-5}$.
We plot the initial condition and the solution profile of velocity $v \( x, t \)$ of the reference solution in \autoref{fig:numerical__SPH_PBC_test23_POD_V_profile}.

The SPH simulation evolves according to equations \autoref{eq:model__SPH__V} and \autoref{eq:model__SPH__D}. 
The spatial step size is set as $\Delta x = 2.5 \e{-3}$, (i.e., $N = 400$ smoothed particles), the SPH smoothing length is set as $h = 1.04 \Delta x$, and the explicit Euler scheme for time integration is used with $\Delta t = 5 \e{-5}$. The relative $L ^{2}$ error in velocity is about $8.488 \e{-2}$, and the relative $L ^{2}$ error in density is about $3.218 \e{-4}$. The errors here are larger compared with the previous case with fewer initial modes. This is due to frequency error caused by the limited number of neighbor particles within the support of the SPH kernel function in the Lagrangian setting. Here, $h = 1.04 \Delta x$ indicates that there are only $2$ neighbors interacting on each side of a smoothing particle. The analysis in \autoref{sec:appendix_derivative} suggests that a larger value of $h / \Delta x$ could help reduce the SPH error. 
However, finding the optimal settings for the SPH simulation is not the objective of this work. Our focus here is on the effectiveness of POD-MOR in SPH simulations, which is demonstrated below.

\subsubsection{\texorpdfstring{\myadd{Results of original POD-MOR}}{A Baseline POD-MOR Case}}

We apply POD-MOR to the SPH simulation. In this example, the solution is periodic with a temporal period of $T' = L \sqrt{\rho _{0} / c} = \sqrt{10}$ in temporal space, and it is symmetric with respect to $T' / 2 \approx 1.6$.

\begin{figure}[!htp]
  \centering
  \includegraphics[width = 0.48\textwidth]{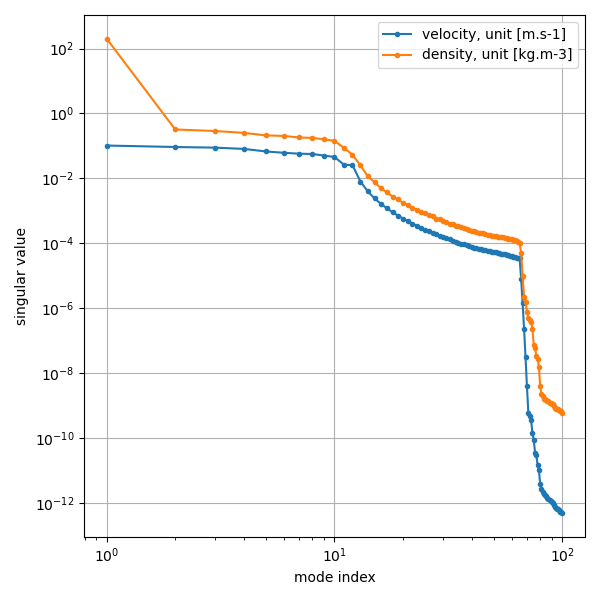}
  \includegraphics[width = 0.48\textwidth]{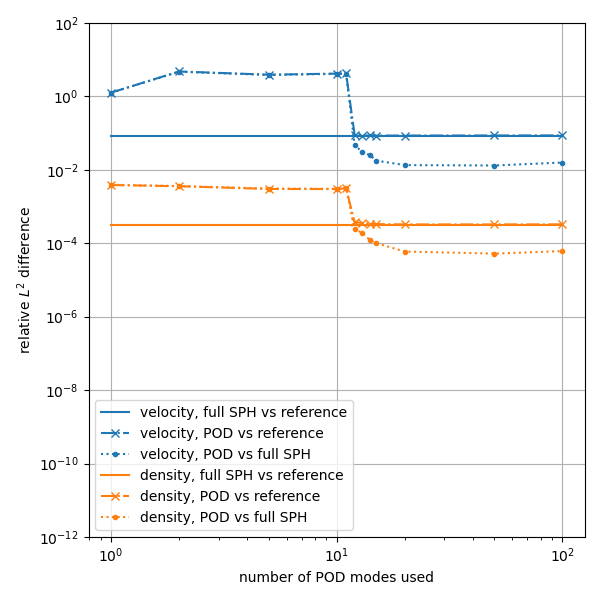}
  \caption{
  The singular value of the POD modes (left figure) and the relative $L ^{2}$ error (right figure) among the reference solution, the full SPH solution, and the POD-MOR of SPH simulation with different number of POD modes used, where
  $100$ snapshots are uniformly collected from $t \in \[ 0, 1.6 \]$ for the POD data.}
  \label{fig:numerical__SPH_PBC_test23_POD_nx400_nt20k_nn1d04_PodDataSph100First160m_error}
\end{figure}

\myadd{(1) \textbf{The baseline case.}} \\
We use the same data collection strategy as before that
a total of $100$ snapshots are uniformly collected from $t \in \[ 0, 1.6 \] $ with spacing $0.016$ from the SPH solution. \autoref{fig:numerical__SPH_PBC_test23_POD_nx400_nt20k_nn1d04_PodDataSph100First160m_error} plots the singular values of the POD modes and the relative $L ^{2}$ error of the POD-MOR \autoref{eq:model__POD__general}. 
The POD error is smaller than the SPH numerical error when using at least $12$ modes, indicating that POD works if more than $12$ modes are used. This is expected since we have at least $10$ different modes in the initial conditions. We observe that the first $12$ singular values are of similar magnitude, with a steep drop in magnitude for the $13$-th singular value (also a steep drop in error). 
This drop indicates that the first $12$ modes are dominant compared to the others and can represent the solutions effectively.

\myadd{(2) \textbf{Other cases with different Snapshots.}}

\begin{figure}[!htp]
  \centering
  \includegraphics[width = 0.48\textwidth]{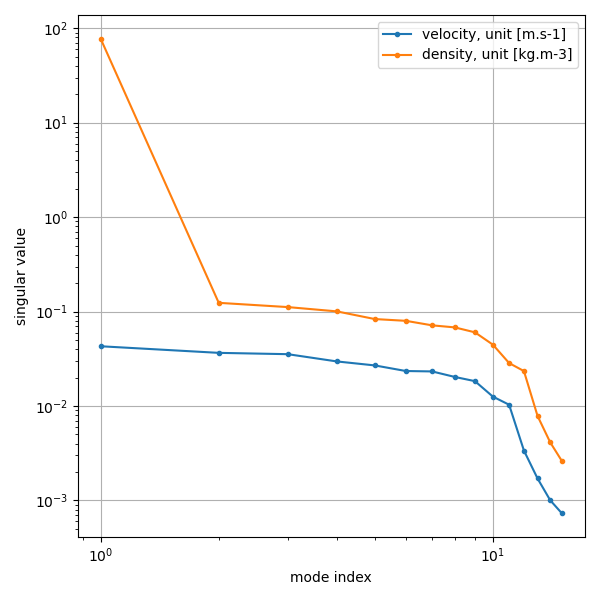}
  \includegraphics[width = 0.48\textwidth]{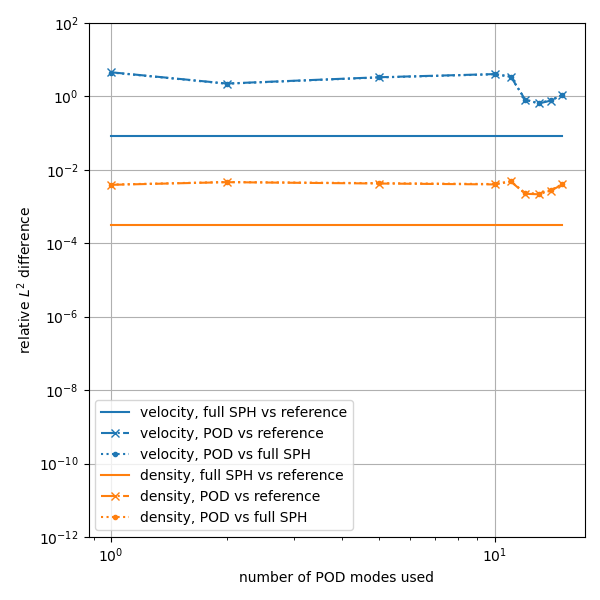}
  \caption{
  The singular value of the POD modes (left figure) and the relative $L ^{2}$ error (right figure) among the reference solution, the full SPH solution, and the POD-MOR of SPH simulation with different number of POD modes used, where
  $15$ snapshots are uniformly collected from $t \in \[ 0, 1.6 \]$ for the POD data.}
  \label{fig:numerical__SPH_PBC_test23_POD_nx400_nt20k_nn1d04_PodDataSph15First150m_error}
\end{figure}

\begin{figure}[!htp]
  \centering
  \includegraphics[width = 0.48\textwidth]{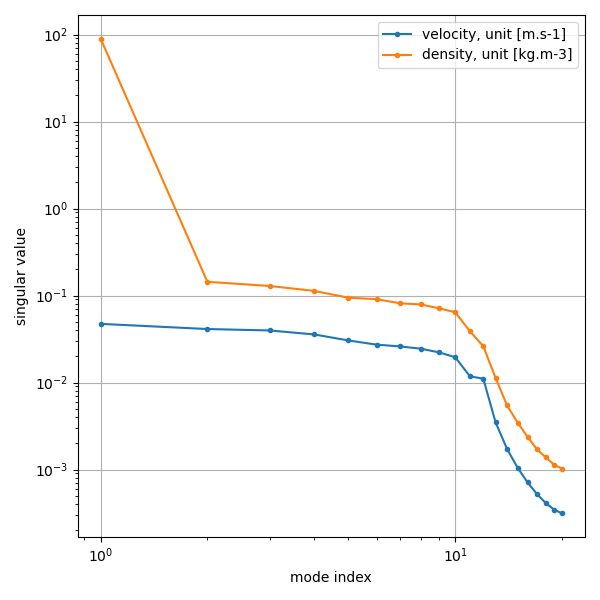}
  \includegraphics[width = 0.48\textwidth]{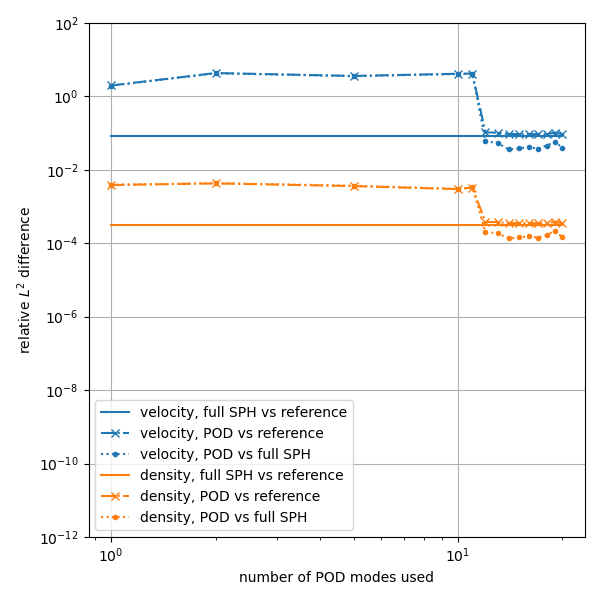}
  \caption{
  The singular value of the POD modes (left figure) and the relative $L ^{2}$ error (right figure) among the reference solution, the full SPH solution, and the POD-MOR of SPH simulation with different number of POD modes used, where
  $20$ snapshots are uniformly collected from $t \in \[ 0, 1.6 \]$ for the POD data.}
\label{fig:numerical__SPH_PBC_test23_POD_nx400_nt20k_nn1d04_PodDataSph20First160m_error}
\end{figure}

\myadd{We further explore other cases with different POD data, i.e., with different numbers of snapshots and small temporal range.
Following the discussion in \autoref{section3_1_2}, we first study the case of selecting relatively few snapshots to cover the half period $T' / 2$.}
In \autoref{fig:numerical__SPH_PBC_test23_POD_nx400_nt20k_nn1d04_PodDataSph15First150m_error}, $15$ snapshots are uniformly collected from $t \in [ 0, 1.6 ]$ of the SPH solution. In \autoref{fig:numerical__SPH_PBC_test23_POD_nx400_nt20k_nn1d04_PodDataSph20First160m_error}, $20$ snapshots are uniformly collected from $t \in [ 0, 1.6 ]$ of the SPH solution. From the errors in the figures, it shows that the POD fails in the case with 15 snapshots while it works in the case with $20$ snapshots. This indicates that enough data is necessary to adequately cover the spatial solution space. By testing the cases in between, we find that $16$ seems to be the critical value of uniform snapshots for the POD to work well in this context.

\begin{figure}[!htp]
  \centering
  \includegraphics[width = 0.48\textwidth]{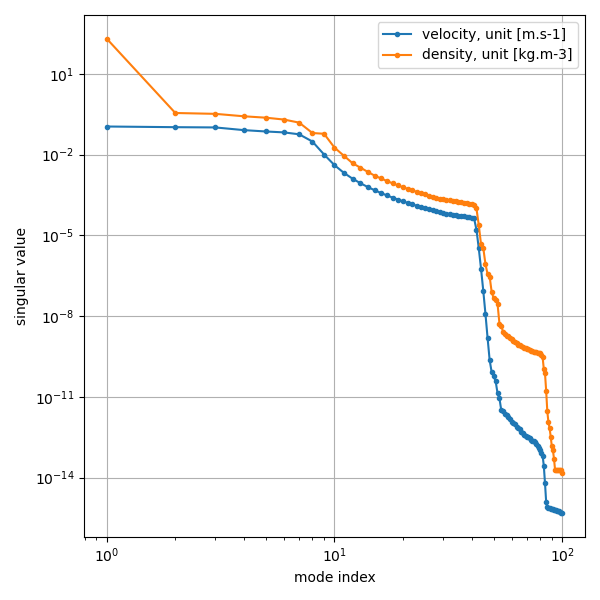}
  \includegraphics[width = 0.48\textwidth]{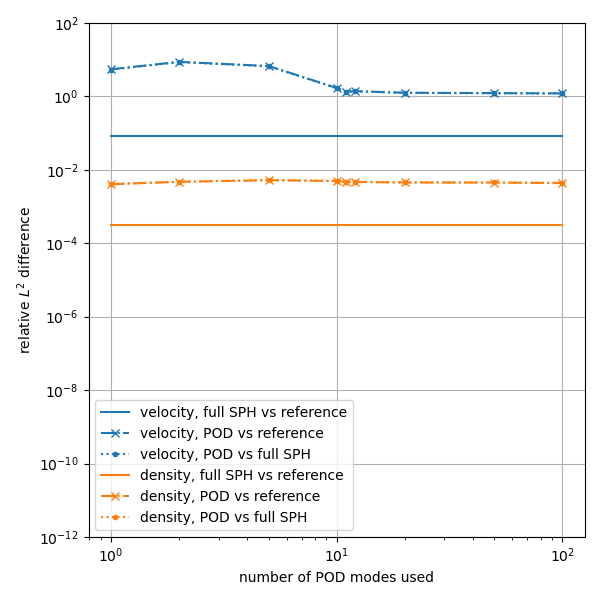}
  \caption{The singular value of the POD modes (left figure) and the relative $L ^{2}$ error (right figure) among the reference solution, the full SPH solution, and the POD-MOR of SPH simulation with different number of POD modes used, where
  $100$ snapshots are uniformly collected from $t \in \[ 0, 1 \]$ for the POD data.}
  \label{fig:numerical__SPH_PBC_test23_POD_nx400_nt20k_nn1d04_PodDataSph100First100m_error}
\end{figure}

Next, \myadd{by following the discussion in \autoref{section3_1_3}}, $100$ snapshots are uniformly collected from $t \in [ 0, 1 ]$ with spacing $0.01$ of the SPH solution. The POD-MOR fails in this case, as indicated by the POD error in \autoref{fig:numerical__SPH_PBC_test23_POD_nx400_nt20k_nn1d04_PodDataSph100First100m_error}. This suggests that using a small temporal range for the POD data is not sufficient to capture all the essential modes in this context with complicated initial conditions.  Based on the properties of the reference solution, it is more effective to collect POD data that covers the range  $\[ 0, T' / 2 \]$.  In practical applications, some prior knowledge is beneficial for selecting an appropriate temporal range.

\subsubsection{\texorpdfstring{Acceleration of POD-MOR}{Acceleration of POD-MOR}}

Now, we examine the acceleration of the POD-MOR in the SPH simulation for this example. 
For the POD data, we use the baseline case that $100$ snapshots are uniformly collected from $t \in \[ 0, 1.6 \]$ of the SPH solution.

\begin{figure}[!htp]
  \centering
  \includegraphics[width = 0.48\textwidth]{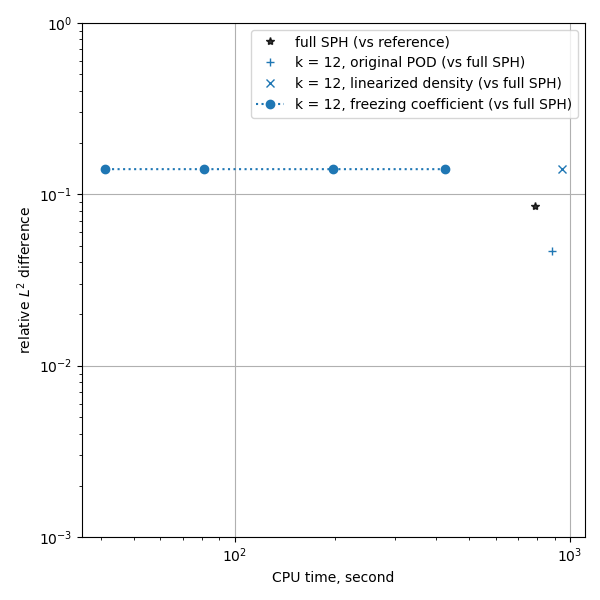}
  \includegraphics[width = 0.48\textwidth]{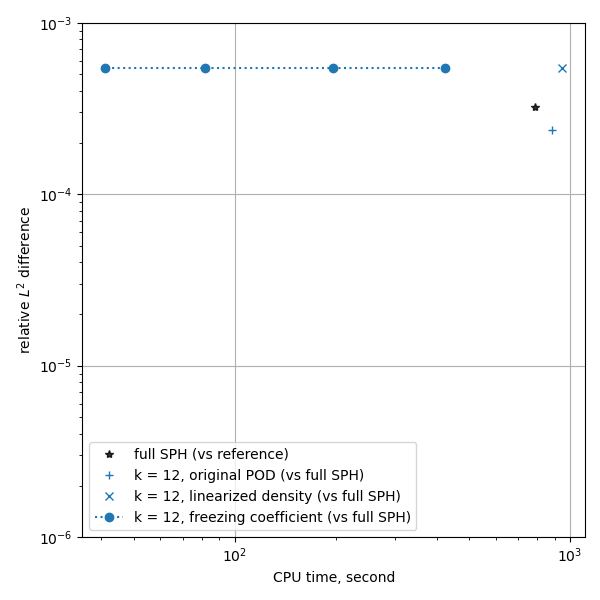}
  \caption{
  The relative $L ^{2}$ error of velocity $v$ (left figures) and density $\rho$ (right figures) as well as the CPU time (in seconds).
  The star marker represents the $L ^{2}$ error and CPU time of the full SPH simulation compared to the reference solution,
  the plus marker represents those of the original POD-MOR compared to the full SPH simulation, and
  the x marker represents those of the POD-MOR with linearized density compared to the full SPH solution.
  The dots (connected with dashed lines) represent the $L ^{2}$ error and CPU time of accelerated POD-MOR with freezing coefficients compared to the full SPH solution, where the freezing step is chosen from $n _{\textup{freeze}} = 1, 2, 5, 10$, plotted from right to the left.}
  \label{fig:numerical__SPH_PBC_test23_POD_nx400_nt20k_nn1d04_PodDataSph100First160m_acceleration}
\end{figure}

The accelerated version of POD-MOR is implemented with freezing coefficients in \autoref{eq:model__acceleration__V} and \autoref{eq:model__acceleration__D}. There is an intermediate step of density linearization \autoref{eq:density_linearization} between the original POD-MOR (\autoref{eq:model__POD__general}, \autoref{eq:model__SPH__D}, \autoref{eq:model__SPH__V}) and final version of POD-MOR with freezing coefficients (\autoref{eq:model__acceleration__V}, \autoref{eq:model__acceleration__D}).  
The freezing step is chosen from $n _{\textup{freeze}} = 1, 2, 5, 10$, and the number of POD modes used is $k = 12$. 
In \autoref{fig:numerical__SPH_PBC_test23_POD_nx400_nt20k_nn1d04_PodDataSph100First160m_acceleration}, we compare the CPU time and error of different numerical solutions, including the full SPH simulation, the original POD-MOR, the POD-MOR with density linearization, and the POD-MOR with freezing coefficients.

The figure shows that both the original POD-MOR and the POD-MOR with density linearization require similar CPU time as the full SPH simulation, since the additional computation mainly involves projection between the solution space and POD space. 
The original POD-MOR achieves a POD error that is lower than the numerical error of the SPH simulation, while the error increases after applying density linearization, making it larger than the numerical error of the SPH method. This is mainly because density variation is still relatively large in the case without damping and with complicated initial conditions, but it will improve with damping effect (shown later). 
The accelerated POD-MOR with freezing coefficients operates in reduced coordinates, so even with $n _{\textup{freeze}} = 1$, the CPU time is lower than that of the original POD, as $12 = k \ll N = 400$. 
As $n _{\textup{freeze}}$ increases, the CPU time decreases significantly while the extra error induced is negligible.  
This indicates that the freezing technique contributes a higher order error compared to the POD error when $n _{\textup{freeze}} \le 10$. 
The reason for this is mainly because the density variation is small in the small freezing time interval. Overall, choosing $n _{\textup{freeze}} = 10$ results in a total POD error that is less than twice the numerical error of the SPH simulation, while achieving a significant speed-up, saving up to $95\%$ of CPU time. 


\subsection{\myadd{Results with damping}}
\label{sec:numerical__SPH_PBC_test22}

In this example, we investigate the damping effect. The initial velocity is defined as in \autoref{eq:numerical__Lagrangian_PBC_test23_V_initial}, 
and the initial density is  $\rho _{0} = 1$. 
The Young's modulus and the external force are the same as before $c = 0.1, f \( t \) = 0$, but the damping coefficient is $d = 10$. 

\begin{figure}[!htp]
  \centering
  \includegraphics[width = 0.48\textwidth]{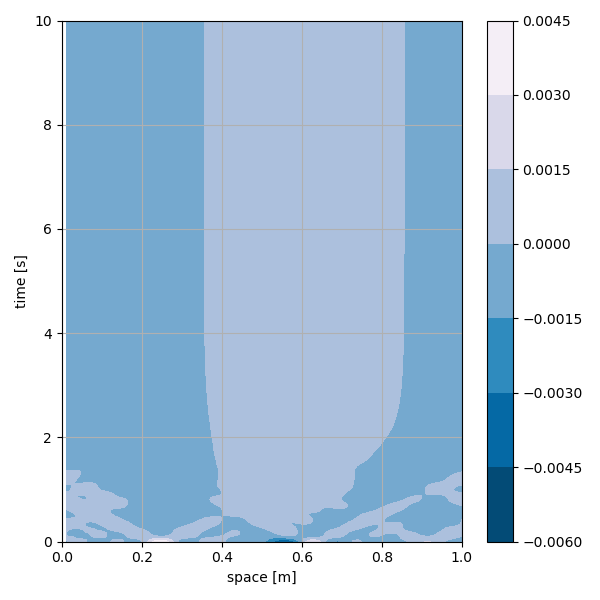}
  \includegraphics[width = 0.48\textwidth]{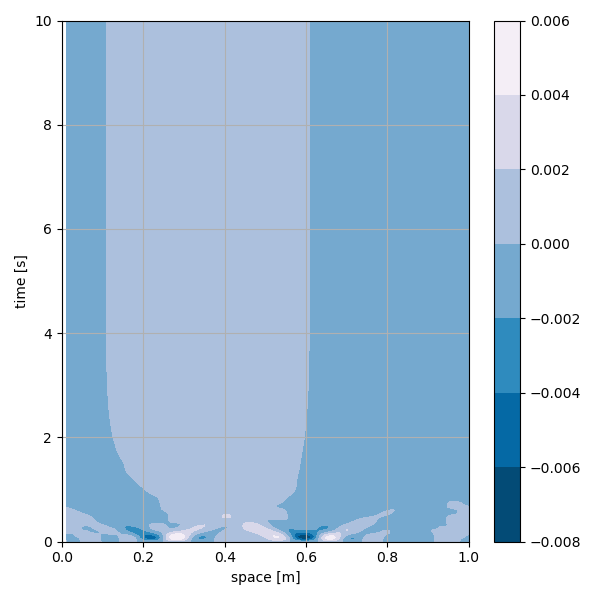}
  \caption{Velocity $v$ (left figure) and density $\rho - \rho _{0}$ (right figure) profile of the reference solution.}
  \label{fig:numerical__SPH_PBC_test22_POD_profile}
\end{figure}

The reference solution for \autoref{eq:model__spring__VD__V} and \autoref{eq:model__spring__VD__D} is generated using a first-order finite difference method in space with $\Delta x = 5 \e{-3}$ and a second-order Runge-Kutta scheme in time with $\Delta t = 1 \e{-4}$. 
The reference solutions for velocity and density are shown in \autoref{fig:numerical__SPH_PBC_test22_POD_profile}.

The SPH simulation evolves according to equations \autoref{eq:model__SPH__V} and \autoref{eq:model__SPH__D}. 
We set the spatial step size as $\Delta x = 1 \e{-2}$, (i.e., $N = 100$ smoothed particles) and 
the SPH smoothing length as $h = 1.04 \Delta x$. 
Using the explicit Euler scheme with $\Delta t = 2 \e{-4}$, the relative $L ^{2}$ error in velocity is approximately $6.456 \e{-2}$, while the relative $L ^{2}$ error in density is about $4.188 \e{-5}$.

\subsubsection{\texorpdfstring{\myadd{Results of original POD-MOR}}{A Baseline POD-MOR Case}}

\begin{figure}[!htp]
  \centering
  \includegraphics[width = 0.48\textwidth]{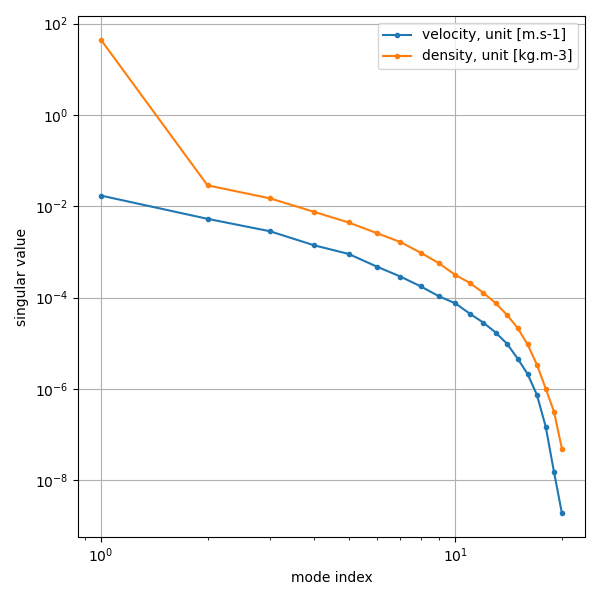}
  \includegraphics[width = 0.48\textwidth]{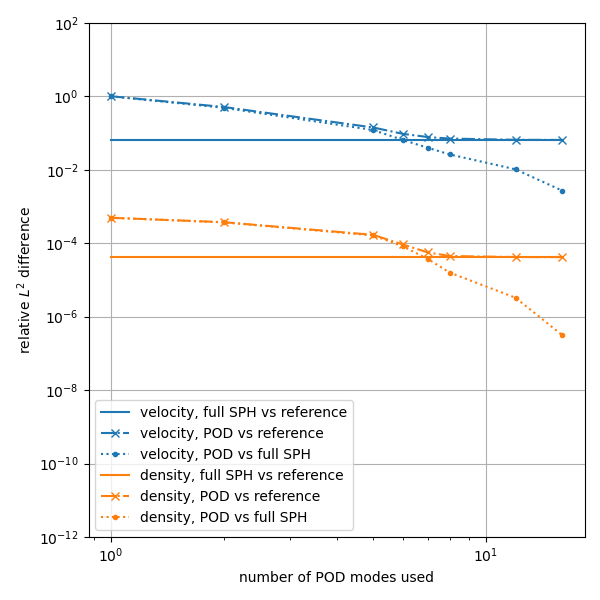}
  \caption{The singular value of the POD modes (left figure) and the relative $L ^{2}$ error (right figure) among the reference solution, the full SPH solution, and the POD-MOR of SPH simulation with different number of POD modes used, where
  $20$ snapshots are uniformly collected from $t \in \[ 0, 1.6 \]$ for the POD data.}
  \label{fig:numerical__SPH_PBC_test22_POD_nx100_nt5k_nn1d04_PodDataSph20First160m_error}
\end{figure}

Now, we apply the POD-MOR to the SPH simulation. Following the previous strategy on snapshots and referring to the case without damping in \autoref{sec:numerical__SPH_PBC_test23}, we first collect $20$ snapshots uniformly from $t \in \[ 0, 1.6 \]$. 
\autoref{fig:numerical__SPH_PBC_test22_POD_nx100_nt5k_nn1d04_PodDataSph20First160m_error} shows the singular values of the POD modes and the relative $L ^{2}$ error of the POD-MOR from \autoref{eq:model__POD__general} in . The POD error is smaller than the SPH numerical error when at least $7$ modes are used.

\begin{figure}[!htp]
  \centering
  \includegraphics[width = 0.48\textwidth]{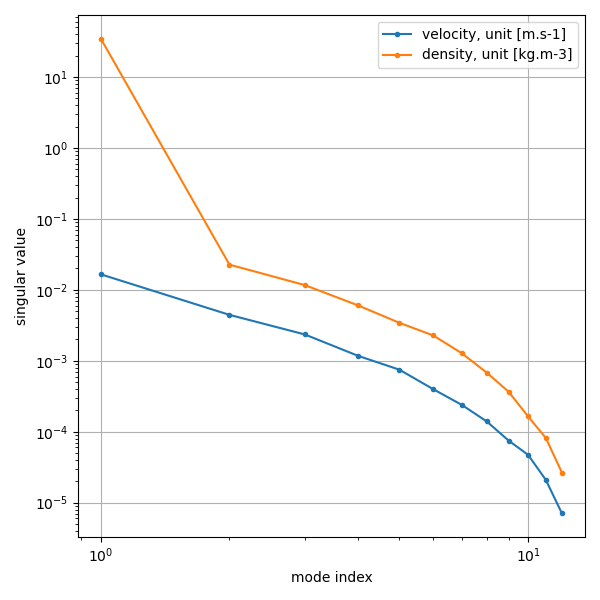}
  \includegraphics[width = 0.48\textwidth]{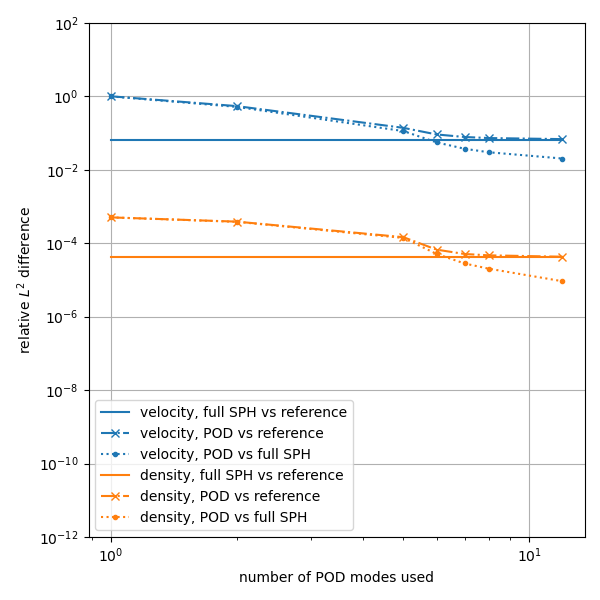}
  \caption{The singular value of the POD modes (left figure) and the relative $L ^{2}$ error (right figure) among the reference solution, the full SPH solution, and the POD-MOR of SPH simulation with different number of POD modes used, where
  $12$ snapshots are uniformly collected from $t \in \[ 0, 1.44 \]$ for the POD data.}
  \label{fig:numerical__SPH_PBC_test22_POD_nx100_nt5k_nn1d04_PodDataSph12First144m_error}
\end{figure}

Comparing \autoref{fig:numerical__SPH_PBC_test22_POD_nx100_nt5k_nn1d04_PodDataSph20First160m_error} with \autoref{fig:numerical__SPH_PBC_test23_POD_nx400_nt20k_nn1d04_PodDataSph20First160m_error}, we notice that the damping effect significantly improves the POD behavior.
In fact, we are able to choose fewer snapshots within a smaller temporal region so that the POD can still work well. This is verified by collecting $12$ snapshots uniformly from $t \in \[ 0, 1.44 \]$. \autoref{fig:numerical__SPH_PBC_test22_POD_nx100_nt5k_nn1d04_PodDataSph12First144m_error} shows the singular values of the POD modes and the relative $L ^{2}$ error of the POD-MOR from \autoref{eq:model__POD__general}. 
The POD error is still smaller than the SPH numerical error when at least $7$ modes are used. It is important to note that in cases without damping, the POD-MOR approach is not effective with this selection of POD data. This indicates that POD-MOR is more likely to achieve a low POD error when damping is present.

In comparison to the scenario without damping, fewer POD modes ($7$ instead of $12$) are needed for effective POD-MOR. 
Moreover, the numerical error of the full SPH solution is smaller, and the POD error is even smaller when damping is applied. 
This can be attributed to the rapid decay of high-frequency modes in the initial condition due to damping, resulting in a simpler spatial structure in the reference solution, as shown in \autoref{fig:numerical__SPH_PBC_test22_POD_profile}.

\subsubsection{\texorpdfstring{\myadd{Acceleration of POD-MOR}}{Acceleration of POD-MOR}}

\begin{figure}[!htp]
  \centering
  \includegraphics[width = 0.48\textwidth]{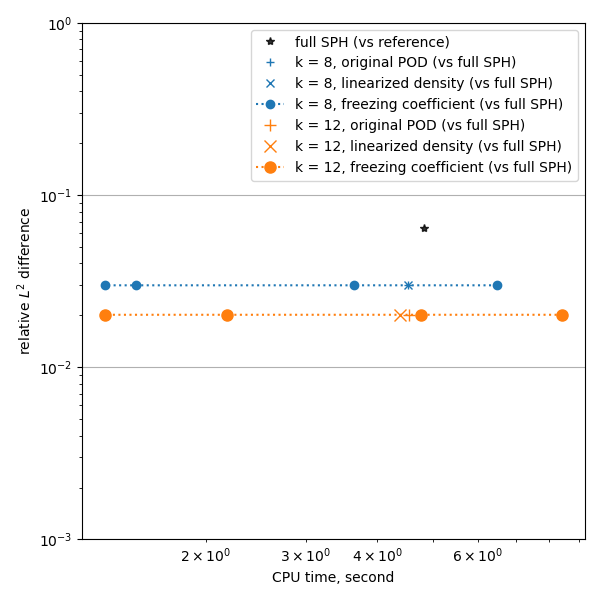}
  \includegraphics[width = 0.48\textwidth]{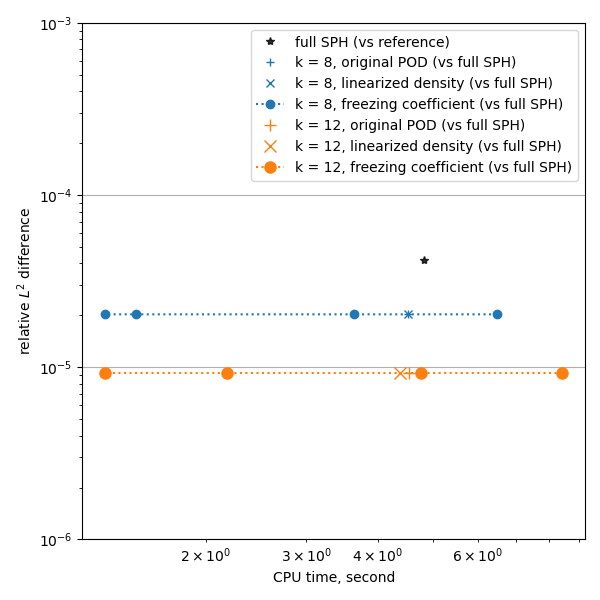}
  \caption{The relative $L ^{2}$ error of velocity $v$ (left figures) and density $\rho$ (right figures) as well as the CPU time (in seconds).
  The dots (connected with dashed lines) represent the $L ^{2}$ error and CPU time of accelerated POD-MOR with freezing coefficients compared to the full SPH solution, where the freezing step is chosen from $n _{\textup{freeze}} = 1, 2, 5, 10$, plotted from right to the left.  
  }
  \label{fig:numerical__SPH_PBC_test22_POD_nx100_nt5k_nn1d04_PodDataSph12First144m_acceleration}
\end{figure}

The accelerated version of POD-MOR, as defined in \autoref{eq:model__acceleration__V} and \autoref{eq:model__acceleration__D}, is implemented with freezing steps chosen from $n _{\textup{freeze}} = 1, 2, 5, 10$, and the number of POD modes set to $k = 8, 12$. 
In \autoref{fig:numerical__SPH_PBC_test22_POD_nx100_nt5k_nn1d04_PodDataSph12First144m_acceleration}, we compare the CPU time and errors of various numerical solutions, including the full SPH simulation, the original POD-MOR, the
POD-MOR with density linearization, and the POD-MOR with freezing coefficients. As the number of POD modes $k$ increases from $8$ to $12$, the POD error decreases, while the CPU time increases slightly.  Similar to previous subsection, both the original POD-MOR and the POD-MOR with density
linearization require similar CPU time as the full SPH simulation. As $n _{\textup{freeze}}$ increases, the CPU time decreases, and the POD error in velocity and density remain relatively stable and much smaller than the error of full SPH solution. 
This suggests that the freezing technique (as well as the POD) introduces a higher order error compared to the error of full SPH solution when $n _{\textup{freeze}} \le 10$. Overall, 
choosing $n _{\textup{freeze}} = 10$ results in a total POD error still smaller than the error of the SPH simulation, while achieving a significant (about $72\%$) saving in CPU time.

\subsection{\myadd{Results with both damping and external Force}}
\label{sec:numerical__SPH_PBC_test21b}

In this example, we explore the combined effects of damping and an external force. The initial velocity is defined as in \autoref{eq:numerical__Lagrangian_PBC_test23_V_initial}.
The initial density $\rho _{0} = 1$,  the Young's modulus is $c = 0.1$, and
the damping coefficient $d = 10$ are the same as previous subsection. But the external force is nonzero and given by
\begin{align}
  f \( t \) = 1\e{-2} \sum _{l = 1, 3, 10} \cos \( 10 l \frac{t}{T} - l ^{3} \).
\end{align}
Another example with mild damping and external force is studied in \autoref{sec:numerical__SPH_PBC_test21}.

\begin{figure}[!htp]
  \centering
  \includegraphics[width = 0.48\textwidth]{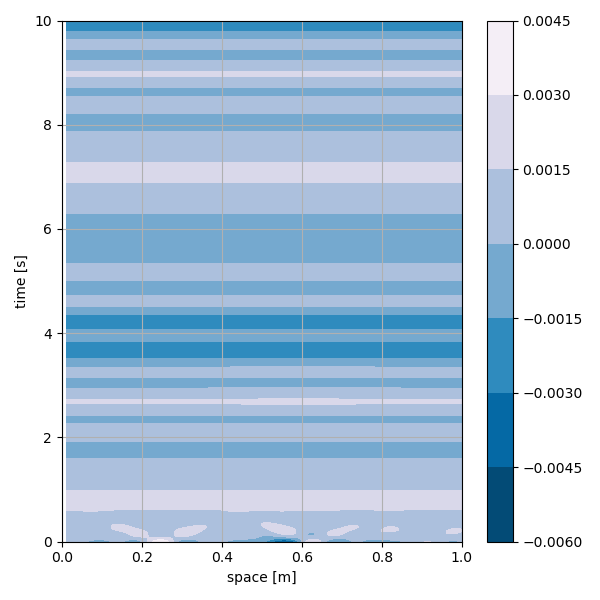}
  \includegraphics[width = 0.48\textwidth]{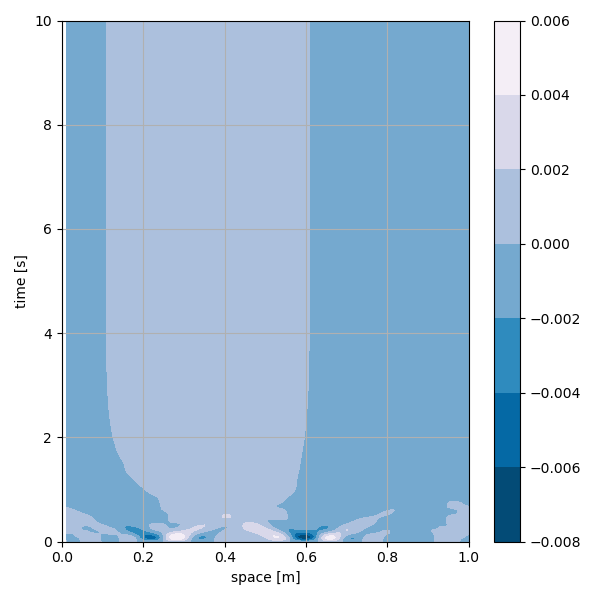}
  \caption{Velocity $v$ (left figure) and density $\rho - \rho _{0}$ (right figure) profile of the reference solution.}
  \label{fig:numerical__SPH_PBC_test21b_POD_profile}
\end{figure}

The reference solution for \autoref{eq:model__spring__VD__V} and \autoref{eq:model__spring__VD__D} is generated using a first-order finite difference method in space with $\Delta x = 5 \e{-3}$ and a second-order Runge-Kutta scheme in time with $\Delta t = 1 \e{-4}$. 
The reference solutions for velocity and density are shown in \autoref{fig:numerical__SPH_PBC_test21b_POD_profile}. The full SPH simulation evolves according to equations \autoref{eq:model__SPH__V} and \autoref{eq:model__SPH__D}. 
We set the spatial step size as $\Delta x = 1 \e{-2}$, (i.e., $N = 100$ smoothed particles) and 
the SPH smoothing length as $h = 1.04 \Delta x$. 
Using the explicit Euler scheme with $\Delta t = 2 \e{-4}$, the relative $L ^{2}$ error in velocity is approximately $6.702 \e{-3}$ and the relative $L ^{2}$ error in density is $4.188 \e{-5}$.

\subsubsection{\texorpdfstring{\myadd{Results of original POD-MOR}}{A Baseline POD-MOR Case}}

\begin{figure}[!htp]
  \centering
  \includegraphics[width = 0.48\textwidth]{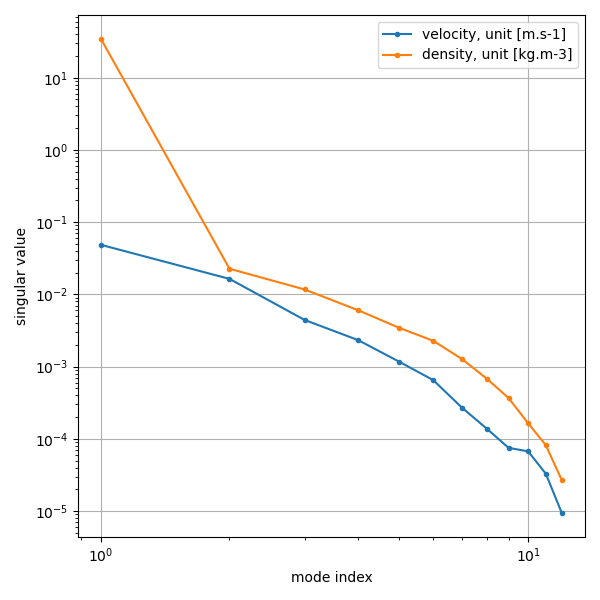}
  \includegraphics[width = 0.48\textwidth]{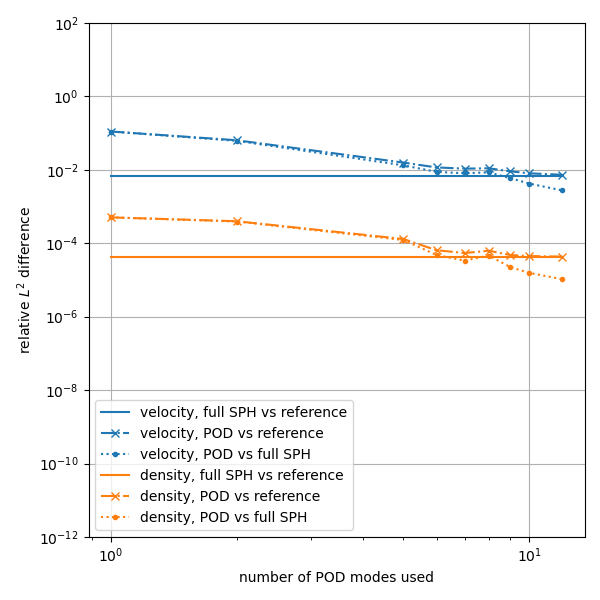}
  \caption{The singular value of the POD modes (left figure) and the relative $L ^{2}$ error (right figure) among the reference solution, the full SPH solution, and the POD-MOR of SPH simulation with different number of POD modes used, where
  $12$ snapshots are uniformly collected from $t \in \[ 0, 1.44 \]$ for the POD data.}
  \label{fig:numerical__SPH_PBC_test21b_POD_nx100_nt5k_nn1d04_PodDataSph12First144m_error}
\end{figure}

We apply the POD-MOR to the SPH simulation, by using collecting $12$ snapshots uniformly from $t \in[ 0, 1.44 ]$. This choice of POD data follows that of the case with damping and without external force, as discussed in \autoref{sec:numerical__SPH_PBC_test22}. \autoref{fig:numerical__SPH_PBC_test21b_POD_nx100_nt5k_nn1d04_PodDataSph12First144m_error} shows the singular values of the POD modes and the relative $L ^{2}$ error of the original POD-MOR from \autoref{eq:model__POD__general}. 
The POD error is smaller than the SPH numerical error when at least $9$ modes are employed. Compared to the scenario without the external force, more POD modes ($9$ instead of $7$) are necessary for effective POD-MOR. However, this is still fewer than the $12$ modes required in the case with no damping and no external forces.

\subsubsection{\texorpdfstring{\myadd{Acceleration of POD-MOR}}{Acceleration of POD-MOR}}

\begin{figure}[!htp]
  \centering
  \includegraphics[width = 0.48\textwidth]{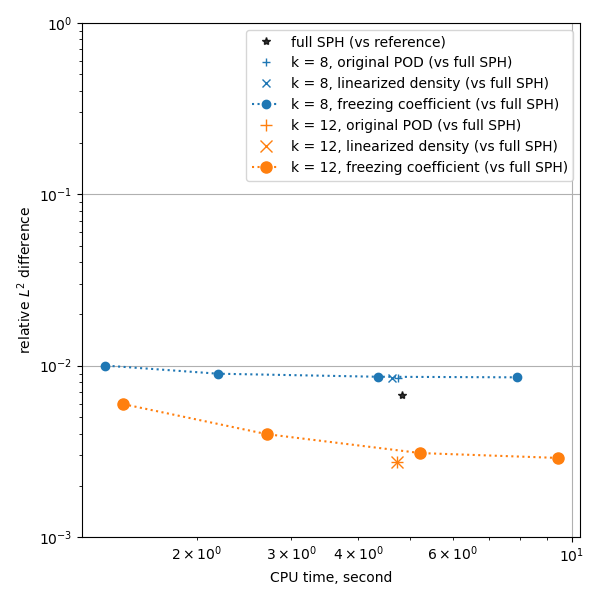}
  \includegraphics[width = 0.48\textwidth]{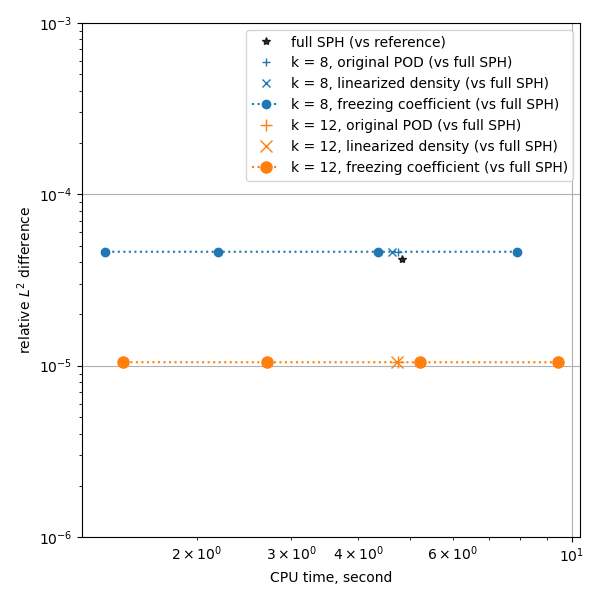}
  \caption{
  The relative $L ^{2}$ error of velocity $v$ (left figures) and density $\rho$ (right figures) as well as the CPU time (in seconds).
  The dots (connected with dashed lines) represent the $L ^{2}$ error and CPU time of accelerated POD-MOR with freezing coefficients compared to the full SPH solution, where the freezing step is chosen from $n _{\textup{freeze}} = 1, 2, 5, 10$, plotted from right to the left.  
  }
  \label{fig:numerical__SPH_PBC_test21b_POD_nx100_nt5k_nn1d04_PodDataSph12First144m_acceleration}
\end{figure}

The accelerated version of POD-MOR, as defined in \autoref{eq:model__acceleration__V} and \autoref{eq:model__acceleration__D}, is implemented with freezing steps chosen from $n _{\textup{freeze}} = 1, 2, 5, 10$, and the number of POD modes set to $k = 8, 12$. \autoref{fig:numerical__SPH_PBC_test21b_POD_nx100_nt5k_nn1d04_PodDataSph12First144m_acceleration} compares the CPU time and errors of various numerical solutions, including the full SPH simulation, the original POD-MOR, the
POD-MOR with density linearization, and the POD-MOR with freezing coefficients. As the number of POD modes $k$ increases from $8$ to $12$, the POD error decreases, while the CPU time increases slightly. 
As $n _{\textup{freeze}}$ increases, CPU time decreases, while the POD error in velocity tends to increase (but below the error of the full SPH simulation), and the POD error in density remains relatively stable and small. The graduate increase in POD error with increasing $n _{\textup{freeze}}$ is different from the case without the external force (see \autoref{fig:numerical__SPH_PBC_test22_POD_nx100_nt5k_nn1d04_PodDataSph12First144m_acceleration}). This is mainly because of the time-dependent external force $\mathbf{f}$ in \autoref{eq:model__acceleration__V} with high frequency temporal mode, which is updated every $n _{\textup{freeze}}$ steps in the acceleration algorithm, resulting in a relatively large error with large $n _{\textup{freeze}}$. Overall, choosing $n _{\textup{freeze}} = 10$ results in the total POD error still smaller than the error of the full SPH simulation, while achieving a significant (about $75\%$) saving in CPU time.

\subsection{\myadd{Results with inhomogeneous Young's Modulus}}

In this example, we examine an inhomogeneous Young's modulus. The initial velocity is given by
\begin{align}
  v _{0} \( x \) = 1 \e{-3} \sin \( 2 \pi l \frac{x}{L} \),
  \label{eq:numerical__Lagrangian_PBC_test26_V_initial}
\end{align}
and the initial density is $\rho _{0} = 1$. 
the damping coefficient is $d = 0$ and The external force is $f \( t \) = 0$. 
The Young's modulus is defined as 
\begin{align}
  c \( x \) 
  = 
  \left\{
  \begin{aligned}
    0.1, & \quad 0 < x / L \leq 1 / 2, \\
    0.2, & \quad 1/2 < x / L < 1.
  \end{aligned}
  \right.
  \label{eq:numerical__SPH_PBC_test26_piecewise_c}
\end{align}

\begin{figure}[!htp]
  \centering
  \includegraphics[width = 0.48\textwidth]{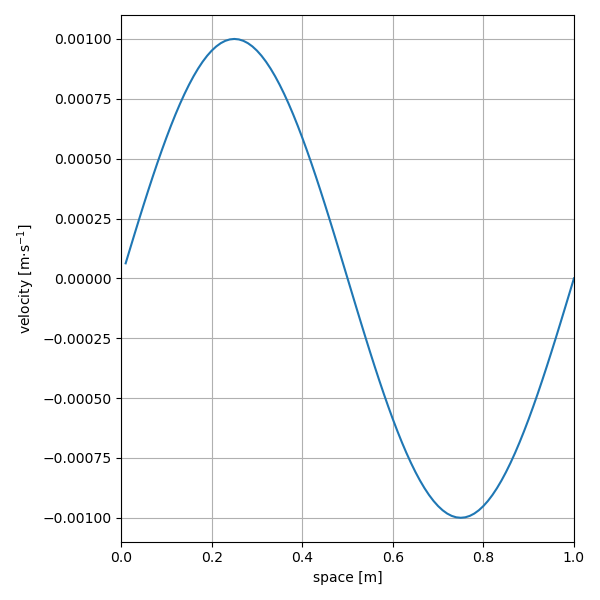}
  \includegraphics[width = 0.48\textwidth]{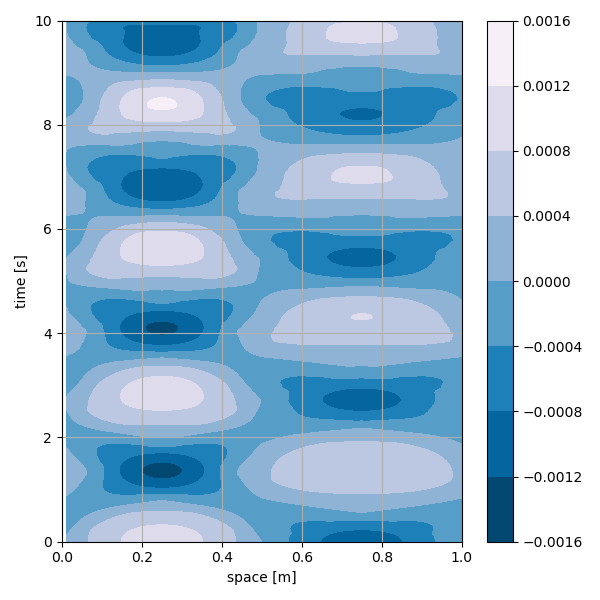}
  \caption{Initial condition (left figure) and profile (right figure) of the reference velocity.}
  \label{fig:numerical__SPH_PBC_test26_POD_V_profile}
\end{figure}

The reference solution for 
\autoref{eq:model__spring__VD__V} and \autoref{eq:model__spring__VD__D} is obtained using a first-order finite difference method in space with $\Delta x = 5 \e{-3}$ and a second-order Runge-Kutta scheme in time with $\Delta t = 1 \e{-4}$.
\autoref{fig:numerical__SPH_PBC_test26_POD_V_profile} shows the initial condition $v _{0} \( x \)$ and the solution profile of velocity $v \( x, t \)$ of the reference solution. The SPH simulation evolves according to equations \autoref{eq:model__SPH__V} and \autoref{eq:model__SPH__D}. 
With a spatial step size of $\Delta x = 1 \e{-2}$ (i.e., $N = 100$ smoothed particles) and a smoothing length of $h = 1.04 \Delta x$, we employ the explicit Euler scheme for time integration with  $\Delta t = 2 \e{-4}$. 
Under these conditions, the relative $L ^{2}$ error in velocity is approximately $6.321 \e{-2}$, while the relative $L ^{2}$ error in density is about $1.358 \e{-4}$.

\subsubsection{\texorpdfstring{\myadd{Results of original POD-MOR}}{A Baseline POD-MOR Case}}

We apply the POD-MOR technique to the SPH simulation, by collecting $40$ snapshots uniformly from $t \in \[ 0, 3.2 \]$. The snapshots and temporal range are doubled here compared with \autoref{sec:numerical__SPH_PBC_test28}, because we expect different behavior of solutions due to piecewise Young's modulus in \autoref{eq:numerical__SPH_PBC_test26_piecewise_c}.

\begin{figure}[!htp]
  \centering
  \includegraphics[width = 0.8\textwidth]{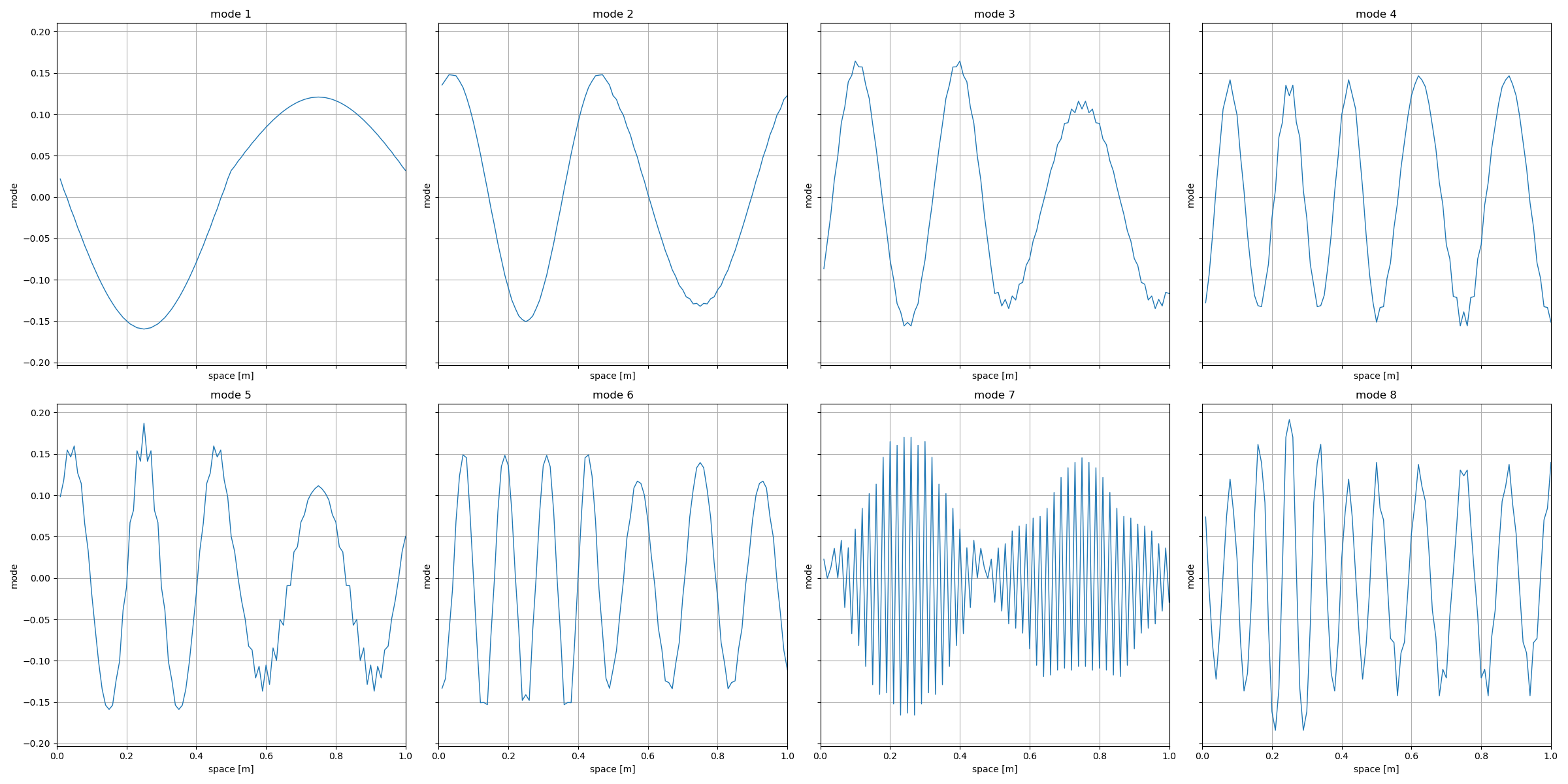}
  \includegraphics[width = 0.8\textwidth]{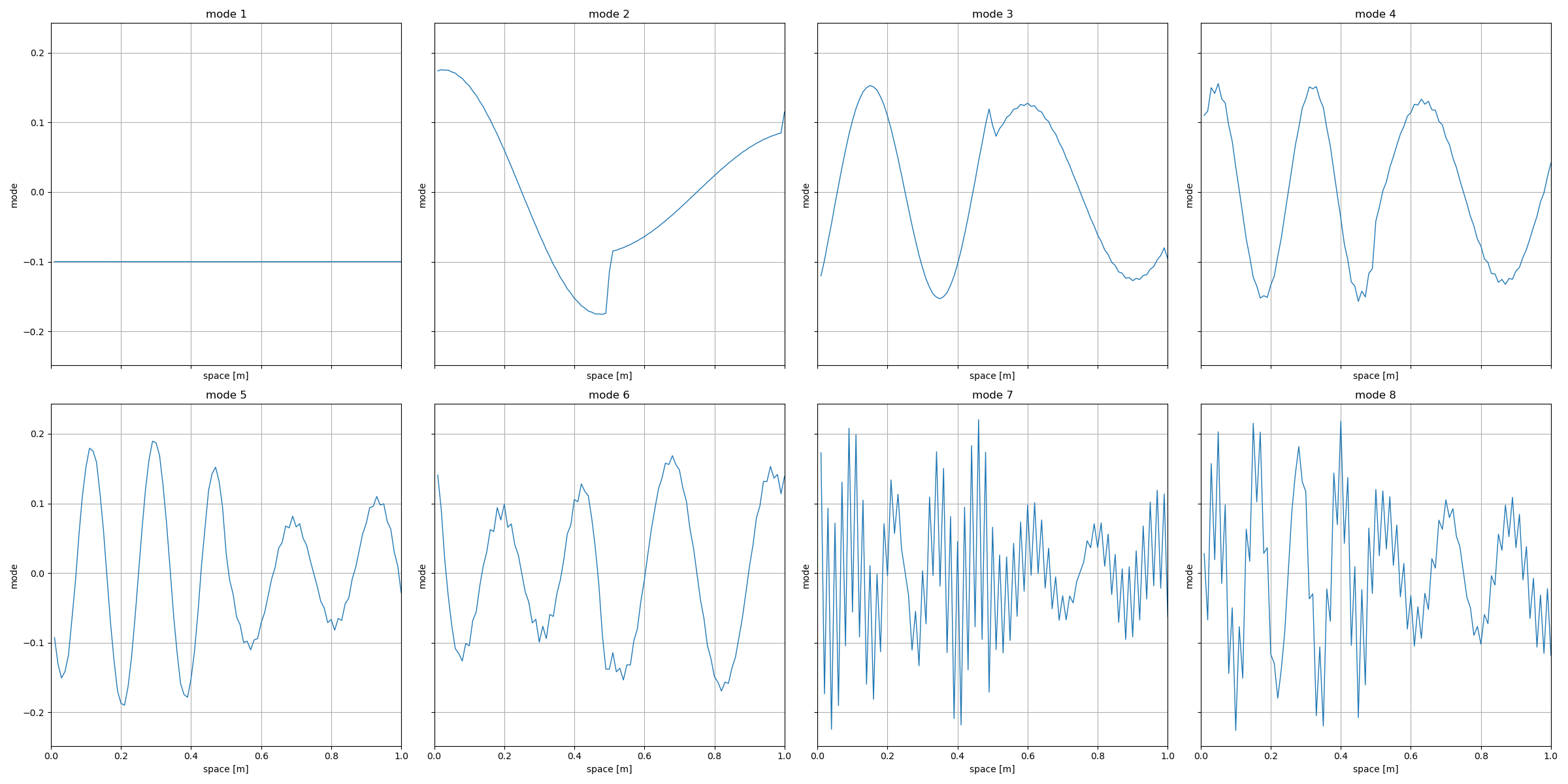}
  \caption{The dominant POD modes of velocity (upper figure) and density (lower figure).}
  \label{fig:numerical__SPH_PBC_test26_POD_nx100_nt5k_nn1d04_PodDataSph40First320m_modes}
\end{figure}

\begin{figure}[!htp]
  \centering
  \includegraphics[width = 0.48\textwidth]{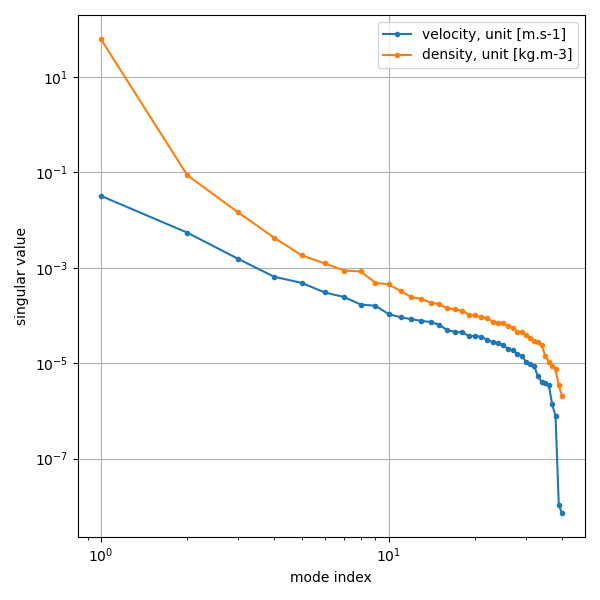}
  \includegraphics[width = 0.48\textwidth]{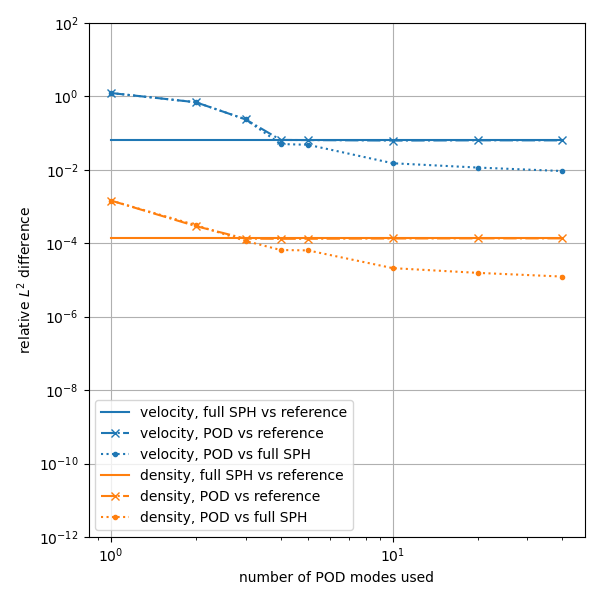}
  \caption{The singular value of the POD modes (left figure) and the relative $L ^{2}$ error (right figure) among the reference solution, the full SPH solution, and the POD-MOR of SPH simulation with different number of POD modes used, where
  $40$ snapshots are uniformly collected from $t \in \[ 0, 3.2 \]$ for the POD data.
  }
  \label{fig:numerical__SPH_PBC_test26_POD_nx100_nt5k_nn1d04_PodDataSph40First320m_error}
\end{figure}

\autoref{fig:numerical__SPH_PBC_test26_POD_nx100_nt5k_nn1d04_PodDataSph40First320m_modes} shows the dominant POD modes for both velocity and density, ordered according to the singular values in \autoref{fig:numerical__SPH_PBC_test26_POD_nx100_nt5k_nn1d04_PodDataSph40First320m_error}(left).  \autoref{fig:numerical__SPH_PBC_test26_POD_nx100_nt5k_nn1d04_PodDataSph40First320m_error}(right) shows the relative $L ^{2}$ error of the POD-MOR from \autoref{eq:model__POD__general}. 
Notably, the POD error is smaller than the SPH numerical error (i.e., POD works) when at least $4$ modes are employed. The differences between the softer region ($0 < x < L / 2$) and the stiffer region ($L / 2 < x < L$) are also reflected in the POD modes.  
We observe that the amplitude of both velocity and density is generally smaller and their spatial frequency tends to be lower in the stiffer region. 
This indicates the POD effectively groups smoothed particles and captures the distinct behaviors in different regions.


\subsubsection{\texorpdfstring{\myadd{Acceleration of POD-MOR}}{Acceleration of POD-MOR}}

\begin{figure}[!htp]
  \centering
  \includegraphics[width = 0.48\textwidth]{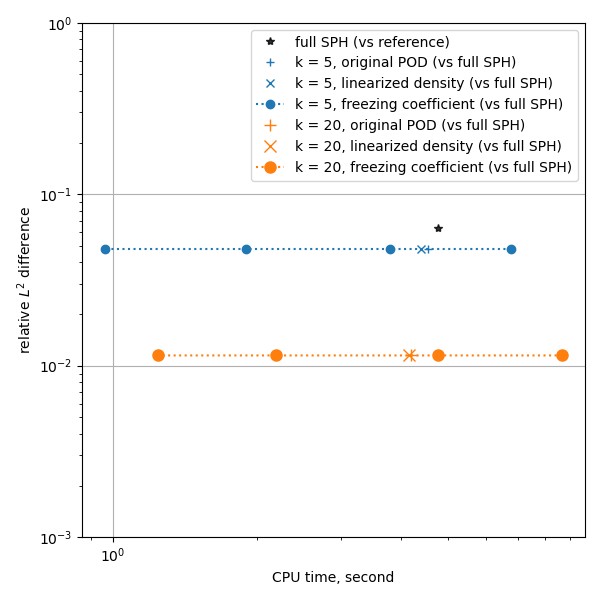}
  \includegraphics[width = 0.48\textwidth]{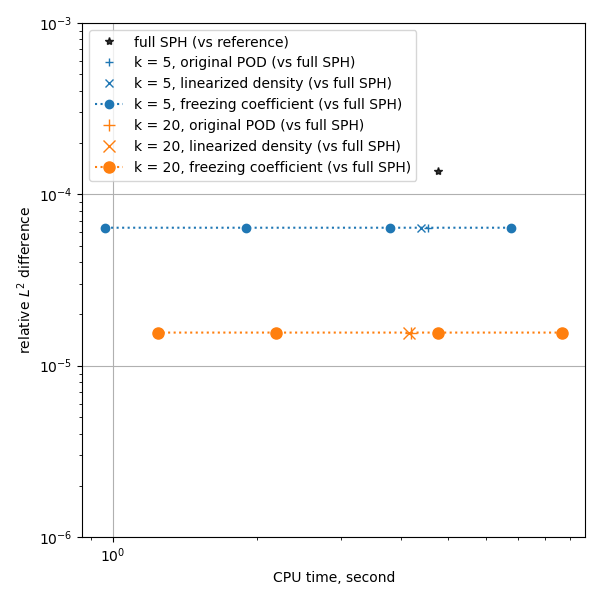}
  \caption{The relative $L ^{2}$ error of velocity $v$ (left figures) and density $\rho$ (right figures) as well as the CPU time (in seconds).
  The dots (connected with dashed lines) represent the $L ^{2}$ error and CPU time of accelerated POD-MOR with freezing coefficients compared to the full SPH solution, where the freezing step is chosen from $n _{\textup{freeze}} = 1, 2, 5, 10$, plotted from right to the left.  
  }
  \label{fig:numerical__SPH_PBC_test26_POD_nx100_nt5k_nn1d04_PodDataSph40First320m_acceleration}
\end{figure}

The accelerated version of POD-MOR, as defined in \autoref{eq:model__acceleration__V} and \autoref{eq:model__acceleration__D}, is implemented, with the freezing step selected from $n _{\textup{freeze}} = 1, 2, 5, 10$ and the number of POD modes set to $k = 5, 20$. 
\autoref{fig:numerical__SPH_PBC_test26_POD_nx100_nt5k_nn1d04_PodDataSph40First320m_acceleration} compares the CPU time and errors of various numerical solutions, including the full SPH simulation, the original POD-MOR, the
POD-MOR with density linearization, and the POD-MOR with freezing coefficients. We observe similar trend and results on computation time and errors, reflecting the effectiveness of the POD acceleration approach. 
When choosing $n _{\textup{freeze}} = 10$, the POD error remains smaller than the numerical error of the SPH simulation, while a significant saving (about $80\%$) in CPU time is observed.

\section{\myadd{Conclusion}}
\label{sec:conclusion}

In this study, we have explored the effectiveness of POD-MOR for Lagrangian SPH simulations. By using the mass-spring-damper system as an example, we demonstrate that POD-MOR significantly reduces computational complexity (or the DoFs) while maintaining accuracy in the SPH simulations.

Our investigation reveals that the dominant POD modes, automatically identified from the POD data, play a crucial role in achieving a robust POD-MOR. 
When selecting POD snapshots with proper strategy based on initial conditions and temporal period, POD-MOR effectively captures the system's behavior using a small number of dominant modes. The accuracy of the POD-MOR improves as the number of included modes increases. Under different contexts, the POD error introduced is smaller than the numerical error of the full SPH simulation.

We have explored the choice and effects of different snapshots used as POD data. In general, the POD work well when the snapshots cover at least half period of dynamical oscillation and the number of snapshots exceeds that of modes in initial condition (i.e., related to inherent dimensionality of the dynamic SPH system).  When the distribution of snapshots is limited in a small temporal region, POD works in some cases when increasing the number of snapshots but it may fail in some cases when the initial condition is very complicated. We also observe that using snapshots with random distribution can lead to better POD performance. The findings are potentially beneficial for real-world data collection.

Furthermore, the effects of various initial conditions as damping and inhomogeneous parameters are revealed. Particularly damping effect is general good for the effectiveness of POD-MOR (even if external forces are present), since high-frequency modes could be gradually removed during dynamics. 
With damping terms, we observe that the required number of POD modes and the POD error are reduced. 
In scenarios with inhomogeneous spring constants, the automatic grouping of smoothed particles and identification of motion patterns by the POD modes demonstrate the adaptability of the POD-MOR technique to various material properties.

We have also investigated the possibility of accelerating the POD-MOR process through linearization and freezing coefficients, finding that the extra error introduced is acceptable in various cases. This approach can lead to substantial savings in computational time without a significant compromise in MOR accuracy.
It is important to note that POD-MOR serves as an effective fast solver in parameter studies or design optimization problems requiring many simulations (e.g., the Friction Stir Welding problem). In these contexts, the bottleneck is the computation time and some mild error is tolerable.

Despite these promising results for effectiveness of POD-MOR in SPH simulations, there are certain limitations in this work. When initial conditions are complicated and there is no damping, it is very challenging for the POR-MOR to reduce the DoFs. The presence of high-frequency components in external forces may reduce the effectiveness of our acceleration techniques. The freezing coefficient method could potentially be improved and extended to an updated Lagrangian setting or applications in more complex systems. Future work will aim to address the identified limitations and and improvements in the acceleration techniques.






\section*{Acknowledgments}

Discussion: Shixin Xu, Faisal Habib

National Research Council of Canada

Fields Institute for Research in Mathematical Sciences

Beijing Normal University at Zhuhai

Canadian Association for Computational Science and Engineering

\bibliographystyle{plain} 
\bibliography{notes_SpringPOD}


\appendix

\section{Model Equivalence}

\label{sec:appendix_equivalence}

In this section, we claim that the mass-spring-damper model \autoref{eq:model__spring__Y_discrete} is an approximation of the velocity-density model \autoref{eq:model__spring__VD__D}, \autoref{eq:model__spring__VD__V}.
To this end, we first define a position-velocity model in the Lagrangian framework for position $z \( x, t \)$, velocity $v \( x, t \)$, and density $\rho \( x, t \)$, $x \in \[ 0, L \]$, $t \in \[ 0, T \]$, and then show that mass-spring-damper model is the discrete version of that, followed by proving the equivalence of  position-velocity model and the velocity-density model.

First, the  position-velocity model is  
\begin{align}
  \partial _{t} z \( x, t \)
  & =
  v \( x, t \)
  ,
  \label{eq:appendix__equivalence__ZV__Z}
  \\
  \rho _{0} \( x \)
  \partial _{t} v \( x, t \)
  & =
  \partial _{x} \[ c \( x, t \) \partial _{x} z \( x, t \) - c \( x, t \) \]
  +
  \rho _{0} \( x \) b \( x, t \)
  ,
  \label{eq:appendix__equivalence__ZV__V}
  \\
  \rho \( x, t \)
  & =
  \rho _{0} \( x \)
  \[ \partial _{x} z \( x, t \) \] ^{-1},
  \label{eq:appendix__equivalence__ZV__D}
\end{align}
where $\rho _{0} \( x \)$ is an given density at $t = 0$, and
the density $\rho$ can be computed after position $z$ and velocity $v$ are solved. Now we show that the mass-spring-damper model \autoref{eq:model__spring__Y_discrete} is a finite difference discretization of \autoref{eq:appendix__equivalence__ZV__Z} and \autoref{eq:appendix__equivalence__ZV__V} using a uniform mesh with $N$ points.
In fact, if we define the grid points as $x _{n} = nL / N$, $n = 0, 1, \cdots, N$, and the unit of area of cross section $a = 1 \[ \mykgms{0}{2}{0} \]$, we have the connections between terms in  \autoref{eq:model__spring__Y_discrete} and those in the above position-velocity model
\begin{align}
  m _{n} \approx \rho _{0} \( x _{n} \) \frac{aL}{N},
  \quad
  d _{n} \( t \) \approx \rho _{0} \( x _{n} \) d \( x _{n}, t \) \frac{aL}{N},
  \quad
  f _{n} \( t \) \approx f \( x _{n}, t \),
  \quad
  y _{n} \( t \) \approx z \( x _{n}, t \) - x _{n}.
\end{align}
Additionally, for the spring constants defined on the half-grid points,
\begin{align}
  k _{n + \frac{1}{2}} \( t \) \approx c \( x _{n + \frac{1}{2}}, t \) \frac{aL}{N}.
\end{align}
The standard central difference is used to approximate the spatial derivatives. 
So we can verify \autoref{eq:model__spring__Y_discrete} is a finite difference approximation of the position-velocity model \autoref{eq:appendix__equivalence__ZV__Z} and \autoref{eq:appendix__equivalence__ZV__V}. 

Next, we define the following velocity-density model,
\begin{align}
  \partial _{t} z \( x, t \)
  & =
  v \( x, t \)
  .
  \label{eq:appendix__equivalence__VD__Z}
  \\
  \rho \( x, 0 \)
  \partial _{t} v \( x, t \)
  & =
  \partial _{x} \[ c \( x, t \) \[ \rho \( x, 0 \) \[ \rho \( x, t \) \] ^{-1} \] - c \( x, t \) \]
  +
  \rho \( x, 0 \) b \( x, t \)
  .
  \label{eq:appendix__equivalence__VD__V}
  \\
  \partial _{t} \rho \( x, t \)
  & =
  - \[ \rho \( x, t \) \] ^{2} \[ \rho \( x, 0 \) \] ^{-1} \partial _{x} v \( x, t \)
  .
  \label{eq:appendix__equivalence__VD__D}
\end{align}
Here, the position $z$ is decoupled and can be solved after velocity $v$ and density $\rho$ are solved. 
So this model is just a copy of \autoref{eq:model__spring__VD__D}, \autoref{eq:model__spring__VD__V} with an additional position variable $z$. Now we show the equivalence between the position-velocity model (\autoref{eq:appendix__equivalence__ZV__Z}, \autoref{eq:appendix__equivalence__ZV__V}, \autoref{eq:appendix__equivalence__ZV__D}) and the velocity-density model (\autoref{eq:appendix__equivalence__VD__Z}, \autoref{eq:appendix__equivalence__VD__V}, \autoref{eq:appendix__equivalence__ZV__D}), when $\rho _{0} \( x \)$ in the position-velocity model is chosen as the initial condition $\rho \( x, 0 \)$ in the velocity-density model.
Since the evolution of $z$ is the same, now we focus on the rest parts.
The key observation here is, in one-dimensional space, we have
\begin{align}
   \partial _{x} z \( x, t \) \rho \( x, t \) = \rho \( x, 0 \).
  \label{eq:appendix__equivalence__motionAndDensity}
\end{align}
where $J \( x, t \) := \partial _{x} z \( x, t \)$ is the Jacobian of the motion in the Lagrangian framework.
On one hand, 
\autoref{eq:appendix__equivalence__ZV__D} can be derived from \autoref{eq:appendix__equivalence__motionAndDensity} and the assumption $\rho \( x, 0 \) = \rho _{0} \( x \)$; 
\autoref{eq:appendix__equivalence__ZV__V} can be derived from \autoref{eq:appendix__equivalence__VD__V}, with \autoref{eq:appendix__equivalence__ZV__D} and the assumption $\rho \( x, 0 \) = \rho _{0} \( x \)$.
On the other hand,
\autoref{eq:appendix__equivalence__VD__V} can be derived from \autoref{eq:appendix__equivalence__ZV__V} and \autoref{eq:appendix__equivalence__ZV__D};
\autoref{eq:appendix__equivalence__VD__D} can be derived from \autoref{eq:appendix__equivalence__motionAndDensity}, \autoref{eq:appendix__equivalence__ZV__D}, and the assumption $\rho \( x, 0 \) = \rho _{0} \( x \)$,
where
\begin{align}
  \partial _{t} \[ \rho \( x, 0 \) \[ \partial _{x} z \( x, t \) \] ^{-1} \]
  =
  \rho \( x, 0 \) \partial _{t} \[ J \( x, t \) \] ^{-1}
  =
  - \rho \( x, 0 \)
  \[ J \( x, t \) \] ^{-2}
  \partial _{x} v \( x, t \).
\end{align}
This concludes the proof of the model equivalence.

\section{Function Derivative in SPH}

\label{sec:appendix_derivative}

We consider the SPH approximation of the spatial derivative on a uniform grid with spacing $\Delta x$. 
We set the SPH smoothing length as $h = \alpha \Delta x$, using the 1D cubic spline kernel function $W _{h}$ defined in \autoref{eq:model__SPH__cubicspline}.

When $2 < 2 \alpha < 3$, there are exactly two neighboring particles on each side of a given particle. Thus, we have:
\begin{align}
  & \quad
  \lsbsb \nabla f \rsbsb \( \mathbf{x} _{i}, t \)
  =
  \nabla \lsbsb f \rsbsb \( \mathbf{x} _{i}, t \)
  \nonumber \\
  & =
  \sum _{j \sim i}
  \frac{m _{j}}{\rho _{j} \( 0 \)}
  f _{j} \( t \)
  \nabla W _{h} \( \mathbf{x} _{ij} \)
  \\
  & =
  a ^{2} \Delta x \[ \[ f _{i + 1} - f _{i - 1} \] \nabla W _{h} \( - \Delta x \) + \[ f _{i + 2} - f _{i - 2} \] \nabla W _{h} \( - 2 \Delta x \) \]
  \label{eq:appendix_derivative_term}
  \\
  & \approx
  a ^{2} \Delta x \[ 2 \[ f _{i}' \Delta x + f _{i}''' \[ \Delta x \] ^{3} / 6 \] \nabla W _{h} \( - \Delta x \) \] 
  \nonumber \\
  & \quad + 
  a ^{2} \Delta x \[ 2 \[ 2 f _{i}' \Delta x + 8 f _{i}''' \[ \Delta x \] ^{3} / 6 \] \nabla W _{h} \( - 2 \Delta x \) \]
  +
  \bigO{\[ \Delta x \] ^{4}}
  \\
  & \approx
  \[ 8 \alpha ^{-2} - 12 \alpha ^{-3} + 5 \alpha ^{-4} \] f _{i}'
  +
  \[ \frac{16}{3} \alpha ^{-2} - 10 \alpha ^{-3} + \frac{29}{6} \alpha ^{-4} \] f _{i}''' \[ \Delta x \] ^{2}
  + 
  \bigO{\[ \Delta x \] ^{4}}
  \\
  & \approx
  \left\{
  \begin{aligned}
    1.003 f _{i}'
    +
    0.173 f _{i}''' \[ \Delta x \] ^{2}
    + 
    \bigO{\[ \Delta x \] ^{4}},
    &
    \qquad
    \alpha = 1.04,
    \\
    1.022 f _{i}'
    +
    0.248 f _{i}''' \[ \Delta x \] ^{2}
    + 
    \bigO{\[ \Delta x \] ^{4}},
    &
    \qquad
    \alpha = 1.2.
  \end{aligned}
  \right.
\end{align}
Here we list two different cases: $\alpha = 1.2$ that is widely used in many other reference such as \cite{fraser2017robust}, and $\alpha = 1.04$ that is used in our main text.
The above calculation means that if we solve a PDE $\partial _{t} f = a \partial _{x} f$ using Lagrangian SPH, we effectively implement:
\begin{align}
  \partial _{t} f _{i} = a \lsbsb \nabla f \rsbsb _{i} \approx a \[ \[ 1 + \epsilon \] f _{i}' + \tilde{\epsilon} f _{i}''' \[ \Delta x \] ^{2} \]
  .
\end{align}
The dispersion term $\tilde{\epsilon} f _{i}''' [ \Delta x ] ^{2}$ vanishes as $\Delta x \to 0$,
however, the coefficient $\epsilon$ in the first term is independent of $\Delta x$, indicating that the inconsistency in the derivative term cannot be improved by simply reducing the spatial spacing. 
We find out in our setup with $\alpha = 1.04$, the error from the term $\lsbsb \nabla f \rsbsb _{i}$ is reduced compared with the case $\alpha = 1.2$.
Another approach to reduce such error is to include more neighbors, where more terms will appear in \autoref{eq:appendix_derivative_term}. Since the optimal SPH setup is not our focus in this paper, we just fix $\alpha = 1.04$ in most of our numerical examples.

\section{More Numerical Results}

\subsection{Gaussian Initial Velocity}
\label{sec:numerical__SPH_PBC_test29}

In this example, we investigate a case with a Gaussian-type initial velocity
\begin{align}
  v _{0} \( x \) = 1 \e{-3} \exp \( - 200 \( \frac{x}{L} - \frac{1}{2} \) ^{2} \),
  \label{eq:numerical__Lagrangian_PBC_test29_V_initial}
\end{align}
with the initial density uniformly set as $\rho _{0} = 1$,
the damping coefficient defined as $d = 0$,
the Young's modulus specified as $c = 0.1$, 
and the external force expressed as $f \( t \) = 0$.

\begin{figure}[!htp]
  \centering
  \includegraphics[width = 0.48\textwidth]{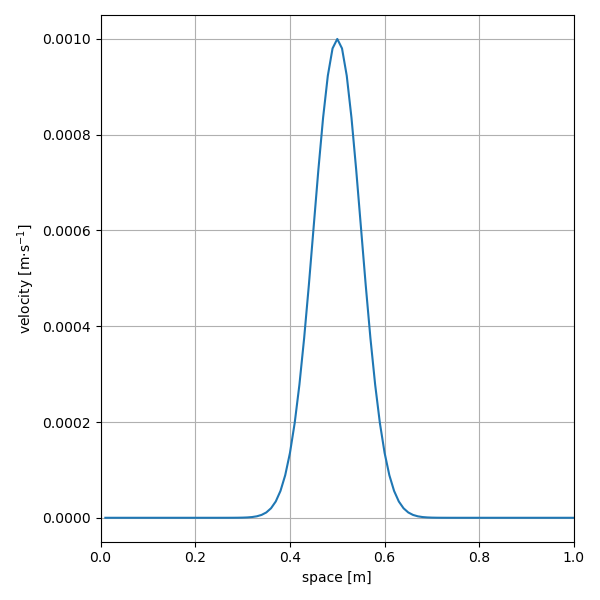}
  \includegraphics[width = 0.48\textwidth]{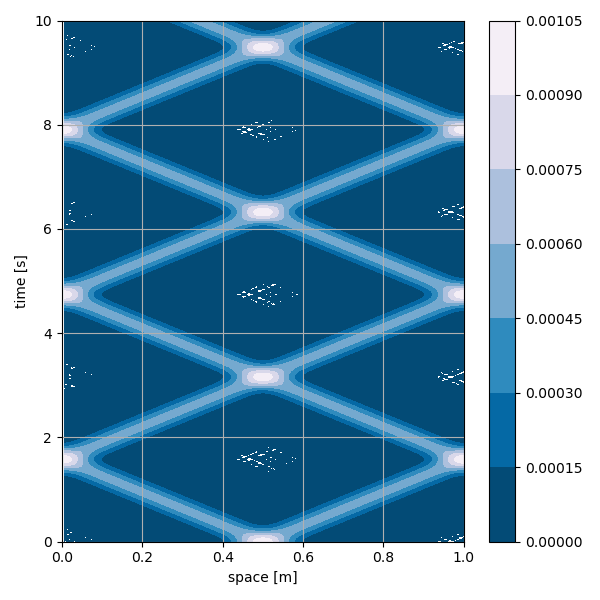}
  \caption{Initial condition (left figure) and profile (right figure) of the reference velocity.}
  \label{fig:numerical__SPH_PBC_test29_POD_V_profile}
\end{figure}

An analytic solution is available through the Fourier transform of the initial velocity. 
We use this analytic solution as the reference for this example, where $v _{\textup{ref}} = v _{\textup{analytic}}$ and $\rho _{\textup{ref}} = \rho _{\textup{analytic}}$. The initial condition and the solution profile of the velocity $v \( x, t \)$ from the analytic solution are plotted in \autoref{fig:numerical__SPH_PBC_test29_POD_V_profile}.

The SPH simulation evolves according to the equations \autoref{eq:model__SPH__V} and \autoref{eq:model__SPH__D}. With a spatial step size of $\Delta x = 1 \e{-2}$ ( corresponding to $N = 100$ smoothed particles) and a SPH smoothing length of $h = 1.06 \Delta x$, we employ the explicit Euler scheme for time integration, using a temporal step size of $\Delta t = 2 \e{-4}$. The relative $L ^{2}$ error in velocity is approximately $1.926 \e{-1}$, while the relative $L ^{2}$ error in density is about $1.416 \e{-4}$.

\begin{figure}[!htp]
  \centering
  \includegraphics[width = 0.48\textwidth]{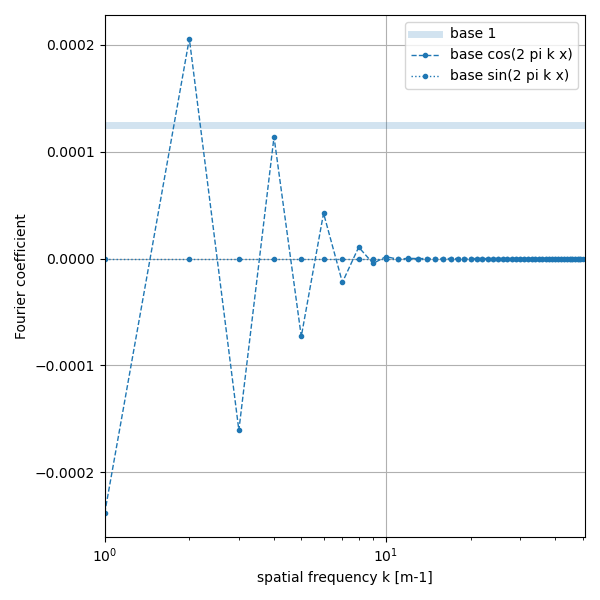}
  \caption{Coefficients of discrete Fourier modes of the initial condition of the reference velocity.}
  \label{fig:numerical__SPH_PBC_test29_POD_V_initialFFT}
\end{figure}

This initial velocity features more Fourier modes than the case in \autoref{sec:numerical__SPH_PBC_test28}, as illustrated in \autoref{fig:numerical__SPH_PBC_test29_POD_V_initialFFT}, particularly including several high-frequency modes. This abundance of modes results in a larger numerical error in the SPH simulation, even with the same discretization. 
Based on these observations, we refine the discretization in the SPH simulation, by setting the spatial step size to $\Delta x = 5 \e{-3}$ (corresponding to $N = 200$ smoothed particles) and the SPH smoothing length to $h = 1.04 \Delta x$. We use the explicit Euler scheme for time integration with a temporal step size of $\Delta t = 1 \e{-4}$. Under these conditions, the relative $L ^{2}$ error in velocity is about $4.838 \e{-2}$, and the relative $L ^{2}$ error in density is approximately $3.695 \e{-5}$.

\begin{figure}[!htp]
  \centering
  \includegraphics[width = 0.8\textwidth]{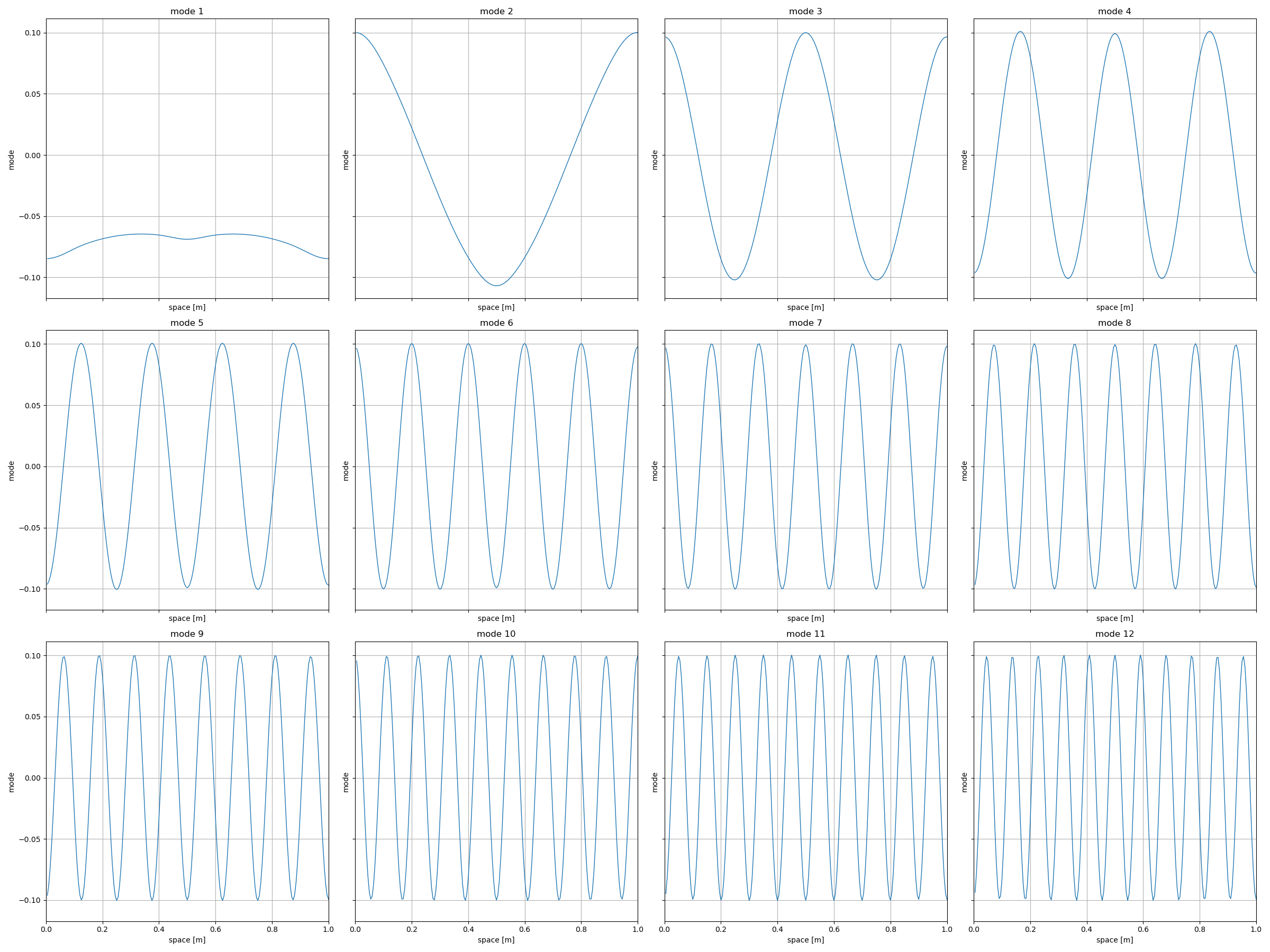}
  \includegraphics[width = 0.8\textwidth]{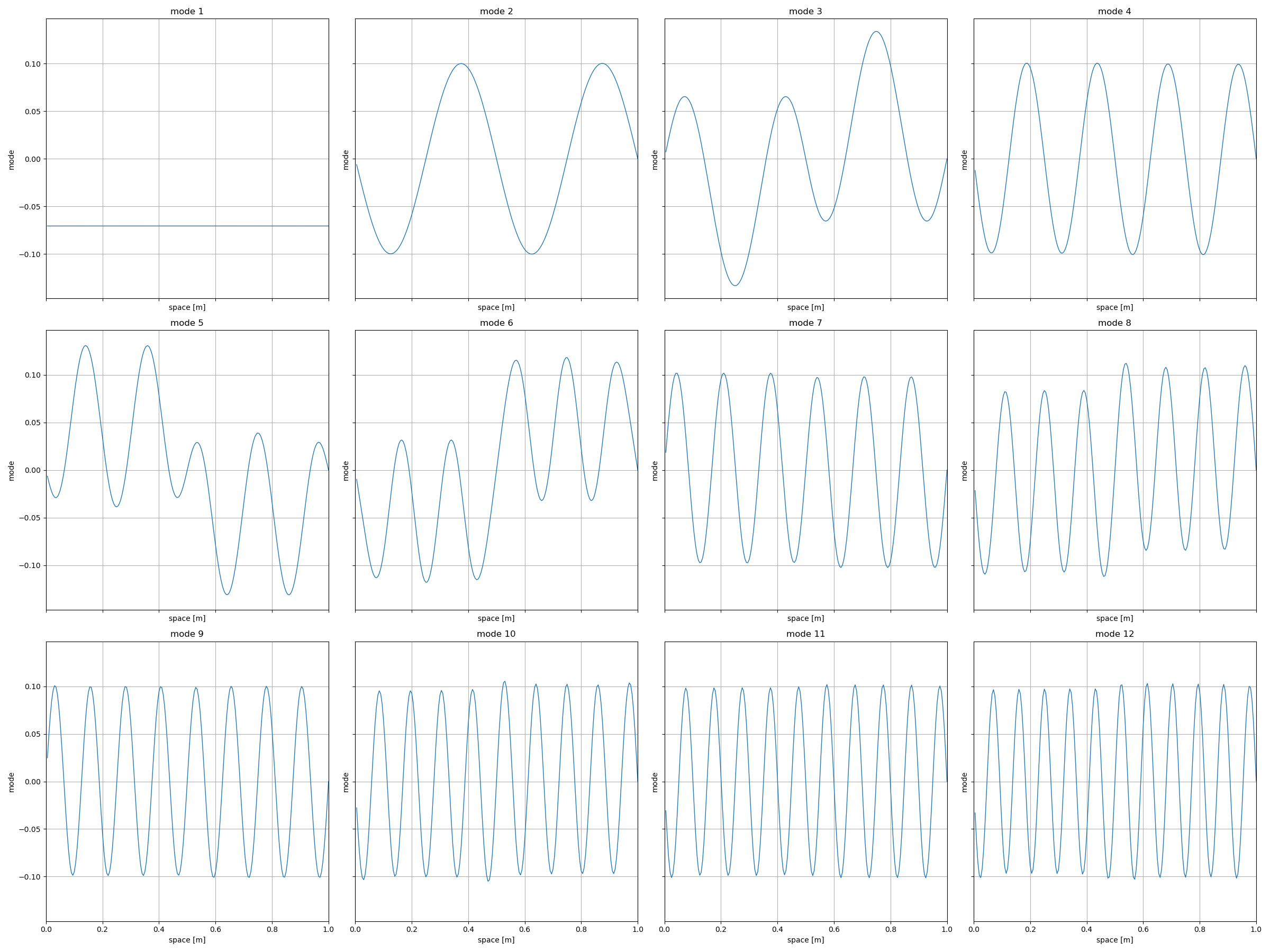}
  \caption{The dominant POD modes of velocity (upper figure) and density (lower figure).}
  \label{fig:numerical__SPH_PBC_test29_POD_nx200_nt10k_nn1d04_PodDataSph100First160m_modes}
\end{figure}

\begin{figure}[!htp]
  \centering
  \includegraphics[width = 0.48\textwidth]{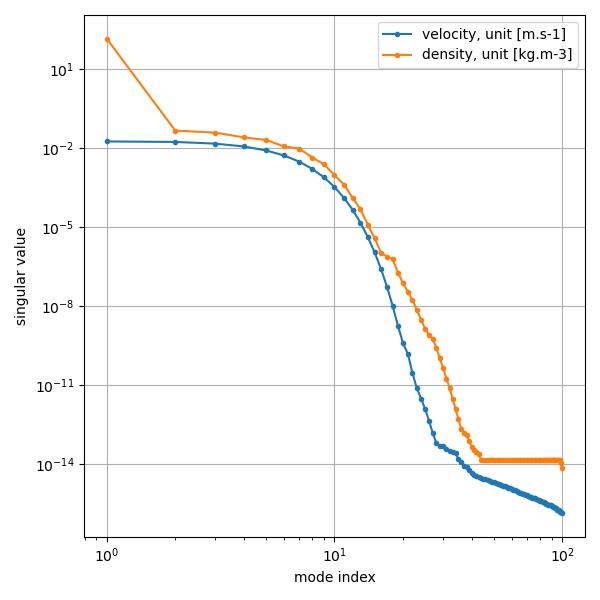}
  \includegraphics[width = 0.48\textwidth]{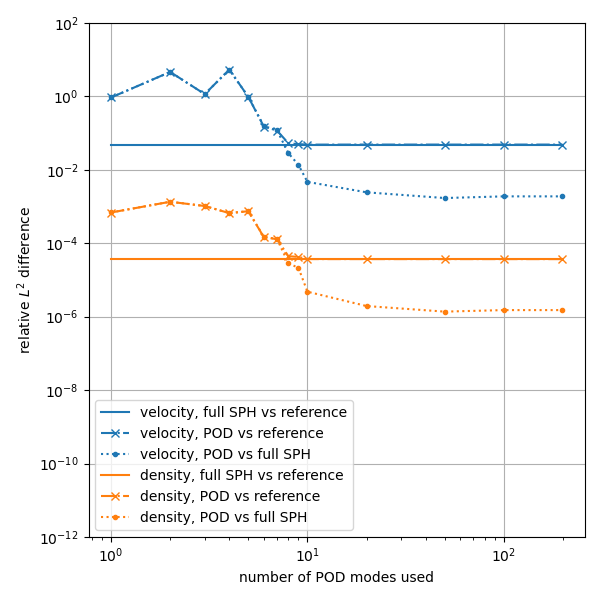}
  \caption{The singular value of the POD modes (left figure) and the relative $L ^{2}$ error (right figure) among the reference solution, the full SPH solution, and the POD-MOR of SPH simulation with different number of POD modes used, where
  $100$ snapshots are uniformly collected from $t \in \[ 0, 1.6 \]$ for the POD data.
}
  \label{fig:numerical__SPH_PBC_test29_POD_nx200_nt10k_nn1d04_PodDataSph100First160m_error}
\end{figure}

Next, we apply the POD-MOR to the SPH simulation, collecting a total of $100$ snapshots uniformly from $t \in \[ 0, 1.6 \]$ of the SPH solution. We plot the dominant POD modes of velocity and density in \autoref{fig:numerical__SPH_PBC_test29_POD_nx200_nt10k_nn1d04_PodDataSph100First160m_modes}, and display the singular values of the POD modes along with the relative $L ^{2}$ error of the original POD-MOR as shown in \autoref{fig:numerical__SPH_PBC_test29_POD_nx200_nt10k_nn1d04_PodDataSph100First160m_error}. The POD error becomes smaller than the SPH numerical error when $8$ POD modes are utilized. A much smaller error is observed in POD-MOR when at least $10$ modes are included. This observation aligns with the results shown in \autoref{fig:numerical__SPH_PBC_test29_POD_V_initialFFT}, where more than $8$ modes are dominant.

\subsection{High-Frequency Initial Velocity}

\label{sec:numerical__SPH_PBC_test25}

In this example, we will verify the analysis in \autoref{sec:appendix_derivative} and test the choice of $\alpha = 1.2$ in the smoothing length. We consider the case of high-frequency initial velocity, which is given by
\begin{align}
  v _{0} \( x \) = 1 \e{-3} \sin \( 20 \pi \frac{x}{L} \),
\end{align}
with the initial density uniformly set as $\rho _{0} = 1$,
the damping coefficient defined as $d = 0$,
the Young's modulus specified as $c = 0.1$, 
and the external force expressed as $f \( t \) = 0$.

\begin{figure}[!htp]
  \begin{center}
    \includegraphics[width = 0.48\textwidth]{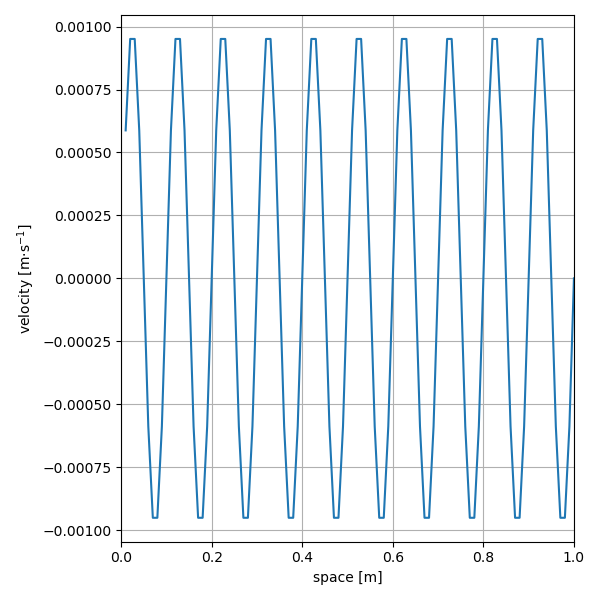}
    \includegraphics[width = 0.48\textwidth]{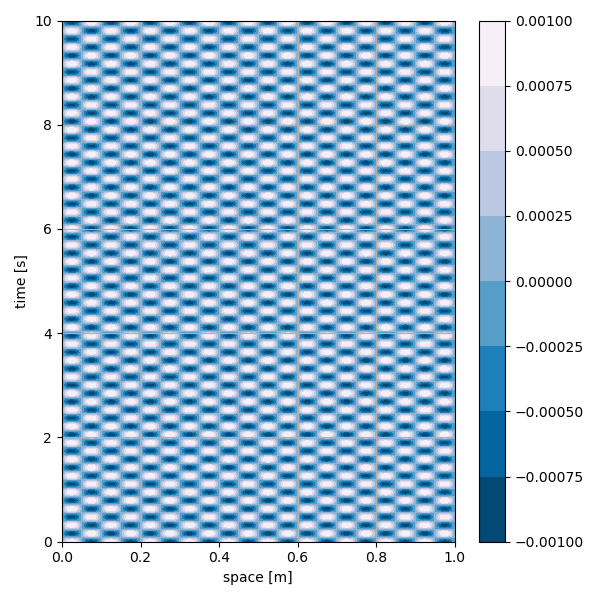}
  \end{center}
  \caption{Initial condition and profile of the analytic velocity.}
  \label{fig:numerical__SPH_PBC_test25_POD_V_profile}
\end{figure}

In this scenario, an analytic solution is available, given by
\begin{align}
  v _{\textup{analytic}} \( x, t \) 
  & = 
  v _{0} \( x \) \cos \( 20 \pi \frac{t}{L} \sqrt{\frac{c}{\rho _{0}}} \)
  ,
  \\
  \rho _{\textup{analytic}} \( x, t \) 
  & =
  \frac{\rho _{0}}{1 + \partial _{x} \int _{0} ^{t} v _{\textup{analytic}} \( x, s \) \mathrm{d} s}
  .
\end{align}
We use this analytic solution as the reference, where $v _{\textup{ref}} = v _{\textup{analytic}}$ and $\rho _{\textup{ref}} = \rho _{\textup{analytic}}$. The initial condition and the solution profile of velocity $v \( x, t \)$ from the analytic solution are displayed in \autoref{fig:numerical__SPH_PBC_test25_POD_V_profile}.

From the reference solution, high frequencies are evident in both the temporal and spatial domains, necessitating a high-resolution mesh grid for the numerical solution. In the numerical setup, the spatial step size is set as $\Delta x = 1 \e{-3}$ (corresponding to $N = 1000$ smoothed particles) with a SPH smoothing length of $h = 1.2 \Delta x$ (i.e., $\alpha = 1.2$ here). The explicit Euler scheme is employed for time integration, using a temporal step size of $\Delta t = 1 \e{-5}$. However, the relative $L ^{2}$ error in velocity is $1.566$, which is very large. This error arises from the inaccuracies in the SPH approximation of derivatives (see \autoref{sec:appendix_derivative}).

\begin{figure}[!htp]
  \begin{center}
    \includegraphics[width = 1\textwidth]{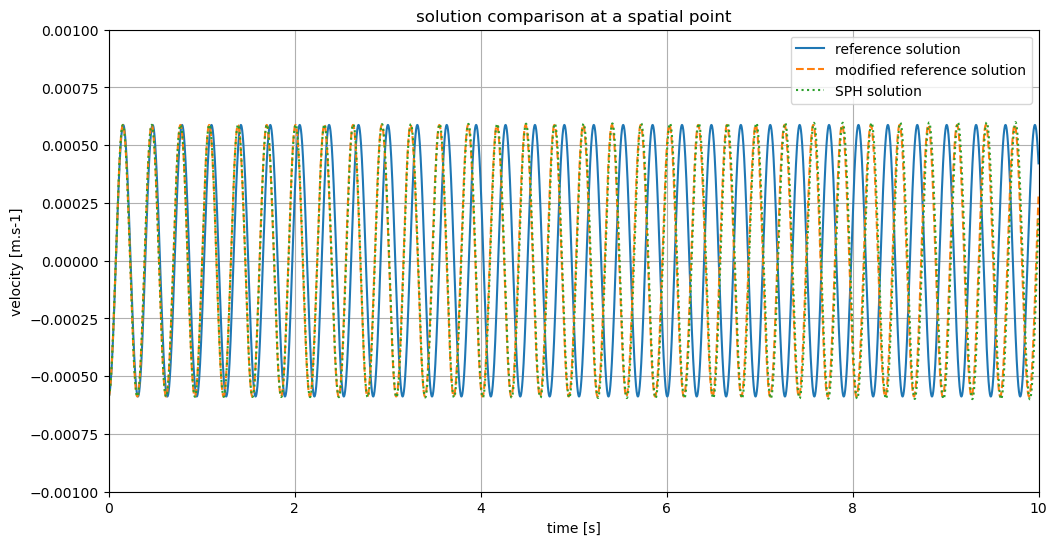}
  \end{center}
  \caption{Velocity of the reference solution and SPH solution at spatial position $0.66L$.}
  \label{fig:numerical__Lagrangian_PBC_test25_V_compare}
\end{figure}

To verify the derived factor in the inconsistency analysis in \autoref{sec:appendix_derivative}, we construct a modified reference solution given by
\begin{align}
  v _{\textup{modified}} \( x, t \) 
  & = 
  v _{0} \( x \) \cos \( 20 \pi \frac{t}{L} \sqrt{\frac{c}{\rho _{0}}} \beta \)
  ,
\end{align}
where $\beta \approx 1.022$ is the coefficient associated with the leading term in the SPH approximation of derivatives (see \autoref{sec:appendix_derivative}). By comparing the SPH solution with this modified reference solution, the relative $L ^{2}$ error in velocity is about $1.139 \e{-1}$. We plot the velocity profile $v \( 0.66, t \)$ for both the reference and SPH solutions in \autoref{fig:numerical__Lagrangian_PBC_test25_V_compare}. Although there is a minor difference in phase between the SPH solution and the modified reference solution, the pointwise errors are larger for the longer time due to the large derivative of the function. Overall, the first few oscillations of the SPH solution matches with the modified reference solution, and thus verifies the analysis in \autoref{sec:appendix_derivative}.

\subsection{Mild Damping and External Force}
\label{sec:numerical__SPH_PBC_test21}

In this example, we examine the case with mild damping and external force. The initial velocity is given by \autoref{eq:numerical__Lagrangian_PBC_test23_V_initial}. The initial density is uniformly set as $\rho _{0} = 1$, the damping coefficient is $d = 1$, the Young's modulus is $c = 0.1$, and the external force is defined as
\begin{align}
  f \( t \) = 1 \e{-3} \sum _{l = 1, 3, 10} \cos \( 10 l \frac{t}{T} - l ^{3} \) .
\end{align}

\begin{figure}[!htp]
  \centering
  \includegraphics[width = 0.48\textwidth]{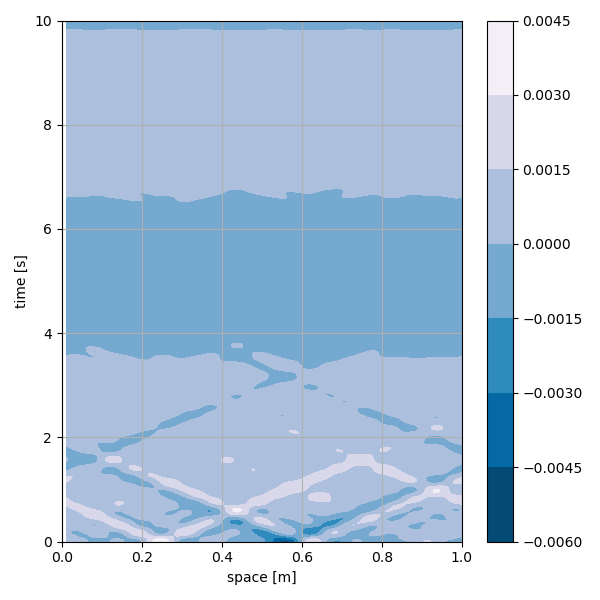}
  \includegraphics[width = 0.48\textwidth]{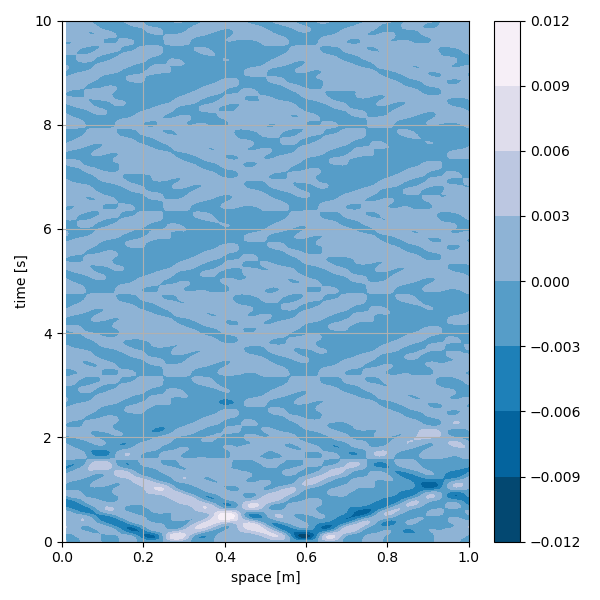}
  \caption{Velocity $v$ (left figure) and density $\rho - \rho _{0}$ (right figure) profile of the reference solution.}
  \label{fig:numerical__SPH_PBC_test21_POD_profile}
\end{figure}

The reference solution for 
\autoref{eq:model__spring__VD__V} and \autoref{eq:model__spring__VD__D} is generated using a first-order finite difference method in space with $\Delta x = 2.5 \e{-3}$ and a second-order Runge-Kutta scheme in time with $\Delta t = 5 \e{-5}$. The reference solutions for velocity and density are shown in \autoref{fig:numerical__SPH_PBC_test21_POD_profile}. The SPH simulation evolves according to equations \autoref{eq:model__SPH__V} and \autoref{eq:model__SPH__D}. With a spatial step size of $\Delta x = 5 \e{-3}$ (i.e., $N = 200$ smoothed particles) and a SPH smoothing length of $h = 1.04 \Delta x$, we use the explicit Euler scheme for time integration with a temporal step size of $\Delta t = 1 \e{-4} \[ \mykgms{0}{0}{1} \]$. The relative $L ^{2}$ error in velocity is about $7.586 \e{-2}$, while the relative $L ^{2}$ error in density is approximately $1.698 \e{-4}$.

\begin{figure}[!htp]
  \centering
  \includegraphics[width = 0.48\textwidth]{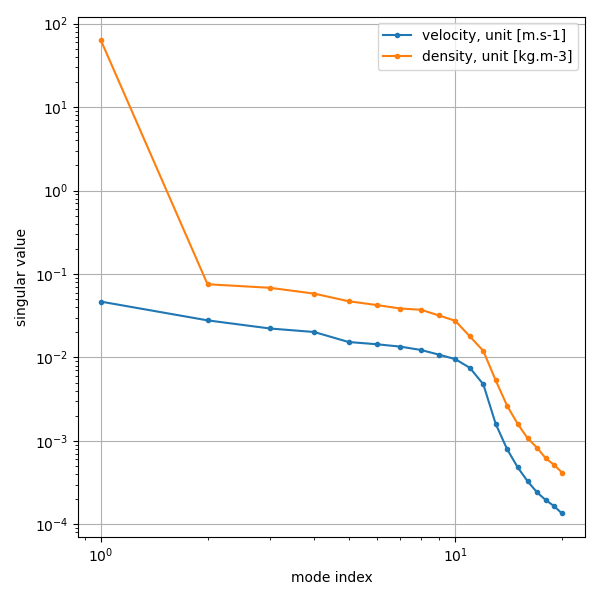}
  \includegraphics[width = 0.48\textwidth]{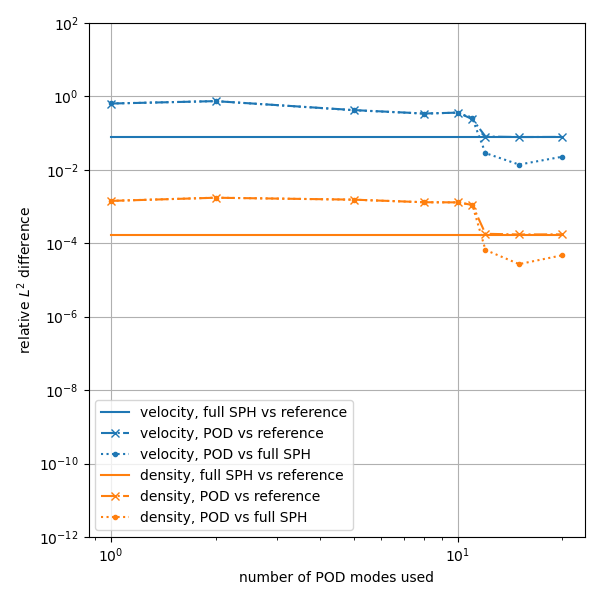}
  \caption{The singular value of the POD modes (left figure) and the relative $L ^{2}$ error (right figure) among the reference solution, the full SPH solution, and the POD-MOR of SPH simulation with different number of POD modes used, where
  $20$ snapshots are uniformly collected from $t \in \[ 0, 1.6 \]$ for the POD data.
  }
  \label{fig:numerical__SPH_PBC_test21_POD_nx200_nt10k_nn1d04_PodDataSph20First160m_error}
\end{figure}

We apply the POD-MOR to the SPH simulation, by collecting a total of $20$ snapshots uniformly from $t \in \[ 0, 1.6 \]$.
The singular values of the POD modes and the relative $L ^{2}$ error of the original POD-MOR \autoref{eq:model__POD__general} are presented in \autoref{fig:numerical__SPH_PBC_test21_POD_nx200_nt10k_nn1d04_PodDataSph20First160m_error}. The POD error is smaller than the SPH numerical error when using at least $12$ modes.

\begin{figure}[!htp]
  \centering
  \includegraphics[width = 0.48\textwidth]{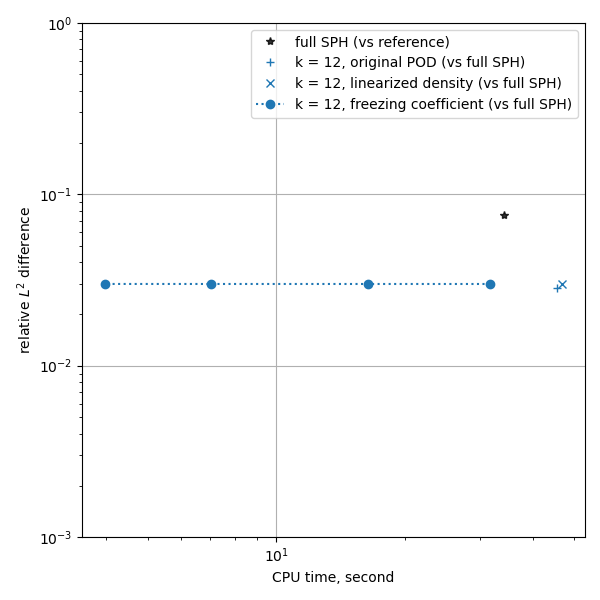}
  \includegraphics[width = 0.48\textwidth]{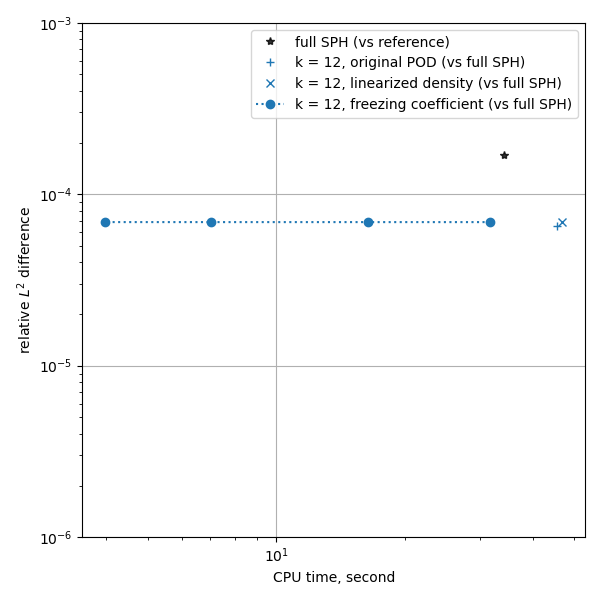}
  \caption{The relative $L ^{2}$ error of velocity $v$ (left figures) and density $\rho$ (right figures) as well as the CPU time (in seconds).
  The dots (connected with dashed lines) represent the $L ^{2}$ error and CPU time of accelerated POD-MOR with freezing coefficients compared to the full SPH solution, where the freezing step is chosen from $n _{\textup{freeze}} =1, 2, 5, 10$, plotted from right to the left.  
  }
  \label{fig:numerical__SPH_PBC_test21_POD_nx200_nt10k_nn1d04_PodDataSph20First160m_acceleration}
\end{figure}

An accelerated version of the POD-MOR \autoref{eq:model__acceleration__V} and \autoref{eq:model__acceleration__D} is implemented, where the freezing step is chosen from $n _{\textup{freeze}} = 1, 2, 5, 10$, and the number of POD modes used is $k = 12$. In \autoref{fig:numerical__SPH_PBC_test21_POD_nx200_nt10k_nn1d04_PodDataSph20First160m_acceleration}, we compare the CPU time and error of different numerical solutions, including the full SPH simulation, the original POD-MOR, the
POD-MOR with density linearization, and the POD-MOR with freezing coefficients. 
When choosing $n _{\textup{freeze}} = 10$, the POD error remains less than the numerical error of the SPH simulation, while we observe a significant speed-up, saving up to $88\%$ of CPU time.

\end{document}